\documentclass[oneside]{sapthesis}

\usepackage{microtype}
\usepackage[english]{babel}
\usepackage[utf8]{inputenx}
\usepackage{graphicx}
\usepackage{skmath}
\usepackage{amsmath,amssymb}
\usepackage{xparse}
\usepackage{wrapfig}
\usepackage{float}
\usepackage{sidecap}
\usepackage{xcolor}
\usepackage{gensymb}
\usepackage{caption}
\usepackage{tabularx}
\usepackage{csvsimple}
\usepackage{subfig}
\usepackage{placeins}
\usepackage{xfrac}
\usepackage[makeroom]{cancel}
\usepackage{verbatim}
\usepackage{longtable}
\usepackage{eucal}
\usepackage{braket}
\usepackage{bbold}
\usepackage{empheq}

\usepackage[numbers]{natbib}

\usepackage{lipsum}
\usepackage{curve2e}

\usepackage{hyperref}
\hypersetup{
    colorlinks=true,
    linkcolor=red,
    citecolor=red,
    urlcolor=blue,
    linktoc=page
}

\hypersetup{
    pdftitle={Beyond Quantum Mechanics: From Nonlinear to Emergent and Solitonic Approaches},
    pdfauthor={Niko Brimi}
}

\providecommand{\doi}[1]{%
    \href{https://doi.org/#1}{doi:\nolinkurl{#1}}%
}

\definecolor{gray}{gray}{0.4}

\title{Beyond Quantum Mechanics: From Nonlinear to Emergent and Solitonic Approaches}
\author{Niko Brimi}
\IDnumber{1850628}
\course{Corso di Laurea Magistrale in Fisica}
\courseorganizer{Facoltà di Scienze Matematiche, Fisiche e Naturali}
\AcademicYear{2025/2026}
\copyyear{2025/2026}
\advisor{prof. Claudio Conti}
\coadvisor{}
\authoremail{brimi.1850628@studenti.uniroma1.it}

\examdate{17 November 2026}

\examiner{Prof. Raffaella Schneider}
\examiner{Prof. Paolo Pani}
\examiner{Prof. Giacomo Artoni}
\examiner{Prof. Adele La Rana}
\examiner{Prof. Valeria Cimini}
\examiner{Prof. Roberta Angelini}
\examiner{Prof. Andrea Ninarell}

\begin{document}

\frontmatter

\maketitle

\dedication{%
In memory of my teacher, Ferruccio Alfonsi.\\
These pages would not have been written without him.\\[1.5em]
\textit{“There is more elegance hidden\\
within a question mark\\
than behind any answer.”}%
}

\newpage
\null
\thispagestyle{empty}
\newpage

\begin{abstract}
In the past century, there have been many efforts to gain a deeper understanding of Quantum Mechanics (QM). This has generated considerable debate within the scientific community. The Copenhagen interpretation is widely regarded as the standard view of QM; however, it has also given rise to significant criticism, particularly among proponents of deterministic approaches. Within this framework, the formalism of QM is considered fundamentally complete, without requiring a detailed account of the underlying mechanisms responsible for quantum phenomena. A variety of alternative interpretations have been proposed in order to provide a more realistic account of quantum phenomena. Many of these approaches have led to conflicting perspectives and remain the subject of ongoing debate.

Among them, one of the most prominent is the Pilot Wave Theory (PWT), originally introduced by Louis de Broglie and later developed by David Bohm, now commonly discussed under the name of Bohmian Mechanics (BM). Although many alternative approaches reproduce the predictions of standard QM, their underlying descriptions can suggest possibilities beyond a purely interpretative reformulation. In particular, the PWT reproduces the usual quantum statistics in quantum equilibrium, while its more general formulation allows for nonequilibrium distributions. The PWT will therefore play a central role in this work, as it offers insights that go beyond its standard interpretative use.

The aim of this work is to explore several theoretical approaches proposed in the past and to use them as a basis for developing a new perspective on QM. In particular, the objective is not to assume QM as a fundamental theory, but rather to investigate whether it can emerge from deeper underlying phenomena. This line of thought has already been explored and criticized in the literature; however, we aim to reassess its viability as a direction for further investigation, shifting the focus from extensions of the quantum formalism to the question of whether QM itself can emerge from deeper underlying dynamics. We consider both the nonlinear formulation of QM proposed by Steven Weinberg and BM as complementary approaches to explore possible extensions beyond standard QM. The former introduces nonlinear effects within a perturbative framework, while the latter provides a dynamical description that offers insight into the underlying structure of quantum phenomena.

We then introduce solitons as candidates for a more fundamental description beyond the notion of point-like particles. In many physical systems, solitons appear as localized wave structures that exhibit particle-like behavior, making them a natural starting point for such an investigation. However, in order to address the problem consistently, the underlying dynamics must be treated as a deeper level of description, from which the observed probabilistic features of QM may emerge as an effective theory. To investigate this possibility, we consider ensembles of solitons evolving in a noisy background and examine whether, and under which conditions, their collective statistics can reproduce the evolution described by the Schr\"odinger equation. This provides a concrete setting for exploring the possible emergence of quantum behavior from an underlying nonlinear field dynamics.
\end{abstract}

    

\tableofcontents


\newpage
\null
\thispagestyle{empty}
\newpage

\mainmatter

\chapter*{Introduction}

Quantum Mechanics (QM) lies at the foundation of modern physics; however, its conceptual interpretation remains a subject of ongoing debate. The standard Copenhagen interpretation assumes the formalism of QM as fundamentally complete and describes particles through a wave function that encodes probabilities of different outcomes, which appear as definite only upon measurement.
Nevertheless, the ontological nature of the quantum particle is not straightforward. Unlike classical point-like objects, quantum entities exhibit properties that cannot be fully captured within a classical framework.
Although the theory is empirically successful, a deeper understanding of the origin of these properties would be desirable, as it may provide the basis for developing more fundamental theoretical frameworks.

The motivation for exploring alternative theoretical frameworks for QM is closely related to the ongoing effort to deepen our understanding of physical reality. In recent years, it has been argued that the limitations in our understanding may not be merely epistemic, i.e. due to experimental constraints, but could instead reflect more fundamental restrictions on how reality can be described. In this view, certain aspects of physical phenomena may not be fully accessible within the standard formalism. Nevertheless, it is often assumed that physical processes are governed by well-defined underlying laws, even if these laws are not directly observable at the macroscopic level. The aim of this thesis is to investigate this perspective by examining previous theoretical approaches that attempt to go beyond the standard formulation of QM, in order to assess whether such a viewpoint can be considered a meaningful direction for further research.

A particularly relevant possibility is to describe particle-like objects as localized configurations of an underlying field. Solitons provide a concrete setting for exploring this idea, and their properties and interactions have been extensively investigated in nonlinear field theories~\cite{MantonSutcliffe_2004,Weinberg_ClassicalSolutions_2012}. Their possible interpretation as elementary particles has been examined both in general discussions and in specific models~\cite{Manton_Particles_2008,Yamamoto_Spinor_1977}. Their physical relevance is also supported by experimental observations, including the recent observation of lump solitons in nonlinear optics~\cite{Dieli_Lump_2026}.

Related questions arise in gravitational physics. Wheeler's geons provide a historical example of an attempt to describe localized bodies through self-gravitating field configurations~\cite{Wheeler_Geons_1955}. More recent investigations include fermion soliton stars and their generalization to scalar potentials with asymmetric vacua~\cite{DelGrosso_FermionStars_2023,DelGrossoPani_Asymmetric_2023}. Solitonic descriptions have also been applied to dark matter condensates subject to the gravitational field of a black hole, with possible laboratory analogs~\cite{DieliConti_DarkMatter_2024}. These studies illustrate the range of physical settings in which localized field configurations can acquire an effective identity as distinct objects. They motivate the more specific question addressed here: whether the collective statistics of such objects can reproduce quantum dynamics.

In our work, in Chapter \textcolor{red}{\ref{chap:1}} we begin by analyzing Weinberg's nonlinear extension of QM as a representative attempt to go beyond the standard formalism while preserving its core assumptions. We then argue that this approach, however refined, remains within a framework in which QM is taken as fundamentally given. This motivates a shift in perspective: rather than extending QM, we investigate whether it can instead be derived from deeper classical-like dynamics.

We then further develop this change of viewpoint, motivated by the fact that the wave function is typically interpreted as a probabilistic object. In particular, we explore the possibility that it may instead represent the statistical behavior of an underlying deterministic dynamics. For this purpose, in Chapter \textcolor{red}{\ref{chap:2}} we introduce the Pilot Wave Theory (PWT), which provides a dynamical framework underlying quantum phenomena, and we use it as a conceptual basis for developing our approach to a more fundamental theory~\cite{Valentini_BeyondQuantum_2025,Valentini_Trouble_2025}. In particular, we examine the ensemble formulation of classical and quantum dynamics, which provides the language needed to compare the PWT with a statistical description of solitonic motion.

In Chapter \textcolor{red}{\ref{chap:3}}, we investigate the behavior of solitons in a fluctuating background, with the aim of understanding whether their collective dynamics can give rise to the statistical evolution described by the Schr\"odinger Equation (SE). We first introduce the Hall--Reginatto exact uncertainty approach~\cite{Hall_Reginatto}, which provides a framework for relating momentum fluctuations to quantum ensemble dynamics. Building on Gomila's work~\cite{Gomila_solitons}, we then consider a Galilean-invariant nonlinear field model in which the soliton and the noisy background evolve as parts of the same deterministic field. We examine the resulting position and momentum statistics and identify the conditions under which the collective description reproduces the SE. The correspondence is established within the Gaussian sector and for effective potentials at most quadratic, providing a concrete setting in which to explore the emergence of quantum statistical behavior from an underlying nonlinear dynamics.
\newpage
\null
\thispagestyle{empty}
\newpage

\chapter{Nonlinear Quantum Mechanics}
\label{chap:1}

In this chapter we discuss the work of Steven Weinberg \textcolor{red}{\cite{Weinberg1989}} on the possibility of introducing nonlinear corrections to standard QM. The motivation is straightforward: QM is a linear theory, while in other wave-based theories, such as optics, linearity typically arises only as an approximation valid in a specific regime (e.g. low intensities), whereas nonlinear effects become relevant at higher intensities. This suggests that linearity may not be fundamental, but rather an emergent feature of a more general framework. This leads to the question addressed by Weinberg: whether QM, being also a wave theory, could similarly be understood as an approximation of a more general nonlinear theory, or whether its linear structure is instead fundamental. It is important to note that Weinberg’s approach is not primarily concerned with the ontology of QM. Rather, he assumes the standard formalism to be fundamentally correct and investigates possible nonlinear generalizations in a purely theoretical and mathematical manner. We emphasize that his work is mainly technical rather than interpretative in nature, providing a mathematically elegant and systematic generalization of QM in a nonlinear framework.

\section{Quantum Mechanics in the Bilinear Formalism}
In order to describe the theory, we first introduce QM in an alternative formalism with respect to the standard operator-based approach. In the conventional formulation, QM is defined within a Hilbert space $\mathcal{H}$, a generic vector space equipped with an inner product, which induces its geometric and probabilistic structure, where physical observables are represented by operators characterized by their eigenstates. The state of a system is described by a wave function, which encodes a probability distribution over these eigenstates, and the expectation values of the corresponding operators define the measurable physical quantities. It is important to stress that the Hilbert space structure does not uniquely determine a preferred basis; rather, different choices of orthonormal bases correspond to different representations of the same physical states, typically selected according to the observables of interest.
Then, the starting point is the SE, written in an abstract mathematical form:
\begin{equation}
i \frac{d}{dt} \ket{\psi(t)} = \hat{H} \ket{\psi(t)}
\label{eq:abstract_SE}
\end{equation}
where we set $\hbar = 1$, $\hat{H}$ is the Hamiltonian operator, and $\ket{\psi(t)} \in \mathcal{H}$  is a generic state vector in the Hilbert space. 
\subsection{From Abstract Operators to Explicit Representations}
At this stage, the formulation is completely basis-independent. Without loss of generality, one can choose a convenient orthonormal basis of $\mathcal{H}$ with respect to which the state and the equation itself can be represented in component form. 
Considering a finite-dimensional Hilbert space 
$\mathcal{H} = (\mathbb{C}^N, \langle \cdot , \cdot \rangle)$, 
endowed with an abstract inner product
\begin{equation*}
\langle \cdot , \cdot \rangle : \mathcal{H} \times \mathcal{H} \to \mathbb{C},
\end{equation*}
satisfying, for all $|\psi\rangle, |\phi\rangle, |\chi\rangle \in \mathcal{H}$ 
and $\alpha, \beta \in \mathbb{C}$,
\begin{align*}
&\langle \psi \mid \alpha \phi + \beta \chi \rangle 
= \alpha \langle \psi \mid \phi \rangle 
+ \beta \langle \psi \mid \chi \rangle, \\
&\langle \psi \mid \phi \rangle 
= \langle \phi \mid \psi \rangle^*, \\
&\langle \psi \mid \psi \rangle \geq 0, \quad 
\langle \psi \mid \psi \rangle = 0 \iff |\psi\rangle = 0.
\end{align*}
In the following, complex conjugation will be denoted by $^*$. By the Gram--Schmidt process, there exists an orthonormal basis 
$\{\ket{i}\}_{i=1}^N \subset \mathcal{H}$ such that
\begin{equation*}
\langle i \mid j \rangle = \delta_{ij}, \qquad 
\sum_{i=1}^N |i\rangle\langle i| = \mathbb{I}.
\end{equation*}
By inserting the completeness relation, any state can be written as
\begin{equation*}
|\psi\rangle = \mathbb{I}|\psi\rangle 
= \sum_{i=1}^N |i\rangle\langle i|\psi\rangle.
\end{equation*}
The components are therefore \textit{defined} as projections via 
the abstract inner product:
\begin{equation*}
\boxed{\psi_i := \langle i \mid \psi \rangle \in \mathbb{C}.}
\end{equation*}
Hence, any state vector $\ket{\psi} \in \mathcal{H}$ can be expanded as
\begin{equation}
\ket{\psi} = \sum_{i=1}^N \psi_i \ket{i}.
\label{eq:state_expansion}
\end{equation}
Inserting the completeness relation into the abstract inner product, one obtains
\begin{equation*}
\langle \psi \mid \phi \rangle 
= \langle \psi \mid \mathbb{I} \mid \phi \rangle 
= \left\langle \psi \,\middle|\, \sum_{i=1}^N |i\rangle\langle i| 
\,\middle|\, \phi \right\rangle 
= \sum_{i=1}^N \langle \psi \mid i \rangle \langle i \mid \phi \rangle 
= \sum_{i=1}^N \psi_i^* \, \phi_i,
\end{equation*}
where in the last step we used 
$\langle \psi \mid i \rangle = \langle i \mid \psi \rangle^* = \psi_i^*$ 
by Hermiticity of the inner product. The final expression
\begin{equation}
\boxed{\langle \psi \mid \phi \rangle = \sum_{i=1}^N \psi_i^* \, \phi_i}
\label{eq:inner_product_components}
\end{equation}
is therefore a \textit{representation} of the abstract inner product in the chosen basis, rather than its definition.
We can also define the components of a linear operator $\hat{A}$ through its matrix elements:
\begin{equation*}
A_{ij} := \langle i \mid \hat{A} \mid j \rangle.
\end{equation*}
The action of the operator on a generic state $\ket{\psi}$ can then be written as
\begin{equation}
\hat{A}\ket{\psi} = \sum_{i,j=1}^N A_{ij} \psi_j \ket{i}.
\label{eq:operator_action}
\end{equation}
Observables are associated with Hermitian operators, satisfying
\begin{equation*}
\hat{A} = \hat{A}^\dagger.
\end{equation*}
Given a state $\ket{\psi}$, the probability of obtaining the outcome associated with the basis vector $\ket{i}$ is
\begin{equation*}
P(i) = |\langle i \mid \psi \rangle|^2 = |\psi_i|^2,
\end{equation*}
with normalization condition
\begin{equation*}
\sum_{i=1}^N P(i) = 1.
\end{equation*}
Finally, the expectation value of the observable $\hat{A}$ in the state $\ket{\psi}$ is defined as
\begin{equation*}
\langle \hat{A} \rangle := \langle \psi \mid \hat{A} \mid \psi \rangle,
\end{equation*}
which, in the chosen basis, takes the form
\begin{equation}
\langle \hat{A} \rangle = \sum_{i,j=1}^N \psi_i^* A_{ij} \psi_j.
\label{eq:expectation}
\end{equation}
If one chooses the eigenbasis $\{\ket{a_n}\}$ of the operator $\hat{A}$, defined by
\begin{equation*}
\hat{A}\ket{a_n} = a_n \ket{a_n},
\end{equation*}
any state $\ket{\psi}$ can be expanded as
\begin{equation*}
\ket{\psi} = \sum_n c_n \ket{a_n}, \qquad c_n = \langle a_n \mid \psi \rangle.
\end{equation*}
In this basis, the expectation value of $\hat{A}$ takes the form
\begin{equation*}
\langle \hat{A} \rangle = \langle \psi \mid \hat{A} \mid \psi \rangle 
= \sum_n a_n |c_n|^2,
\end{equation*}
which admits a direct probabilistic interpretation, where
\begin{equation*}
P(a_n) = |c_n|^2
\end{equation*}
represents the probability of obtaining the outcome $a_n$ upon measurement.
The choice of basis is purely a mathematical convenience and does not affect the physical measurement process. While an observable is always measured through its eigenvalues, expressing the system in a different basis introduces off-diagonal terms that encode quantum coherence, without altering the physical content of the theory.
A crucial point is that the probabilistic structure of QM is invariant under rescaling of state vectors. Physical states are not represented by vectors in $\mathcal{H}$ themselves, but by rays, i.e. equivalence classes under multiplication by a non-zero complex scalar.
Indeed, if $|\psi\rangle$ is replaced by $|\psi'\rangle = \lambda |\psi\rangle$, with $\lambda \in \mathbb{C}\setminus\{0\}$, then expectation values of observables remain unchanged after proper normalization:
\begin{equation*}
\frac{\langle \psi' | \hat{A} | \psi' \rangle}{\langle \psi' | \psi' \rangle}
=\frac{|\lambda|^2 \langle \psi | \hat{A} | \psi \rangle}{|\lambda|^2 \langle \psi | \psi \rangle}
=\frac{\langle \psi | \hat{A} | \psi \rangle}{\langle \psi | \psi \rangle}.
\end{equation*}
This shows that the physical content of the theory depends only on the direction of the state vector in Hilbert space, not on its norm. Consequently, one may either work with normalized vectors (unit vectors) or with arbitrary non-normalized vectors, provided that expectation values and probabilities are defined consistently.
In particular, probabilities and expectation values are invariant under the identification of physically equivalent states (rays), ensuring the consistency of the probabilistic interpretation.

\subsection{Bilinear Forms}

Now that we have established the standard Hilbert space framework,
we can introduce its reformulation in terms of bilinear forms.
Starting from the expectation value in
Eq.~(\textcolor{red}{\ref{eq:expectation}}), we associate with a
bounded linear operator $\hat A$ the sesquilinear form
\begin{equation*}
    a:\mathcal H\times\mathcal H\longrightarrow\mathbb C,
    \qquad
    a(|\psi\rangle,|\phi\rangle):=\langle\psi|\hat A|\phi\rangle.
\end{equation*}
Conversely, any bounded sesquilinear form on $\mathcal H$
uniquely determines a bounded linear operator as a consequence
of the Riesz representation theorem  \cite{Teschl2014}. This establishes a one-to-one
correspondence between bounded linear operators and bounded
sesquilinear forms.

With the convention used here, the form is antilinear in its
first argument and linear in its second. Thus, for all
$|\psi\rangle,|\phi\rangle,|\chi\rangle\in\mathcal H$ and $\alpha,\beta\in\mathbb C$,
\begin{align*}
    a(\alpha|\psi\rangle+\beta|\phi\rangle,|\chi\rangle)
    &=
    \alpha^*a(|\psi\rangle,|\chi\rangle)+\beta^*a(|\phi\rangle,|\chi\rangle),\\
    a(|\psi\rangle,\alpha|\phi\rangle+\beta|\chi\rangle)
    &=
    \alpha a(|\psi\rangle,|\phi\rangle)+\beta a(|\psi\rangle,|\chi\rangle).
\end{align*}
Given an orthonormal basis $\{|i\rangle\}$, the form is
determined by its values on the basis vectors:
\begin{equation*}
    A_{ik}
    :=
    a(|i\rangle,|k\rangle)
    =
    \langle i|\hat A|k\rangle.
\end{equation*}
Expanding two arbitrary vectors as in
Eq.~(\textcolor{red}{\ref{eq:state_expansion}}), we obtain
\begin{equation*}
    a(|\psi\rangle,|\phi\rangle)
    =
    \sum_{i,k}\psi_i^*A_{ik}\phi_k.
\end{equation*}

To express this structure as a bilinear form, we introduce
the continuous dual $\mathcal H^\vee$, whose elements are
continuous linear functionals on $\mathcal H$.
Each ket $|\psi\rangle\in\mathcal H$ defines a bra
$\langle\psi|\in\mathcal H^\vee$ through its action
\begin{equation*}
    \langle\psi|:\mathcal H\longrightarrow\mathbb C,
    \qquad
    \langle\psi|\bigl(|\chi\rangle\bigr)
    :=
    \langle\psi|\chi\rangle.
\end{equation*}
The Riesz representation theorem guarantees that every
element of $\mathcal H^\vee$ is associated in this way
with a unique vector of $\mathcal H$ \cite{Teschl2014}.
The correspondence
\begin{equation*}
    \mathcal R:\mathcal H\longrightarrow\mathcal H^\vee,
    \qquad
    \mathcal R(|\psi\rangle)=\langle\psi|
\end{equation*}
is an antilinear bijection:
\begin{equation*}
    \mathcal R\bigl(\alpha|\psi\rangle+\beta|\phi\rangle\bigr)
    =
    \alpha^*\langle\psi|+\beta^*\langle\phi|.
\end{equation*}
Thus, each bra acts linearly on its ket argument, while the
correspondence from a ket to its associated bra is antilinear.

In particular, starting from
\begin{equation*}
    |\psi\rangle=\sum_i\psi_i|i\rangle,
    \qquad
    \psi_i=\langle i|\psi\rangle,
\end{equation*}
we obtain
\begin{equation*}
    \langle\psi|
    =
    \mathcal R\Bigl(\sum_i\psi_i|i\rangle\Bigr)
    =
    \sum_i\psi_i^*\mathcal R(|i\rangle)
    =
    \sum_i\psi_i^*\langle i|.
\end{equation*}

We now associate with the sesquilinear form $a$ a bilinear form
\begin{equation*}
    B:\mathcal H^\vee\times\mathcal H\longrightarrow\mathbb C,
    \qquad
    B(\langle\psi|,|\phi\rangle):=a(|\psi\rangle,|\phi\rangle).
\end{equation*}
To verify linearity in the first argument, consider the bra
$\alpha\langle\psi_1|+\beta\langle\psi_2|$.
By the antilinearity of $\mathcal R$, this is the bra associated
with the ket
$\alpha^*|\psi_1\rangle+\beta^*|\psi_2\rangle$.
Consequently,
\begin{align*}
    B\bigl(\alpha\langle\psi_1|+\beta\langle\psi_2|,
           |\phi\rangle\bigr)
    &=
    a(\alpha^*|\psi_1\rangle+\beta^*|\psi_2\rangle,|\phi\rangle)\\
    &=
    \alpha a(|\psi_1\rangle,|\phi\rangle)+\beta a(|\psi_2\rangle,|\phi\rangle)\\
    &=
    \alpha B(\langle\psi_1|,|\phi\rangle)
    +\beta B(\langle\psi_2|,|\phi\rangle).
\end{align*}
The second conjugation, arising from the antilinearity of $a$
in its first argument, restores the original coefficients
$\alpha$ and $\beta$.

Linearity in the second argument follows directly from the
corresponding property of $a$:
\begin{align*}
    B\bigl(\langle\psi|,
           \alpha|\phi_1\rangle+\beta|\phi_2\rangle\bigr)
    &=
    a(|\psi\rangle,\alpha|\phi_1\rangle+\beta|\phi_2\rangle)\\
    &=
    \alpha B(\langle\psi|,|\phi_1\rangle)
    +\beta B(\langle\psi|,|\phi_2\rangle).
\end{align*}
Therefore, $B$ is linear in both its bra and ket arguments.

Using the basis expansions of the bra and ket, we find
\begin{align*}
    B(\langle\psi|,|\phi\rangle)
    &=
    B\Bigl(
        \sum_i\psi_i^*\langle i|,
        \sum_k\phi_k|k\rangle
    \Bigr)\\
    &=
    \sum_{i,k}\psi_i^*\phi_k
    B(\langle i|,|k\rangle)\\
    &=
    \sum_{i,k}\psi_i^*A_{ik}\phi_k.
\end{align*}
Once the basis is fixed, the ket is specified by its components
$\{\phi_k\}$ and the bra by its components $\{\psi_i^*\}$.
We can therefore express the same bilinear form directly
in terms of these components:
\begin{equation*}
    \widetilde a\bigl(\{\phi_k\},\{\psi_i^*\}\bigr)
    :=
    B(\langle\psi|,|\phi\rangle)
    =
    \sum_{i,k}\psi_i^*A_{ik}\phi_k.
\end{equation*}
We abbreviate this notation as $\widetilde a(\phi,\psi^*)$,
where $\phi$ denotes the ket amplitudes and $\psi^*$ denotes
the conjugate amplitudes appearing in the bra expansion.
The ordering places the ket amplitudes first and the bra
coefficients second, following the notation adopted in
Weinberg's formulation.

Setting $\phi=\psi$, the bilinear form reproduces the
expectation value of normalized states:
\begin{equation}
    \widetilde a(\psi,\psi^*)
    =
    \sum_{i,k}\psi_i^*A_{ik}\psi_k
    =
    \langle\psi|\hat A|\psi\rangle.
    \label{eq:bilinear_form}
\end{equation}
In the following, we omit the tilde and denote this expression
by $a(\psi,\psi^*)$.

This reformulation expresses the sesquilinear structure on
$\mathcal H\times\mathcal H$ as a bilinear form on
$\mathcal H^\vee\times\mathcal H$.
Its coordinate expression provides the amplitude--conjugate
notation used in the following formulation of QM.

We can now write the SE component-wise, using
Eq.~(\textcolor{red}{\ref{eq:operator_action}}):


\begin{equation*}
i \frac{d}{dt}\sum_j \psi_j(t)\, \ket{j}
= \sum_{j,k=1}^N H_{jk}\, \psi_k(t)\, \ket{j}.
\end{equation*}
Taking the time derivative inside the sum on the left-hand side, and relabeling dummy indices on the right-hand side, we obtain
\begin{equation*}
\sum_j i \frac{d\psi_j(t)}{dt}\, \ket{j}
= \sum_j \left( \sum_k H_{jk}\, \psi_k(t) \right) \ket{j}.
\end{equation*}
Using the linear independence of the basis vectors $\{\ket{j}\}$, we can equate the coefficients of each $\ket{j}$, which yields the component form of the equation:
\begin{equation}
i \frac{d\psi_j(t)}{dt} = \sum_k H_{jk}\, \psi_k(t).
\label{eq:comp_SE}
\end{equation}
We now multiply both sides by $\psi_j^*(t)$ and sum over $j$:
\begin{equation*}
i \sum_j \psi_j^*(t)\, \frac{d\psi_j(t)}{dt}
= \sum_{j,k} \psi_j^*(t)\, H_{jk}\, \psi_k(t).
\end{equation*}
The right-hand side defines a bilinear form in the amplitudes $\psi$ and their complex conjugates $\psi^*$. We introduce the associated Hamiltonian function (in the sense of Weinberg)
\begin{equation}
h(\psi,\psi^*) := \sum_{j,k} \psi_j^*\, H_{jk}\, \psi_k.
\label{eq:hamiltonian_function}
\end{equation}
This allows us to rewrite the previous equation as
\begin{equation*}
i \sum_j \psi_j^*\, \frac{d\psi_j}{dt} = h(\psi,\psi^*).
\end{equation*}
Moreover, taking the partial derivative of $h$ with respect to $\psi_j^*$, we obtain
\begin{equation*}
\frac{\partial h}{\partial \psi_j^*} = \sum_k H_{jk}\, \psi_k.
\end{equation*}
Comparing with the SE in Eq.~\textcolor{red}{\eqref{eq:comp_SE}}, we finally arrive at
\begin{equation}
\boxed{i \frac{d\psi_j}{dt} = \frac{\partial h}{\partial \psi_j^*}}.
\label{eq:weinberg_eom_discrete}
\end{equation}
Since $h(\psi,\psi^*)$ is real-valued, taking the complex conjugate of Eq.~\textcolor{red}{\eqref{eq:weinberg_eom_discrete}}, we obtain
\begin{equation*}
- i \frac{d\psi_j^*}{dt} = \left( \frac{\partial h}{\partial \psi_j^*} \right)^*.
\end{equation*}
Using the fact that $h$ is real, it follows that
\begin{equation*}
\left( \frac{\partial h}{\partial \psi_j^*} \right)^* = \frac{\partial h}{\partial \psi_j},
\end{equation*}
and therefore
\begin{equation}
\boxed{ i \frac{d\psi_j^*}{dt} = - \frac{\partial h}{\partial \psi_j}}.
\label{eq:weinberg_eom_conjugate}
\end{equation}

\subsection{The Eigenvalue Problem}
Now, let us choose as basis the set of eigenvectors of $\hat{H}$, satisfying
\begin{equation*}
\hat{H} \ket{i} = E_i \ket{i}, \qquad \forall i.
\end{equation*}
In this basis, the matrix elements of the Hamiltonian are diagonal:
\begin{equation*}
H_{jk} = E_j \delta_{jk}.
\end{equation*}
Substituting this into Eq.~(\textcolor{red}{\ref{eq:comp_SE}}), we obtain
\begin{equation*}
i \frac{d\psi_j(t)}{dt} = \sum_k E_j \delta_{jk}\, \psi_k(t) = E_j \psi_j(t).
\end{equation*}
Therefore, the SE decouples into independent differential equations for each component. The solution is then
\begin{equation}
\psi_j(t) = \psi_j(0)\, e^{-i E_j t}.
\label{eq:linear_energy_mode_evolution}
\end{equation}
In order to express the eigenvalue equation in terms of bilinear
forms, consider a linear operator $\hat A$ satisfying
\begin{equation*}
    \hat A|\psi\rangle=\alpha|\psi\rangle.
\end{equation*}
Projecting the eigenvalue equation onto $\langle k|$ gives
\begin{equation*}
    \langle k|\hat A|\psi\rangle
    =\sum_j A_{kj}\psi_j
    =\alpha\psi_k.
\end{equation*}
so that the eigenvalue equation becomes
\begin{equation*}
\sum_j A_{kj}\psi_j=\alpha \psi_k.
\end{equation*}
Multiplying both sides by $\psi_k^*$ and summing over $k$, we obtain
\begin{equation*}
\sum_{k,j}\psi_k^* A_{kj}\psi_j=\alpha \sum_k \psi_k^*\psi_k.
\end{equation*}
The left-hand side defines a bilinear form, which we denote as
\begin{equation}
a(\psi,\psi^*):=\sum_{k,j}\psi_k^* A_{kj}\psi_j,
\label{eq:observable_bilinear_components}
\end{equation}
so that the equation becomes
\begin{equation*}
a(\psi,\psi^*)=\alpha \sum_k |\psi_k|^2.
\end{equation*}

We introduce the bilinear form associated with the norm
\begin{equation}
n(\psi,\psi^*):=\sum_k \psi_k^*\psi_k.
\label{eq:norm}
\end{equation}
Then the eigenvalue problem is equivalent to 
\begin{equation}
\boxed{\alpha=\frac{a(\psi,\psi^*)}{n(\psi,\psi^*)}}.
\label{eq:ev_fraction}
\end{equation}
This relation has been obtained using bilinear property.
We now recover this using only the differential eigenvalue equation and the homogeneity identities. This identifies the properties needed to retain the same relation in the nonlinear
generalization, where $a$ need not be bilinear.

Taking the derivative of $a$ with respect to $\psi_k^*$, we obtain
\begin{equation}
\frac{\partial a}{\partial \psi_k^*}=\sum_j A_{kj}\psi_j.
\label{eq:bilinear_observable_derivative}
\end{equation}
Therefore, the eigenvalue equation can be written as
\begin{equation}
\boxed{\frac{\partial a}{\partial \psi_k^*}=\alpha \psi_k}.
\label{eq:ev_eq}
\end{equation}
Taking the complex conjugate of the previous equation, and assuming $a(\psi,\psi^*) \in \mathbb{R}$, we obtain
\begin{equation*}
\left(\frac{\partial a}{\partial \psi_k^*}\right)^*=\alpha^* \psi_k^*.
\end{equation*}
Using the reality of $a$ and that $\alpha \in \mathbb{R}$ for Hermitian operators, it follows that
\begin{equation}
\boxed{\frac{\partial a}{\partial \psi_k}=\alpha \psi_k^*}.
\label{eq:conj_ev_eq}
\end{equation}
Both $a(\psi,\psi^*)$ and $n(\psi,\psi^*)$ satisfy the homogeneity property
\begin{equation}
a(z\psi,z^*\psi^*)=|z|^2 a(\psi,\psi^*).
\label{eq:hom_1}
\end{equation}
This implies separate homogeneity relations, which can be written in differential form as
\begin{equation}
\sum_k \psi_k^*\frac{\partial a}{\partial \psi_k^*}=a(\psi,\psi^*),\qquad
\sum_k \psi_k\frac{\partial a}{\partial \psi_k}=a(\psi,\psi^*).
\label{eq:hom_2}
\end{equation}
These homogeneity conditions ensure that the normalized quantity
$a/n$ is invariant under multiplication of the state vector by
any nonzero complex scalar:
\begin{equation*}
    \frac{a(z\psi,z^*\psi^*)}{n(z\psi,z^*\psi^*)}
    =
    \frac{|z|^2a(\psi,\psi^*)}{|z|^2n(\psi,\psi^*)}
    =
    \frac{a(\psi,\psi^*)}{n(\psi,\psi^*)}.
\end{equation*}
Thus, expectation values depend only on the ray representing
the physical state, rather than on the normalization or global
phase of its representative vector. For an eigenstate, this
ratio equals the corresponding eigenvalue, which is therefore
also independent of the chosen representative. These conditions
play a fundamental role in the nonlinear formulation, where
they preserve this invariance even when observables are no
longer represented by bilinear functions.

Starting from the eigenvalue Equation~(\textcolor{red}{\ref{eq:ev_eq}}) we multiply both sides by $\psi_k^*$ and sum over $k$:
\begin{equation*}
\sum_k \psi_k^*\frac{\partial a}{\partial \psi_k^*}=\alpha \sum_k \psi_k^*\psi_k.
\end{equation*}
Using the homogeneity relation, we obtain
\begin{equation*}
a(\psi,\psi^*)=\alpha\, n(\psi,\psi^*),
\end{equation*}
and therefore we obtain again the relation~(\textcolor{red}{\ref{eq:ev_fraction}}). Thus, the relation $\alpha=a/n$ does not require bilinearity:
it follows from the differential eigenvalue equation together
with the homogeneity identity.
Note that this relation holds only when $\psi$ is an eigenvector of the operator $\hat{A}$. In that case, the ratio $a/n$ coincides with the corresponding eigenvalue.

We now extend this construction by defining the function 
\begin{equation*}
\tilde a(\psi,\psi^*):=\frac{a(\psi,\psi^*)}{n(\psi,\psi^*)},
\end{equation*}
for arbitrary $\psi$, thus promoting it to a scalar function on the space of states.
We formulate a variational problem and look for stationary points of $\tilde a$. In this framework, $\psi$ and $\psi^*$ are treated as independent variables, and therefore stationarity requires
\begin{equation*}
\frac{\partial \tilde a}{\partial \psi_k^*}=0,
\qquad
\frac{\partial \tilde a}{\partial \psi_k}=0.
\end{equation*}
Although $\psi$ and $\psi^*$ are treated as independent variables in the variational formulation, they are related by complex conjugation at the level of physical states. Imposing stationarity with respect to both variables ensures not only the eigenvalue equation, but also the consistency of the solution with the complex structure of the Hilbert space.
Using the expression for the derivative of $\tilde a$, we obtain
\begin{equation*}
\frac{\partial \tilde a}{\partial \psi_k^*}
=\frac{1}{n}\frac{\partial a}{\partial \psi_k^*}-\frac{a}{n^2}\psi_k,
\qquad
\frac{\partial \tilde a}{\partial \psi_k}
=\frac{1}{n}\frac{\partial a}{\partial \psi_k}-\frac{a}{n^2}\psi_k^*.
\end{equation*}
Let $\psi^{(s)}$ be a stationary point. Evaluating the above expressions at $\psi^{(s)}$ and imposing stationarity, we find
\begin{equation*}
\left.\frac{\partial a}{\partial \psi_k^*}\right|_{\psi^{(s)}}
=\left.\frac{a}{n}\right|_{\psi^{(s)}}\psi_k^{(s)},
\qquad
\left.\frac{\partial a}{\partial \psi_k}\right|_{\psi^{(s)}}
=\left.\frac{a}{n}\right|_{\psi^{(s)}}\psi_k^{*(s)}.
\end{equation*}
Defining
\begin{equation*}
\alpha:=\left.\frac{a(\psi,\psi^*)}{n(\psi,\psi^*)}\right|_{\psi^{(s)}},
\end{equation*}
we obtain
\begin{equation*}
\left.\frac{\partial a}{\partial \psi_k^*}\right|_{\psi^{(s)}}=\alpha \psi_k^{(s)},
\qquad
\left.\frac{\partial a}{\partial \psi_k}\right|_{\psi^{(s)}}=\alpha \psi_k^{*(s)}.
\end{equation*}
These two equations correspond to the eigenvalue equation and its complex conjugate, and must be satisfied simultaneously by the same state $\psi^{(s)}$. Therefore, stationary points of $\tilde a$ are in one-to-one correspondence with eigenvectors of $\hat{A}$, and the corresponding values of $\tilde a$ yield the associated eigenvalues.

\subsection{Symmetries}

In standard QM, a continuous unitary symmetry acts on the state
$|\psi\rangle$ as
\begin{equation*}
    |\psi'\rangle=\hat X(\varepsilon)|\psi\rangle.
\end{equation*}
For transformations close to the identity, we write
\begin{equation*}
    \hat X(\varepsilon)
    =
    \hat{\mathbb I}-i\varepsilon\hat A
    +O(\varepsilon^2),
\end{equation*}
where $\hat A$ is the Hermitian generator of the transformation.
The transformed state is therefore
\begin{equation*}
    |\psi'\rangle
    =
    |\psi\rangle-i\varepsilon\hat A|\psi\rangle
    +O(\varepsilon^2).
\end{equation*}
The infinitesimal variation is therefore
\begin{equation*}
    \varepsilon\,\delta_{\hat A}|\psi\rangle
    =
    |\psi'\rangle-|\psi\rangle
    =
    -i\varepsilon\hat A|\psi\rangle,
\end{equation*}
where terms of order $\varepsilon^2$ have been neglected. Then
\begin{equation*}
    \delta_{\hat A}|\psi\rangle=-i\hat A|\psi\rangle.
\end{equation*}

In an orthonormal basis $\{|k\rangle\}$, this becomes
\begin{equation*}
    \delta_{\hat A}\psi_k
    =
    \langle k|\delta_{\hat A}|\psi\rangle
    =
    -i\sum_l A_{kl}\psi_l.
\end{equation*}
This is the standard infinitesimal action of a symmetry
generator in linear QM.

In the functional formulation, this expression and its complex
conjugate are generalized by introducing a real differentiable
generator $a(\psi,\psi^*)$ and defining
\begin{equation}
    \delta_a\psi_k
    =
    -i\frac{\partial a}{\partial\psi_k^*},
    \qquad
    \delta_a\psi_k^*
    =
    i\frac{\partial a}{\partial\psi_k}.
    \label{eq:inf_trasform}
\end{equation}
These expressions reduce to the standard linear QM
transformation when
\begin{equation*}
    a(\psi,\psi^*)
    =
    \langle\psi|\hat A|\psi\rangle
    =
    \sum_{k,l}\psi_k^*A_{kl}\psi_l,
\end{equation*}
as in
Eq.~\textcolor{red}{\eqref{eq:observable_bilinear_components}}.
Indeed, using
Eq.~\textcolor{red}{\eqref{eq:bilinear_observable_derivative}},
we recover
\begin{equation*}
    \delta_a\psi_k
    =
    -i\sum_l A_{kl}\psi_l
    =
    \delta_{\hat A}\psi_k.
\end{equation*}
Thus, for a bilinear generator, the functional formulation
reproduces the standard linear action of the Hermitian operator
$\hat A$ on quantum states.

\subsection{Composite Systems}
When describing a composite system made of two subsystems, the state space is not given by a simple juxtaposition of the individual state spaces, but by their tensor product. This construction is necessary in order to account for all possible states of the system, including superpositions that cannot be factorized into states of the individual subsystems (entangled states).
Given two Hilbert spaces $\mathcal{H}_1$ and $\mathcal{H}_2$, the tensor product $\mathcal{H}_1 \otimes \mathcal{H}_2$ is defined as the vector space generated by formal linear combinations of elements of the form
\begin{equation*}
|\psi\rangle \otimes |\phi\rangle, \qquad |\psi\rangle \in \mathcal{H}_1, \; |\phi\rangle \in \mathcal{H}_2,
\end{equation*}
subject to the linearity relations
\begin{equation*}
(a|\psi\rangle) \otimes |\phi\rangle = |\psi\rangle \otimes (a|\phi\rangle) = a(|\psi\rangle \otimes |\phi\rangle),
\end{equation*}
\begin{equation*}
(|\psi_1\rangle + |\psi_2\rangle) \otimes |\phi\rangle = |\psi_1\rangle \otimes |\phi\rangle + |\psi_2\rangle \otimes |\phi\rangle,
\end{equation*}
and similarly in the second slot.
If $\{|i\rangle\}$ and $\{|k\rangle\}$ are orthonormal bases of $\mathcal{H}_1$ and $\mathcal{H}_2$, respectively, then the set
\begin{equation*}
\{|i\rangle \otimes |k\rangle\}
\end{equation*}
forms an orthonormal basis of $\mathcal{H}_1 \otimes \mathcal{H}_2$, and any state of the composite system can be written as
\begin{equation*}
|\Psi\rangle = \sum_{i,k} \Psi_{ik}\, |i\rangle \otimes |k\rangle.
\end{equation*}
The coefficients $\Psi_{ik}$ define the components of the state in the product basis. States for which $\Psi_{ik} = \psi_i \phi_k$ factorize as $|\psi\rangle \otimes |\phi\rangle$, while generic states do not admit such a decomposition and are called entangled.

The inner product on the tensor product space is defined by
requiring the product basis
$|i,k\rangle:=|i\rangle\otimes|k\rangle$ to be orthonormal:
\begin{equation*}
    \langle i,k|j,\ell\rangle
    =
    \delta_{ij}\delta_{k\ell}.
\end{equation*}
Extending this prescription antilinearly in the first argument
and linearly in the second, for arbitrary states
$|\Psi\rangle=\sum_{i,k}\Psi_{ik}|i,k\rangle$ and
$|\Phi\rangle=\sum_{j,\ell}\Phi_{j\ell}|j,\ell\rangle$ we obtain
\begin{equation*}
    \langle\Psi|\Phi\rangle
    =
    \sum_{i,k,j,\ell}
    \Psi_{ik}^*\Phi_{j\ell}\delta_{ij}\delta_{k\ell}
    =
    \sum_{i,k}\Psi_{ik}^*\Phi_{ik}.
\end{equation*}
This expression applies to both product and entangled states.
For product states, it reduces to
\begin{equation*}
    \begin{aligned}
        &(\langle\psi_1|\otimes\langle\phi_1|)
        (|\psi_2\rangle\otimes|\phi_2\rangle)\\
        &\quad=
        \sum_{i,k}\psi_{1,i}^*\phi_{1,k}^*
        \psi_{2,i}\phi_{2,k}\\
        &\quad=
        \langle\psi_1|\psi_2\rangle
        \langle\phi_1|\phi_2\rangle.
    \end{aligned}
\end{equation*}

For a system of two non-interacting subsystems with Hamiltonians $\hat{H}_1$ and $\hat{H}_2$, the total Hamiltonian acts on $\mathcal{H}_1 \otimes \mathcal{H}_2$ as
\begin{equation}
\hat{H} = \hat{H}_1 \otimes \mathbb{I} + \mathbb{I} \otimes \hat{H}_2.
\label{eq:tensor_hamiltonian}
\end{equation}
In the product basis, its matrix elements are
\begin{equation*}
(H)_{ik,j\ell} = (H_1)_{ij}\,\delta_{k\ell} + \delta_{ij}\,(H_2)_{k\ell}.
\end{equation*}
A generic state evolves according to
\begin{equation*}
i\frac{d\Psi_{ik}}{dt} = \sum_{j} (H_1)_{ij}\,\Psi_{jk} + \sum_{\ell} (H_2)_{k\ell}\,\Psi_{i\ell}.
\end{equation*}
We multiply both sides by $\Psi_{ik}^*$ and sum over $i,k$, obtaining
\begin{equation*}
i \sum_{i,k} \Psi_{ik}^* \frac{d\Psi_{ik}}{dt}
= \sum_{i,j,k} \Psi_{ik}^* (H_1)_{ij} \Psi_{jk}
+ \sum_{i,k,\ell} \Psi_{ik}^* (H_2)_{k\ell} \Psi_{i\ell}.
\end{equation*}
This naturally suggests defining the Hamiltonian functional as
\begin{equation*}
h(\Psi,\Psi^*) = \sum_k h_1^{(k)} + \sum_i h_2^{(i)},
\end{equation*}
where
\begin{equation*}
h_1^{(k)} := \sum_{i,j} \Psi_{ik}^* (H_1)_{ij} \Psi_{jk},
\qquad
h_2^{(i)} := \sum_{k,\ell} \Psi_{ik}^* (H_2)_{k\ell} \Psi_{i\ell}.
\end{equation*}
In this decomposition, $h_1^{(k)}$ depends only on the components with fixed $k$ (i.e. on each column of the matrix $\Psi_{ik}$), while $h_2^{(i)}$ depends only on the components with fixed $i$ (i.e. on each row).
To make this structure explicit, we introduce for each fixed $k$ the column vectors
\begin{equation*}
\phi^{(k)}_j := \Psi_{jk},
\end{equation*}
and for each fixed $i$ the row vectors
\begin{equation*}
\xi^{(i)}_k := \Psi_{ik}.
\end{equation*}
In terms of these variables, the Hamiltonian functional can be written as
\begin{equation}
\boxed{h(\Psi,\Psi^*) = \sum_k h_1(\phi^{(k)},\phi^{(k)*}) + \sum_i h_2(\xi^{(i)},\xi^{(i)*})},
\label{eq:composite_hamiltonian}
\end{equation}
where each term depends only on a fixed column or a fixed row, respectively.
When taking derivatives with respect to $\Psi_{ik}^*$, one observes that:
\begin{itemize}
\item the terms $h_1(\phi^{(k)},\phi^{(k)*})$ depend on $\Psi_{ik}^*$ only through the column $\phi^{(k)}$,
\item the terms $h_2(\xi^{(i)},\xi^{(i)*})$ depend on $\Psi_{ik}^*$ only through the row $\xi^{(i)}$.
\end{itemize}
Therefore, the derivative splits as
\begin{equation*}
\frac{\partial h}{\partial \Psi_{ik}^*}
= \frac{\partial h_1(\phi^{(k)},\phi^{(k)*})}{\partial \phi^{(k)*}_i}
+ \frac{\partial h_2(\xi^{(i)},\xi^{(i)*})}{\partial \xi^{(i)*}_k}.
\end{equation*}
Evaluating the two contributions explicitly, we recover
\begin{equation*}
\frac{\partial h}{\partial \Psi_{ik}^*}
= \sum_j (H_1)_{ij}\,\Psi_{jk}
+ \sum_{\ell} (H_2)_{k\ell}\,\Psi_{i\ell}.
\end{equation*}
Therefore, using the general form of the SE,
\begin{equation*}
i\frac{d\Psi_{ik}}{dt} = \frac{\partial h}{\partial \Psi_{ik}^*},
\end{equation*}
we can equivalently write
\begin{equation}
\boxed{
 i\frac{d\Psi_{ik}}{dt} =
\frac{\partial h_1(\phi^{(k)},\phi^{(k)*})}{\partial \phi^{(k)*}_i}
+
\frac{\partial h_2(\xi^{(i)},\xi^{(i)*})}{\partial \xi^{(i)*}_k}
}
\label{eq:composite_eom_nonlinear_form}
\end{equation}
We have thus obtained the bilinear Hamiltonian functional for a
composite system of two non-interacting subsystems, together with
the corresponding equations of motion.

\section{Introduction of Nonlinearities}
We now aim to extend the standard framework of QM by introducing nonlinearities, while preserving its fundamental structural properties. In the linear theory, observables are represented by bilinear (more precisely, sesquilinear) forms of the form~(\textcolor{red}{\ref{eq:bilinear_form}}), which correspond to linear operators on the Hilbert space. In the nonlinear generalization, we promote observables to more general functions
\begin{equation*}
a(\psi,\psi^*),
\end{equation*}
which are not necessarily bilinear. However, in order to preserve the physical interpretation of states, we require that observables satisfy the homogeneity condition~(\textcolor{red}{\ref{eq:hom_1}}) or equivalently~(\textcolor{red}{\ref{eq:hom_2}}), for any complex number $z$. This condition ensures that physically equivalent states, differing only by an overall complex factor, yield the same expectation values after normalization. In particular, it guarantees that observables depend only on the ray defined by $\psi$, consistently with the standard formulation of QM.
We then define how observables combine:
The sum and scalar multiplication are given in the natural way
\begin{equation*}
(a+b)(\psi,\psi^*) = a(\psi,\psi^*) + b(\psi,\psi^*), 
\qquad
(sa)(\psi,\psi^*) = s\,a(\psi,\psi^*),
\end{equation*}
for any complex scalar $s$.
The nontrivial structure arises in the definition of the product between two observables. Following Weinberg, we introduce the product
\begin{equation}
(a*b)(\psi,\psi^*) := \sum_j \frac{\partial a}{\partial \psi_j}\frac{\partial b}{\partial \psi_j^*}.
\label{eq:weinberg_product}
\end{equation}
This definition is motivated by the requirement that, when $a$ and $b$ are bilinear forms associated with linear operators $\hat{A}$ and $\hat{B}$, the product $a*b$ reproduces the composition of operators. In this sense, the nonlinear structure is constructed so as to retain the algebraic features of standard QM.
Therefore, the entire construction is guided by the principle that the generalized formalism must reproduce the key properties of the linear theory, while allowing for a broader class of observables beyond bilinear forms.

We can define, in complete analogy with QM, the norm~(\textcolor{red}{\ref{eq:norm}}). Since this corresponds to the identity operator, it satisfies
\begin{equation*}
n * a = a * n = a,
\end{equation*}
for any observable $a(\psi,\psi^*)$.

\subsection{Symmetries and Hamiltonian Structure}
We now consider infinitesimal symmetry transformations. In analogy with the standard QM case, the variation generated by an observable $a$ follows from Eq.~(\textcolor{red}{\ref{eq:inf_trasform}}). 
Let $b(\psi,\psi^*)$ be another observable. Its variation under the transformation generated by $a$ is obtained using the chain rule:
\begin{equation*}
\delta_a b 
= \sum_k \left(
\frac{\partial b}{\partial \psi_k}\delta_a \psi_k
+
\frac{\partial b}{\partial \psi_k^*}\delta_a \psi_k^*
\right).
\end{equation*}
Substituting the transformation laws, we find
\begin{equation*}
\delta_a b 
= -i \sum_k \frac{\partial b}{\partial \psi_k}\frac{\partial a}{\partial \psi_k^*}
+ i \sum_k \frac{\partial b}{\partial \psi_k^*}\frac{\partial a}{\partial \psi_k}.
\end{equation*}
This expression defines a Poisson-like bracket between observables:
\begin{equation}
[a,b]
:= \sum_k\left(
\frac{\partial a}{\partial\psi_k}\frac{\partial b}{\partial\psi_k^*}
-
\frac{\partial b}{\partial\psi_k}\frac{\partial a}{\partial\psi_k^*}
\right).
\label{eq:bracket_def}
\end{equation}
Using the product introduced above, this can be written as
\begin{equation}
[a,b]=(a*b-b*a).
\label{eq:bracket_product}
\end{equation}
Therefore the infinitesimal variation of an observable $b$ generated by $a$ takes the form
\begin{equation*}
\delta_a b=i[a,b].
\end{equation*}
In the special case in which $a$ and $b$ are bilinear forms associated with operators $\hat A$ and $\hat B$, this reduces to the standard QM expression
\begin{equation*}
[a,b]\longrightarrow\langle\psi|\,[\hat A,\hat B]\,|\psi\rangle.
\end{equation*}
Finally, using the fact that the norm corresponds to the identity observable, one finds
\begin{equation*}
[a,n]=(a*n-n*a)=0,
\end{equation*}
which shows that the norm is invariant under all infinitesimal symmetry transformations.

We have seen that the bracket can be written as~(\textcolor{red}{\ref{eq:bracket_product}}) which is manifestly antisymmetric,
\begin{equation*}
[a,b]=-[b,a],
\end{equation*}
but the product $*$ is not associative. In general one has
\begin{equation*}
(a*b)*c\neq a*(b*c),
\end{equation*}
since derivatives act on different arguments and do not distribute as an associative product.
However, despite the non-associativity of $*$, the bracket satisfies the Jacobi identity:
\begin{equation*}
[a,[b,c]]+[b,[c,a]]+[c,[a,b]]=0.
\end{equation*}
Therefore, the space of observables endowed with this bracket has the structure of a (generally nonlinear) Lie algebra.
\newline A symmetry of central importance is time translation, generated by the Hamiltonian functional \(h(\psi,\psi^*)\). The infinitesimal time evolution of the state under this symmetry can be written as
\begin{equation*}
\psi_k(t+dt)=\psi_k(t)+dt\,\delta_h\psi_k(t),
\end{equation*}
where \(\delta_h\psi_k\) denotes the variation generated by \(h\). Using (\textcolor{red}{\ref{eq:inf_trasform}}), and dividing by \(dt\) followed by the limit \(dt\to 0\), we obtain the equation of motion, with the same form as Eq.~\textcolor{red}{\eqref{eq:weinberg_eom_discrete}},
\begin{equation*}
\frac{d\psi_k}{dt}=-i\frac{\partial h}{\partial\psi_k^*}.
\end{equation*}
Using the bracket introduced in~(\textcolor{red}{\ref{eq:bracket_product}}), the time evolution of any observable \(a(\psi,\psi^*)\) is given by
\begin{equation}
\frac{da}{dt}=-i[a,h].
\label{eq:observable_time_evolution}
\end{equation}
To make the underlying Hamiltonian structure explicit, we introduce real canonical variables
\begin{equation}
\psi_k=\frac{q_k+ip_k}{\sqrt{2}},\qquad
\psi_k^*=\frac{q_k-ip_k}{\sqrt{2}},
\label{eq:weinberg_real_canonical_variables}
\end{equation}
with \(q_k,p_k\in\mathbb{R}\). In terms of these variables, the equations of motion become
\begin{equation}
\frac{dq_k}{dt}=\frac{\partial h}{\partial p_k},\qquad
\frac{dp_k}{dt}=-\frac{\partial h}{\partial q_k},
\label{eq:weinberg_real_hamilton_equations}
\end{equation}
which are precisely Hamilton's equations.
Expressing the bracket in terms of the real variables \((q_k,p_k)\), and using the chain rule, one finds
\begin{equation*}
[a,b]
=i\sum_k\left(
\frac{\partial a}{\partial q_k}\frac{\partial b}{\partial p_k}
-
\frac{\partial a}{\partial p_k}\frac{\partial b}{\partial q_k}
\right)
=i\{a,b\}_{\rm PB},
\end{equation*}
where \(\{\cdot,\cdot\}_{\rm PB}\) denotes the canonical Poisson bracket (PB).
In particular, the fundamental relations are
\begin{equation*}
\{q_j,p_k\}_{\rm PB}=\delta_{jk},\qquad
\{q_j,q_k\}_{\rm PB}=0,\qquad
\{p_j,p_k\}_{\rm PB}=0.
\end{equation*}
Therefore, the evolution equation~\textcolor{red}{\eqref{eq:observable_time_evolution}} becomes
\begin{equation*}
\frac{da}{dt}=\{a,h\}_{\rm PB}.
\end{equation*}
We conclude that the theory is equivalent to a classical Hamiltonian system on the phase space coordinatized by \((q_k,p_k)\), with Hamiltonian function \(h(q,p)\). The algebraic and dynamical structure coincides with that of classical mechanics (CM); the difference lies in the interpretation of the variables, which originate from the complex amplitudes of quantum states.

\subsection{Eigenvalues and Perturbation Theory}
Following the construction introduced in the first part of this thesis for the linear theory, we now extend the notion of eigenvalues to the nonlinear framework of Weinberg. In this setting, observables are represented by homogeneous functions $a(\psi,\psi^*)$, and a state $\psi_k$ is said to have a definite value $\alpha$ for the observable $a$ if it satisfies the condition of Eq.~\textcolor{red}{\eqref{eq:ev_eq}},
\begin{equation*}
\frac{\partial a}{\partial \psi_k^*}=\alpha\,\psi_k,
\end{equation*}
together with its complex conjugate. This condition generalizes the usual eigenvalue equation of QM.
An important property of this definition is that the eigenvalue can be expressed directly in terms of the function $a$. Contracting the above equation with $\psi_k^*$ and using homogeneity, one recovers Eq.~\textcolor{red}{\eqref{eq:ev_fraction}}, with the norm defined in Eq.~\textcolor{red}{\eqref{eq:norm}}.
The algebraic structure of observables is partially preserved. If a state has definite values $\alpha$ and $\beta$ for two observables $a$ and $b$, then it trivially has definite value $\xi\alpha+\eta\beta$ for $\xi a+\eta b$. Less trivially, using the product of observables in Eq.~\textcolor{red}{\eqref{eq:weinberg_product}} and the homogeneity properties in Eq.~\textcolor{red}{\eqref{eq:hom_2}}, one can show that the same state has definite value $\alpha\beta$ for the product $a*b$. Therefore, despite the nonlinearity, the eigenvalue structure remains compatible with the algebra of observables.
We now consider the effect of a small perturbation. Let the observable be written as
\begin{equation*}
a=a_0+a_1,
\end{equation*}
where $a_0$ is the unperturbed part and $a_1$ is a small nonlinear correction. We look for an eigenvector and eigenvalue of the form
\begin{equation*}
\psi=\psi_0+\psi_1,
\qquad
\alpha=\alpha_0+\alpha_1,
\end{equation*}
where $\psi_0$ satisfies the unperturbed equation
\begin{equation*}
\frac{\partial a_0}{\partial \psi_k^*}\bigg|_{\psi_0}
=\alpha_0\,\psi_{0k}.
\end{equation*}

Expanding the full eigenvalue equation to first order, one obtains
\begin{equation*}
\sum_l
\frac{\partial^2 a_0}{\partial \psi_k^*\partial \psi_l}\bigg|_{\psi_0}\psi_{1l}
+
\sum_l
\frac{\partial^2 a_0}{\partial \psi_k^*\partial \psi_l^*}\bigg|_{\psi_0}\psi_{1l}^*
+
\frac{\partial a_1}{\partial \psi_k^*}\bigg|_{\psi_0}
=
\alpha_0\psi_{1k}
+\alpha_1\psi_{0k}.
\end{equation*}
To extract the correction $\alpha_1$, we multiply this equation by $\psi_{0k}^*$ and sum over $k$. The terms involving second derivatives of $a_0$ combine in such a way that they reproduce the unperturbed eigenvalue equation, yielding
\begin{equation*}
\sum_k\psi_{0k}^*
\left[
\sum_l
\frac{\partial^2 a_0}{\partial \psi_k^*\partial \psi_l}\psi_{1l}
+
\sum_l
\frac{\partial^2 a_0}{\partial \psi_k^*\partial \psi_l^*}\psi_{1l}^*
\right]
=
\alpha_0\sum_k\psi_{0k}^*\psi_{1k}.
\end{equation*}
As a result, these contributions cancel against the term on the right-hand side proportional to $\alpha_0$, and we are left with
\begin{equation*}
\sum_k\psi_{0k}^*
\frac{\partial a_1}{\partial \psi_k^*}\bigg|_{\psi_0}
=
\alpha_1\sum_k|\psi_{0k}|^2.
\end{equation*}
At this point, the homogeneity of $a_1$ plays a crucial role. For any homogeneous function of degree one in $\psi$ and $\psi^*$, one has the identity
\begin{equation*}
\sum_k\psi_k^*\frac{\partial a_1}{\partial \psi_k^*}=a_1(\psi,\psi^*).
\end{equation*}
Applying this to $\psi_0$, we obtain
\begin{equation*}
a_1(\psi_0,\psi_0^*)=\alpha_1\,n(\psi_0,\psi_0^*).
\end{equation*}
Therefore, the first-order correction to the eigenvalue is
\begin{equation}
\alpha_1=\frac{a_1(\psi_0,\psi_0^*)}{n(\psi_0,\psi_0^*)}.
\label{eq:weinberg_first_order_eigenvalue_shift}
\end{equation}
This result is the direct analogue of first-order perturbation
theory in ordinary QM: in both cases, the correction is evaluated
on the unperturbed eigenstate. The distinctive feature of the
nonlinear generalization lies instead in the dynamics generated
by the Hamiltonian functional. For a fixed time-independent
linear Hamiltonian, changing the amplitudes of an initial
superposition changes the weights of its components, but not
the frequencies associated with energy differences. Nonlinear
terms can make these frequencies depend on the amplitudes
themselves, and hence on the initial preparation. Such
preparation-dependent frequency shifts provide a possible route
to experimental tests of nonlinear QM.

\subsection{Separated Systems and Separability}
We now use the same construction introduced in Eq.~(\textcolor{red}{\ref{eq:composite_hamiltonian}}) for the Hamiltonian of two non-interacting systems. We write
\begin{equation*}
h(\Psi,\Psi^*)=\sum_j h_1(\phi^{(j)},\phi^{(j)*})+\sum_i h_2(\xi^{(i)},\xi^{(i)*}),
\end{equation*}
where
\begin{equation*}
\phi_i^{(j)}:=\Psi_{ij},\qquad \xi_j^{(i)}:=\Psi_{ij}.
\end{equation*}
Therefore the equation of motion of the composite system is Eq.~\textcolor{red}{\eqref{eq:composite_eom_nonlinear_form}}, with the second subsystem index denoted by $j$.
We now consider a factorized state
\begin{equation*}
\Psi_{ij}(t)=\phi_i(t)\xi_j(t).
\end{equation*}
Then, for fixed \(j\), the column vector \(\phi^{(j)}\) is just the vector \(\phi\) multiplied by the scalar \(\xi_j\):
\begin{equation*}
\phi^{(j)}=\xi_j\phi.
\end{equation*}
Similarly, for fixed \(i\), the row vector \(\xi^{(i)}\) is the vector \(\xi\) multiplied by the scalar \(\phi_i\):
\begin{equation*}
\xi^{(i)}=\phi_i\xi.
\end{equation*}
We now use the homogeneity property. Since
\begin{equation*}
h_1(z\phi,z^*\phi^*)=|z|^2h_1(\phi,\phi^*),
\end{equation*}
we set \(z=\xi_j\) and obtain
\begin{equation*}
h_1(\xi_j\phi,\xi_j^*\phi^*)=|\xi_j|^2h_1(\phi,\phi^*).
\end{equation*}
Differentiating both sides with respect to \(\phi_i^*\), and using the chain rule, gives
\begin{equation*}
\sum_l
\frac{\partial h_1(\phi^{(j)},\phi^{(j)*})}{\partial \phi_l^{(j)*}}
\frac{\partial \phi_l^{(j)*}}{\partial \phi_i^*}
=
|\xi_j|^2\frac{\partial h_1(\phi,\phi^*)}{\partial\phi_i^*}.
\end{equation*}
Since \(\phi_l^{(j)*}=\xi_j^*\phi_l^*\), one has
\begin{equation*}
\frac{\partial \phi_l^{(j)*}}{\partial \phi_i^*}
=
\xi_j^*\delta_{li},
\end{equation*}
and therefore
\begin{equation*}
\xi_j^*
\frac{\partial h_1(\phi^{(j)},\phi^{(j)*})}{\partial \phi_i^{(j)*}}
=
|\xi_j|^2\frac{\partial h_1(\phi,\phi^*)}{\partial\phi_i^*}.
\end{equation*}
Dividing by \(\xi_j^*\), we obtain
\begin{equation*}
\frac{\partial h_1(\phi^{(j)},\phi^{(j)*})}{\partial \phi_i^{(j)*}}
=
\xi_j\frac{\partial h_1(\phi,\phi^*)}{\partial\phi_i^*}.
\end{equation*}
In the same way, using \(\xi_j^{(i)}=\phi_i\xi_j\), one finds
\begin{equation*}
\frac{\partial h_2(\xi^{(i)},\xi^{(i)*})}{\partial \xi_j^{(i)*}}
=
\phi_i\frac{\partial h_2(\xi,\xi^*)}{\partial\xi_j^*}.
\end{equation*}
Substituting these two identities into the equation of motion of the composite system, we get
\begin{equation*}
i\frac{d\Psi_{ij}}{dt}
=
\xi_j\frac{\partial h_1(\phi,\phi^*)}{\partial\phi_i^*}
+
\phi_i\frac{\partial h_2(\xi,\xi^*)}{\partial\xi_j^*}.
\end{equation*}
On the other hand, differentiating the factorized ansatz \(\Psi_{ij}=\phi_i\xi_j\) gives
\begin{equation*}
i\frac{d\Psi_{ij}}{dt}
=
i\xi_j\frac{d\phi_i}{dt}
+
i\phi_i\frac{d\xi_j}{dt}.
\end{equation*}
Comparing the two expressions, we obtain
\begin{equation*}
\xi_j\left(
i\frac{d\phi_i}{dt}
-
\frac{\partial h_1(\phi,\phi^*)}{\partial\phi_i^*}
\right)
+
\phi_i\left(
i\frac{d\xi_j}{dt}
-
\frac{\partial h_2(\xi,\xi^*)}{\partial\xi_j^*}
\right)
=0.
\end{equation*}
Defining
\begin{equation*}
F_i:=
i\frac{d\phi_i}{dt}
-
\frac{\partial h_1(\phi,\phi^*)}{\partial\phi_i^*},
\qquad
G_j:=
i\frac{d\xi_j}{dt}
-
\frac{\partial h_2(\xi,\xi^*)}{\partial\xi_j^*},
\end{equation*}
the previous equation becomes
\begin{equation*}
\xi_j F_i+\phi_i G_j=0.
\end{equation*}
For nonzero components $\phi_i$ and $\xi_j$, this can be written as
\begin{equation*}
\frac{F_i}{\phi_i}=-\frac{G_j}{\xi_j}.
\end{equation*}
The left-hand side depends only on the index $i$, while the right-hand side depends only on the index $j$. Since the equality must hold for all $i$ and $j$, both sides must be equal to the same constant, say $\lambda$:
\begin{equation*}
\frac{F_i}{\phi_i}=\lambda,
\qquad
-\frac{G_j}{\xi_j}=\lambda.
\end{equation*}
Thus
\begin{equation*}
F_i=\lambda\phi_i,
\qquad
G_j=-\lambda\xi_j.
\end{equation*}
The constant $\lambda$ only produces opposite overall phase rotations of the two factors and therefore cancels in the product $\Psi_{ij}=\phi_i\xi_j$. It can be absorbed into a redefinition of the phases of $\phi$ and $\xi$. We may therefore set it to zero without loss of generality, obtaining
\begin{equation*}
i\frac{d\phi_i}{dt}
=
\frac{\partial h_1(\phi,\phi^*)}{\partial\phi_i^*},
\qquad
i\frac{d\xi_j}{dt}
=
\frac{\partial h_2(\xi,\xi^*)}{\partial\xi_j^*}.
\end{equation*}
Therefore, if the initial state is factorized, the nonlinear equations preserve this factorized form, and each subsystem evolves independently according to its own Hamiltonian function. This shows that separability of non-interacting systems is preserved by Weinberg's construction.
This shows that the nonlinear composition rule preserves the separability of noninteracting systems: if the initial state is factorized, it remains factorized and each subsystem evolves according to its own Hamiltonian function.
This property is essential. A nonlinear theory that failed to preserve separability would introduce unphysical correlations between independent systems. In this sense, separability is one of the structural requirements that must be retained from ordinary QM. By contrast, the superposition principle is not imposed as a fundamental dynamical requirement: in a nonlinear theory, linear combinations of solutions are not in general solutions. Thus Weinberg's construction preserves the tensor-product separability of independent systems, while allowing the dynamical superposition principle to fail.

\subsection{Integrability and Time Evolution}
We have shown that Weinberg's nonlinear QM can be written as a Hamiltonian system on the phase space coordinatized by the real variables $(q_k,p_k)$ defined in Eq.~\textcolor{red}{\eqref{eq:weinberg_real_canonical_variables}}.
It is therefore natural to analyze the time evolution using the language of classical Hamiltonian mechanics. The essential question is whether the nonlinear equations of motion can be integrated explicitly, and how this differs from the standard linear case.

Let us first recall why the usual solution of standard QM takes the form of a superposition of energy eigenstates. In the linear theory, the Hamiltonian functional has the bilinear form of Eq.~\textcolor{red}{\eqref{eq:hamiltonian_function}}, where $H$ is Hermitian. By the spectral theorem, one can choose an orthonormal basis of eigenvectors of $\hat H$.
In this basis, the equations of motion decouple and give the component evolution in Eq.~\textcolor{red}{\eqref{eq:linear_energy_mode_evolution}}.
The time evolution of each component is therefore
\begin{equation*}
    \psi_k(t)=c_k e^{-iE_kt},
    \qquad
    c_k=\psi_k(0),
\end{equation*}
and the complete state is
\begin{equation*}
    |\psi(t)\rangle
    =
    \sum_k c_k e^{-iE_kt}|k\rangle.
\end{equation*}

This result relies on the diagonalizability of the Hermitian
Hamiltonian and the linearity of the equation of motion.
In the energy eigenbasis, each component evolves independently:
its evolution is unchanged by the presence or absence of other
nonzero components in the initial state. Thus, the solutions
obtained by exciting each energy eigenstate separately can be
combined to construct the solution for an arbitrary initial state,
including one in which all components are nonzero.

The same structure can be understood from the point of view of conserved quantities. In the linear theory, if $\hat H$ is conserved, then all its powers
\begin{equation*}
\hat H,\ \hat H^2,\ \dots,\ \hat H^N
\end{equation*}
are also conserved, due to the associativity of operator multiplication.
In a finite-dimensional system with non-degenerate eigenvalues, the spectral projectors can be written as polynomials in $\hat H$:
\begin{equation*}
\hat P_j=\prod_{\ell\neq j}\frac{\hat H-E_\ell \mathbb{I}}{E_j-E_\ell}.
\end{equation*}
Therefore, the conservation of the powers of $\hat H$ implies the conservation of the projectors $\hat P_j$, and hence of the quantities
\begin{equation*}
I_j=\langle\psi|\hat P_j|\psi\rangle=|c_j|^2.
\end{equation*}
These conserved quantities determine the amplitudes $|c_j|$, while the equations of motion determine the phases. This is why the general solution reduces to a superposition of independent phase rotations.

In the nonlinear theory, the Hamiltonian is a general function
\begin{equation*}
h(\psi,\psi^*),
\end{equation*}
and the equations of motion retain the form of Eq.~\textcolor{red}{\eqref{eq:weinberg_eom_discrete}}.
In general, the equations are nonlinear and do not decouple.
More importantly, the algebraic mechanism that underlies the linear case is lost. The product $*$ is not associative, and therefore the conservation of $h$ does not imply the conservation of higher products such as $h*h$. As a consequence, there is no automatic construction of a complete set of independent conserved quantities in involution, and the dynamics is therefore no longer guaranteed to be integrable.

The appropriate notion in the nonlinear case is Liouville integrability ~\textcolor{red}{\cite{CannasDaSilva2007}}. Since the system has $N$ complex variables, or equivalently $2N$ real canonical variables, integrability requires the existence of $N$ functionally independent conserved quantities in involution,
\begin{equation*}
I_1,\dots,I_N,
\qquad
\{I_r,I_s\}_{\rm PB}=0.
\end{equation*}
If a complete set of $N$ independent conserved quantities in
involution exists, then, near a compact connected regular
invariant torus, one can introduce canonical action-angle
variables $(I_r,\theta_r)$. The actions are constant on each
torus, while the angles describe the position on it.
In these coordinates the Hamiltonian depends only on the
actions, and Hamilton's equations become
\begin{equation*}
    \dot I_r=-\frac{\partial h}{\partial\theta_r}=0,
    \qquad
    \dot\theta_r=\frac{\partial h}{\partial I_r}
    =\omega_r(I).
\end{equation*}
Hence
\begin{equation*}
    \theta_r(t)=\theta_r(0)+\omega_r(I)t.
\end{equation*}
Using the definition in
Eq.~\textcolor{red}{\eqref{eq:weinberg_real_canonical_variables}},
the original amplitudes are expressed in these coordinates as
\begin{equation*}
    \psi_k(I,\theta)
    =
    \frac{q_k(I,\theta)+ip_k(I,\theta)}{\sqrt{2}}.
\end{equation*}
Since each angle is defined modulo $2\pi$, these functions
are periodic in every angle. At fixed actions, they can
therefore be expanded in Fourier modes on the invariant torus:
\begin{equation*}
\psi_k(t)=\sum_{\mathbf n\in\mathbb{Z}^N}c_k(\mathbf n;I)\,
\operatorname{exp}\!\left[-i\,\mathbf n\cdot\boldsymbol{\theta}(t)\right].
\end{equation*}
Using the linear evolution of the angles, this becomes
\begin{equation*}
\psi_k(t)=\sum_{\mathbf n\in\mathbb{Z}^N}c_k(\mathbf n;I)\,
\operatorname{exp}\!\left[-i\,\mathbf n\cdot\boldsymbol{\theta}(0)\right]
\operatorname{exp}\!\left[-i\,\mathbf n\cdot\boldsymbol{\omega}(I)t\right].
\end{equation*}
Equivalently, absorbing the initial phases into the coefficients,
\begin{equation}
\psi_k(t)=\sum_{\mathbf n\in\mathbb{Z}^N}\tilde c_k(\mathbf n;I)\,
\operatorname{exp}\!\left[-i\,\mathbf n\cdot\boldsymbol{\omega}(I)t\right].
\label{eq:integrable_quasiperiodic_evolution}
\end{equation}
This is the general quasi-periodic solution of an integrable Hamiltonian system. The important point is that, in the nonlinear case, the frequencies $\omega_r(I)$ generally depend on the actions, and therefore on the initial conditions.

We now show explicitly how the usual QM solution is recovered as a special case of the quasi-periodic form.
In the linear theory, choosing the energy eigenbasis, the
Hamiltonian functional is
\begin{equation*}
    h=\sum_j E_j|\psi_j|^2
    =\frac12\sum_j E_j(q_j^2+p_j^2).
\end{equation*}
For each nonzero component, we introduce the action and angle
through the canonical change of variables
\begin{equation*}
    q_j=\sqrt{2I_j}\cos\theta_j,
    \qquad
    p_j=-\sqrt{2I_j}\sin\theta_j.
\end{equation*}
Here $\theta_j$ is defined modulo $2\pi$ and specifies the
position on the circle $q_j^2+p_j^2=2I_j$. Using
Eq.~\textcolor{red}{\eqref{eq:weinberg_real_canonical_variables}},
we obtain
\begin{equation*}
    \psi_j=\frac{q_j+ip_j}{\sqrt2}
    =\sqrt{I_j}\,e^{-i\theta_j},
    \qquad
    I_j=|\psi_j|^2.
\end{equation*}
Thus, the action is the population of the corresponding
energy eigenstate, while the angle is the negative of
the phase of its amplitude.
In these variables,
\begin{equation}
    h(I)=\sum_j E_j I_j.
    \label{eq:linear_hamiltonian_actions}
\end{equation}
Hamilton's equations therefore give
\begin{equation*}
    \dot I_j=-\frac{\partial h}{\partial\theta_j}=0,
\end{equation*}
and
\begin{equation}
    \dot\theta_j=\omega_j(I)
    =\frac{\partial h}{\partial I_j}=E_j.
    \label{eq:linear_action_frequencies}
\end{equation}
Thus, in the linear case the frequencies are constant and
do not depend on the initial state.
Substituting this into the action-angle solution gives
\begin{equation*}
\psi_j(t)=\sqrt{I_j}\,e^{-i\theta_j(t)}
=\sqrt{I_j}\,e^{-i\theta_j(0)}e^{-iE_jt}.
\end{equation*}
Since
\begin{equation*}
\psi_j(0)=\sqrt{I_j}\,e^{-i\theta_j(0)},
\end{equation*}
we recover Eq.~\textcolor{red}{\eqref{eq:linear_energy_mode_evolution}}.
In terms of the Fourier expansion, this means that for each component $\psi_j$ only one Fourier mode is present. More explicitly, the general quasi-periodic expression reduces to
\begin{equation*}
\psi_j(t)=\sum_{\mathbf n\in\mathbb{Z}^N}\tilde c_j(\mathbf n;I)
e^{-i\mathbf n\cdot\boldsymbol{\omega}t}
\quad\longrightarrow\quad
\psi_j(t)=\tilde c_j(\mathbf e_j;I)e^{-i\omega_jt},
\end{equation*}
where $\mathbf e_j$ is the integer vector with components $(\mathbf e_j)_r=\delta_{jr}$. Since $\omega_j=E_j$, this again gives Eq.~\textcolor{red}{\eqref{eq:linear_energy_mode_evolution}}. Combining these components with the corresponding energy
eigenvectors gives the complete time evolution
\begin{equation*}
    |\psi(t)\rangle
    =
    \sum_k \psi_k(0)e^{-iE_kt}|k\rangle.
\end{equation*}
Thus, the standard QM evolution is recovered because the linear Hamiltonian is diagonalizable, the action variables are simply the populations $I_j=|c_j|^2$, and the Fourier expansion collapses to single-frequency modes.

In the nonlinear theory, the Hamiltonian is not generally linear in the actions:
\begin{equation*}
h\neq \sum_j E_j I_j.
\end{equation*}
Instead, one typically has
\begin{equation*}
h=h(I_1,\dots,I_N),
\end{equation*}
so that
\begin{equation*}
\omega_j(I)=\frac{\partial h}{\partial I_j}
\end{equation*}
depends on the conserved quantities and hence on the initial state.
Moreover, the amplitudes $\psi_k$ are generally nonlinear functions of all the angles $\theta_r$, not just of a single angle. Therefore, the Fourier expansion in Eq.~\textcolor{red}{\eqref{eq:integrable_quasiperiodic_evolution}} does not collapse to one mode, but contains many harmonics.
This is why, even in the integrable nonlinear case, the solution is not a simple superposition of energy eigenstates with fixed frequencies.
Finally, if the nonlinear system is not integrable, even this quasi-periodic form is not guaranteed. In that case there may not exist enough independent conserved quantities to introduce action-angle variables, and the motion can be considerably more complicated.

Even in the nonlinear case, stationary states are still defined by
\begin{equation*}
\frac{\partial h}{\partial\psi_k^*}=E\psi_k.
\end{equation*}
For such states, the equations of motion reduce to
\begin{equation*}
i\frac{d\psi_k}{dt}=E\psi_k,
\end{equation*}
and therefore
\begin{equation*}
\psi_k(t)=\psi_k(0)e^{-iEt}.
\end{equation*}
Thus, a state with definite energy evolves only by an overall phase. However, unlike in the linear theory, stationary states do not generate the general solution by superposition.

The linear theory is special because the Hamiltonian is linear in the action variables, as in Eq.~\textcolor{red}{\eqref{eq:linear_hamiltonian_actions}}, so that the frequencies are constant according to Eq.~\textcolor{red}{\eqref{eq:linear_action_frequencies}}, and the Fourier expansion reduces to single modes. This gives the standard QM solution as a superposition of independently rotating energy components.
In the nonlinear theory, the norm is still conserved, but this alone is not sufficient for integrability. A complete set of independent conserved quantities in involution is required. If such a set exists, the motion is quasi-periodic; if not, the dynamics can be considerably more complex.
A crucial conceptual difference now emerges. In the linear theory, the superposition principle ensures that an arbitrary initial state can be decomposed into energy eigenstates, each evolving independently with a simple phase. In the nonlinear case, this mechanism is lost: although stationary states still evolve with a single phase factor, a generic state cannot be written as a linear superposition of such solutions with independently evolving phases.
In this sense, the superposition principle is no longer a dynamical property of the theory, but at most a statement about initial conditions.

\subsection{Small Nonlinear Perturbations}

In the following, we are mainly interested in nonlinear theories which represent small deviations from standard QM. Therefore, we assume that the Hamiltonian functional can be written as
\begin{equation*}
h=h_0+h_1,
\end{equation*}
where $h_0$ is the integrable Hamiltonian corresponding to the linear theory, while $h_1$ is a small nonlinear perturbation.
Since $h_0$ is integrable, we can introduce action-angle variables $(J_k,\theta_k)$ for the unperturbed system. In these variables,
\begin{equation*}
h_0=h_0(J),
\end{equation*}
and the equations of motion are
\begin{equation*}
\dot J_k=0, \qquad \dot\theta_k=\omega_k(J),
\end{equation*}
with
\begin{equation*}
\omega_k(J)=\frac{\partial h_0}{\partial J_k}.
\end{equation*}
In the linear theory considered here, $\omega_k(J)=E_k$
is constant. We retain the notation $\omega_k(J)$ below,
although there is no actual dependence on the actions
in this case.
\newline We now express the full Hamiltonian in the same variables,
\begin{equation*}
h(J,\theta)=h_0(J)+h_1(J,\theta),
\end{equation*}
so that the equations of motion become
\begin{equation*}
\dot J_k=-\frac{\partial h_1}{\partial\theta_k}, \qquad
\dot\theta_k=\omega_k(J)+\frac{\partial h_1}{\partial J_k}.
\end{equation*}
Since $h_1$ is small, these equations exhibit a separation of time scales. Indeed, one has
\begin{equation*}
\dot J_k=O(h_1),
\end{equation*}
while
\begin{equation*}
\dot\theta_k=\omega_k(J)+O(h_1).
\end{equation*}
Thus, the actions vary slowly, whereas the angles evolve rapidly
with frequencies of order $\omega_k$. For rationally independent
frequencies, the angular motion at fixed actions is ergodic on
the torus, so its long-time averages coincide with uniform
averages over the angles. If this averaging takes place on a
time scale short compared with the evolution of the actions,
we can treat $J_k$ as approximately constant and describe the
leading effect of the perturbation through its angular average.
This motivates introducing the average of $h_1$ over the angles:
\begin{equation*}
\langle h_1\rangle(J)
:=
\frac{1}{(2\pi)^N}
\int_0^{2\pi}d\theta_1\cdots\int_0^{2\pi}d\theta_N\, h_1(J,\theta).
\end{equation*}
This average is in general nonzero, since periodicity does not imply vanishing mean value. On the other hand, derivatives with respect to the angle variables have vanishing average on the torus. Indeed, let $f(\theta)$ be smooth and $2\pi$-periodic in each angle. Then
\begin{equation*}
\frac{1}{(2\pi)^N}\int_{\mathbb{T}^N} d^N\theta\,\frac{\partial f}{\partial\theta_k}
=
\frac{1}{(2\pi)^N}\int_0^{2\pi}\prod_{r\neq k}d\theta_r
\left[\int_0^{2\pi}d\theta_k\,\frac{\partial f}{\partial\theta_k}\right].
\end{equation*}
The inner integral is
\begin{equation*}
\int_0^{2\pi}d\theta_k\,\frac{\partial f}{\partial\theta_k}
=
f(\theta_1,\dots,2\pi,\dots,\theta_N)
-
f(\theta_1,\dots,0,\dots,\theta_N)=0,
\end{equation*}
where the last equality follows from periodicity in $\theta_k$. Hence
\begin{equation*}
\left\langle\frac{\partial f}{\partial\theta_k}\right\rangle=0.
\end{equation*}
In contrast, the average
\begin{equation*}
\langle f\rangle=
\frac{1}{(2\pi)^N}\int_{\mathbb{T}^N}d^N\theta\,f(\theta)
\end{equation*}
is not constrained to vanish. Therefore $h_1(J,\theta)$ may contain a nonzero mean part which survives the averaging procedure.
Because the integration is performed over a full period and the boundary terms cancel. This reflects the periodicity of the motion on the torus.
Averaging the equations of motion, one then obtains
\begin{equation}
\dot J_k=0, \qquad
\dot\theta_k=\omega_k(J)+\frac{\partial\langle h_1\rangle}{\partial J_k}.
\label{eq:averaged_action_angle_dynamics}
\end{equation}
These equations have precisely the canonical form of an integrable system in action-angle variables, with effective Hamiltonian
\begin{equation*}
h_{\rm eff}(J)=h_0(J)+\langle h_1\rangle(J).
\end{equation*}
In fact, the equations
\begin{equation*}
\dot J_k=0, \qquad
\dot\theta_k=\frac{\partial h_{\rm eff}}{\partial J_k}
\end{equation*}
show that the averaged dynamics coincides with that generated by $h_{\rm eff}$. In this sense, the original system is effectively equivalent, for long times, to an integrable system with renormalized Hamiltonian.
The physical reason for this equivalence is that the fast oscillations in the angle variables average out over timescales large compared to $1/\omega_k$, so that only their mean effect influences the slow evolution. The rapid variations do not accumulate coherently, while the averaged contribution produces a secular effect on the dynamics.
The net result is a shift in the frequencies,
\begin{equation*}
\omega_k(J)\longrightarrow
\omega_k^{\rm eff}(J)
=
\omega_k(J)+\frac{\partial\langle h_1\rangle}{\partial J_k},
\end{equation*}
which now depend on the actions $J_k$.
This is a crucial difference with respect to the linear theory. In standard QM the frequencies are fixed by the energy eigenvalues and are independent of the state. In the nonlinear case, even when the motion remains approximately quasi-periodic, the effective frequencies depend on the initial conditions through the actions.

\subsection{Galilean Invariance and One-Particle Realizations}
We now discuss how time translations can be embedded into the larger Galilean symmetry group. Up to this point the wave function was described by a finite number of complex amplitudes $\psi_k$. In order to describe a single particle, we now also introduce the position variable $x$, so that the wave function becomes
\begin{equation*}
\Psi_k(x),
\end{equation*}
where $k$ labels the internal degrees of freedom, such as spin. Equivalently, the state may be regarded as living in the tensor product between the internal space and the position space, with basis vectors of the form
\begin{equation*}
\ket{k}\otimes\ket{x}.
\end{equation*}
In this notation, sums over the discrete index $k$ are accompanied by integrals over space, and ordinary derivatives with respect to $\psi_k$ are replaced by functional derivatives with respect to $\Psi_k(x)$. For example, the product between two observables becomes
\begin{equation*}
a*b=\sum_k\int d^3x\,
\frac{\delta a}{\delta\Psi_k(x)}
\frac{\delta b}{\delta\Psi_k^*(x)}.
\end{equation*}
All the algebraic structures introduced above therefore generalize directly to this continuous case.
For a free particle, we require invariance under the Galilean group. This means that the theory must admit generators of time translations, spatial translations, rotations and boosts, denoted respectively by
\begin{equation*}
h,\qquad p,\qquad j,\qquad k,
\end{equation*}
together with the central element $\mathcal M$. These generators must satisfy the Galilean commutation relations
\begin{equation}
\begin{aligned}
[p_i,h]&=0,\qquad [j_i,h]=0,\qquad [k_i,k_j]=0,\\
[k_i,h]&=ip_i,\qquad [k_i,p_j]=i\delta_{ij}\mathcal M,\\
[j_i,p_j]&=i\epsilon_{ijk}p_k,\qquad [j_i,k_j]=i\epsilon_{ijk}k_k,\\
[j_i,j_j]&=i\epsilon_{ijk}j_k.
\end{aligned}
\label{eq:weinberg_galilean_algebra}
\end{equation}
These relations encode the usual meaning of the generators: $p$ generates translations, $j$ rotations, $k$ boosts, and $h$ time translations. Following Weinberg, we keep the standard bilinear realization of $p,j,k$ and $\mathcal M$. This is natural because these are kinematical generators with a fixed geometrical meaning. In contrast, the Hamiltonian is the dynamical object of the theory, and it is precisely here that nonlinear corrections can be introduced.
The conventional realizations are
\begin{equation*}
p=-i\sum_k\int d^3x\,\Psi_k^*(x)\nabla\Psi_k(x),
\end{equation*}
\begin{equation*}
k=m\sum_k\int d^3x\,\Psi_k^*(x)\,x\,\Psi_k(x),
\end{equation*}
\begin{equation*}
j=\sum_{k,l}\int d^3x\,\Psi_k^*(x)
\left[-i(x\times\nabla)\delta_{kl}+S_{kl}\right]\Psi_l(x),
\end{equation*}
and
\begin{equation*}
\mathcal M=m\sum_k\int d^3x\,\Psi_k^*(x)\Psi_k(x)=mn.
\end{equation*}
Here $S_{kl}$ are the spin matrices acting on the internal index $k$. These bilinear functions give the standard realization of translations, rotations and boosts. Since $h$ does not appear on the right-hand side of the commutation relations in a way that uniquely fixes its form, one can add nonlinear contributions to the Hamiltonian, provided that the full set of commutation relations remains satisfied.
For a free particle, the Hamiltonian can then be written as
\begin{equation*}
h=-\frac{1}{2m}\sum_k\int d^3x\,\Psi_k^*(x)\nabla^2\Psi_k(x)+h',
\end{equation*}
where the first term is the usual bilinear kinetic energy and $h'$ is an internal energy functional. Since we are describing a free system, no external field is present and no spatial direction can be preferred. Therefore $h'$ must commute with the generators of translations, boosts and rotations:
\begin{equation*}
[h',p_i]=0,\qquad [h',k_i]=0,\qquad [h',j_i]=0.
\end{equation*}
These conditions are not only physically motivated, but also required for consistency. Indeed, the total Hamiltonian $h=h_0+h'$ must satisfy the Galilean commutation relations in Eq.~\textcolor{red}{\eqref{eq:weinberg_galilean_algebra}}, and since the kinetic term $h_0$ already satisfies them, any additional contribution $h'$ must commute with the generators $p_i$, $k_i$ and $j_i$ in order not to spoil the algebra. The first two conditions imply that $h'$ is an internal contribution, independent of the center-of-mass motion. The last condition is the requirement of rotational invariance: $h'$ may describe self-interactions among the internal components of the wave function, but it must be a scalar under rotations.
A simple local choice is
\begin{equation*}
h'=\int d^3x\,\mathcal H(\Psi(x),\Psi^*(x)),
\end{equation*}
where $\mathcal H$ is a scalar function of the spin components of the wave function at the same point $x$. Rotational invariance means that $\mathcal H$ must be invariant under infinitesimal spin rotations. If the internal components transform with spin matrices $S_i$, this condition can be written as
\begin{equation*}
0=
\frac{\partial\mathcal H}{\partial\Psi_k}(S_i)_{kl}\Psi_l
-
\frac{\partial\mathcal H}{\partial\Psi_k^*}(S_i)_{lk}\Psi_l^*,
\end{equation*}
for each generator $S_i$. This equation states that the change of $\mathcal H$ under an infinitesimal rotation of the spin components vanishes.
The homogeneity requirement is the same as before:
\begin{equation*}
\sum_k\frac{\partial\mathcal H}{\partial\Psi_k}\Psi_k
=
\sum_k\frac{\partial\mathcal H}{\partial\Psi_k^*}\Psi_k^*
=
\mathcal H.
\end{equation*}
These conditions ensure that the physical state is still represented by a ray, so that multiplying the wave function by an overall complex factor does not change the physical content of the theory.
To construct the most general local nonlinear terms, one may expand $\mathcal H$ as a sum of homogeneous scalar contributions. Let
\begin{equation*}
\rho(x)=\sum_k\Psi_k^*(x)\Psi_k(x).
\end{equation*}
A term containing $1+s$ factors of $\Psi$ and $1+s$ factors of $\Psi^*$ has degree $1+s$ in each variable. Dividing by $\rho^s$, which has degree $s$ in both $\Psi$ and $\Psi^*$, gives a term homogeneous of degree one in both variables. Thus one can write
\begin{equation*}
\mathcal H(\Psi,\Psi^*)=\sum_{s=0}^{\infty}\rho^{-s}\mathcal H_s(\Psi,\Psi^*),
\end{equation*}
where $\mathcal H_s$ is built from $1+s$ factors of $\Psi$ and $1+s$ factors of $\Psi^*$, with the coefficients chosen so that each $\mathcal H_s$ is a rotational scalar.
For example, the bilinear term has the form
\begin{equation*}
\mathcal H_0=\sum_{k,l}\Psi_k^*H_{kl}\Psi_l,
\end{equation*}
while the first genuinely nonlinear contribution is quartic in the wave function and divided by $\rho$:
\begin{equation*}
\rho^{-1}\mathcal H_1
=
\rho^{-1}\sum_{k,l,m,n}
G_{klmn}\Psi_k^*\Psi_l^*\Psi_m\Psi_n.
\end{equation*}
The role of the factor $\rho^{-s}$ is precisely to preserve the homogeneity required by the ray structure of the theory. In this way, the nonlinear internal Hamiltonian is built as a local homogeneous rotational scalar, while the kinematical generators of translations, rotations and boosts retain their standard QM form. If the spin matrices furnish an irreducible representation,
Schur's lemma~\textcolor{red}{\cite{Woit2017}} implies that
the matrix defining the bilinear term is proportional to
the identity. This term therefore contributes only a
constant shift of the energy, which can be absorbed into
the linear Hamiltonian or removed by a choice of the energy
origin, since it affects only the global phase of the state. 
\newline It should be noted that this is not the most general way to construct a homogeneous Hamiltonian. However, it is the most general local realization compatible with the symmetry and homogeneity requirements.
The time evolution of the wave function is then obtained from the Hamiltonian functional through the functional derivative, according to
\begin{equation*}
i\frac{\partial \Psi_k(x)}{\partial t}
=
\frac{\delta h}{\delta \Psi_k^*(x)}.
\end{equation*}
Substituting the explicit form of the Hamiltonian, one finds
\begin{equation}
i\frac{\partial \Psi_k(x)}{\partial t}
=
-\frac{1}{2m}\nabla^2 \Psi_k(x)
+
\frac{\partial \mathcal H(\Psi(x),\Psi^*(x))}{\partial \Psi_k^*(x)}.
\label{eq:continuous_nonlinear_se}
\end{equation}
This equation represents the nonlinear generalization of the SE in Weinberg's formalism. The first term is the usual kinetic contribution, while the second term encodes the nonlinear self-interaction of the wave function through the local functional $\mathcal H$.
The separability mechanism described in Eq.~(\textcolor{red}{\ref{eq:composite_eom_nonlinear_form}}) explains the separability of the one-particle wave function with spatial and internal degrees of freedom. If the wave function is written as $\Psi_k(x,t)$, one may regard the label $x$ and the discrete index $k$ as playing the role of two different degrees of freedom. For Hamiltonians of the local homogeneous form introduced above, the ansatz
\begin{equation*}
\Psi_k(x,t)=\varphi(x,t)\psi_k(t)
\end{equation*}
leads to two separated equations:
\begin{equation*}
i\frac{\partial\varphi}{\partial t}
=
-\frac{1}{2m}\nabla^2\varphi,
\end{equation*}
and
\begin{equation*}
i\frac{d\psi_k}{dt}
=
\frac{\partial \mathcal H(\psi,\psi^*)}{\partial\psi_k^*}.
\end{equation*}
The first equation describes the free translational motion, while the second contains the nonlinear internal dynamics. Therefore, after establishing this separation, one can consistently ignore the translational degree of freedom and use the finite-dimensional formalism developed in the first part of this thesis for systems with a finite number of degrees of freedom, with the Hamiltonian function replaced by the internal Hamiltonian density $\mathcal H$.

\section{Entropy and Non-Unitarity}
We now discuss an important objection raised by Peres against nonlinear variants of the SE \textcolor{red}{\cite{Peres1989}}. The central point is that, if the evolution of pure states is nonlinear, then the scalar product between two different states is not preserved in general. This may lead to a decrease of the entropy of a statistical mixture, in violation of the second law of thermodynamics.
Let us consider two normalized states $\phi$ and $\psi$. Their scalar product is given by Eq.~\textcolor{red}{\eqref{eq:inner_product_components}},
\begin{equation*}
\langle\phi|\psi\rangle=\sum_k\phi_k^*\psi_k.
\end{equation*}
In the linear theory, the time evolution is unitary and therefore
\begin{equation*}
|\langle\phi(t)|\psi(t)\rangle|^2=|\langle\phi(0)|\psi(0)\rangle|^2.
\end{equation*}
In the nonlinear theory, however, the two states evolve according to equations whose right-hand side depends on the state itself. Therefore the cancellation which guarantees the conservation of the scalar product in the linear case no longer occurs in general.More explicitly, using
Eqs.~\textcolor{red}{\eqref{eq:weinberg_eom_conjugate}}
and~\textcolor{red}{\eqref{eq:weinberg_eom_discrete}},
we have
\begin{equation*}
    \frac{d\phi_k^*}{dt}
    =
    i\frac{\partial h(\phi,\phi^*)}{\partial\phi_k},
    \qquad
    \frac{d\psi_k}{dt}
    =
    -i\frac{\partial h(\psi,\psi^*)}{\partial\psi_k^*}.
\end{equation*}
Differentiating the scalar product therefore gives
\begin{align*}
    \frac{d}{dt}\langle\phi|\psi\rangle
    &=
    \sum_k\left(
        \frac{d\phi_k^*}{dt}\psi_k
        +
        \phi_k^*\frac{d\psi_k}{dt}
    \right)\\
    &=
    i\sum_k
    \frac{\partial h(\phi,\phi^*)}{\partial\phi_k}\psi_k
    -
    i\sum_k
    \phi_k^*
    \frac{\partial h(\psi,\psi^*)}{\partial\psi_k^*}.
\end{align*}
For a bilinear Hamiltonian this expression vanishes because the same linear Hermitian operator acts on both states. For a nonlinear Hamiltonian, instead, the functional derivatives are evaluated on different states, $\phi$ and $\psi$, and there is no general reason for the two terms to cancel. Thus the scalar product between different states is not preserved in general.
We now consider a statistical mixture of two pure states. This is described by the density matrix
\begin{equation*}
\rho=\lambda P_\phi+(1-\lambda)P_\psi,
\qquad
0<\lambda<1,
\end{equation*}
where
\begin{equation*}
P_\phi=|\phi\rangle\langle\phi|,
\qquad
P_\psi=|\psi\rangle\langle\psi|.
\end{equation*}
This density matrix represents an ensemble in which a fraction $\lambda$ of the systems is prepared in the state $\phi$, while the remaining fraction $1-\lambda$ is prepared in the state $\psi$. The nonzero eigenvalues of $\rho$ are
\begin{equation*}
w_\pm=
\frac{1}{2}\left[
1\pm
\sqrt{1-4\lambda(1-\lambda)(1-x)}
\right],
\end{equation*}
where
\begin{equation*}
x:=|\langle\phi|\psi\rangle|^2.
\end{equation*}
The derivation of these eigenvalues is given in Appendix~\ref{app:entropy}.
The entropy of the mixture is
\begin{equation*}
S=-k_B\sum_{\pm}w_\pm\operatorname{ln}( w_\pm).
\end{equation*}
Since the entropy decreases as the overlap $x$ increases, a closed evolution cannot increase $|\langle\phi|\psi\rangle|^2$ without decreasing the entropy of the mixture. Therefore, in order to avoid a violation of the second law, one must require
\begin{equation*}
|\langle\phi(t)|\psi(t)\rangle|^2
\leq
|\langle\phi(0)|\psi(0)\rangle|^2
\end{equation*}
for every pair of states.
However, this condition cannot be satisfied consistently unless all scalar products are exactly preserved. 

To see this, consider a complete orthonormal basis
$\{|\phi_n(0)\rangle\}$.
Any two distinct basis states define an equal-weight mixture
whose entropy would decrease if they became nonorthogonal.
Thus, requiring that the entropy of no isolated mixture
decrease, together with norm conservation, ensures that
$\{|\phi_n(t)\rangle\}$ remains an orthonormal basis at every time.
For every normalized state $|\psi(t)\rangle$, one therefore has
\begin{equation*}
    \sum_n|\langle\phi_n(t)|\psi(t)\rangle|^2=1.
\end{equation*}
If during the evolution one component decreases,
\begin{equation*}
|\langle\phi_m(t)|\psi(t)\rangle|^2
<
|\langle\phi_m(0)|\psi(0)\rangle|^2,
\end{equation*}
then, since the sum over all components must remain equal to one, at least one other component must increase:
\begin{equation*}
|\langle\phi_n(t)|\psi(t)\rangle|^2
>
|\langle\phi_n(0)|\psi(0)\rangle|^2.
\end{equation*}
But this increase implies that the entropy of a mixture of $\phi_n$ and $\psi$ decreases. Hence, avoiding entropy decrease for every possible mixture
requires each overlap with the evolving basis states to remain
constant. Since any normalized initial state $|\phi(0)\rangle$
can be chosen as the first vector of a complete orthonormal
basis, the same argument applies to any pair of normalized states
$|\phi(0)\rangle$ and $|\psi(0)\rangle$. Therefore,
\begin{equation*}
|\langle\phi(t)|\psi(t)\rangle|^2
=
|\langle\phi(0)|\psi(0)\rangle|^2
\end{equation*}
for all pairs of states.

This condition is extremely strong. By Wigner's theorem,
any bijective transformation preserving transition probabilities
between all pairs of rays can be represented by either a unitary
or an antiunitary operator ~\textcolor{red}{\cite{PreskillNotes}}.
Since time evolution is continuously connected to the identity,
the antiunitary possibility is excluded.
The dynamics is therefore equivalent, up to unobservable overall
phases, to a linear unitary evolution. This contradicts a genuinely
nonlinear dynamics of physical states. Hence, under the usual
statistical and thermodynamic assumptions, nonlinear evolution
can decrease the entropy of an isolated mixture.
This also clarifies the connection with superluminal signaling. In standard QM, two different decompositions of the same density matrix are physically indistinguishable. For an entangled bipartite system, a measurement performed by Alice can prepare Bob's subsystem in different ensembles which nevertheless correspond to the same reduced density matrix. If the subsequent evolution is linear, Bob cannot distinguish which ensemble Alice prepared. In a nonlinear theory, however, the evolution can depend on the particular pure-state decomposition, not only on the density matrix itself. Bob could then detect Alice's choice by observing the later evolution of his subsystem, allowing information to be transmitted without any classical signal.
Thus, although Weinberg's nonlinear theory is mathematically elegant and preserves several structural aspects of QM, Peres' argument shows that nonlinear dynamics is deeply problematic once the standard statistical interpretation is kept. The failure to preserve scalar products leads to entropy-decreasing processes and opens the door to superluminal signaling. Together with the absence of experimental evidence for such effects, this strongly constrains possible nonlinear corrections to QM. If they exist at all, they must occur at scales far beyond present experimental accessibility.

\newpage
\null
\thispagestyle{empty}
\newpage

\chapter{Pilot Wave Theory}
\label{chap:2}

In Chapter~1, we investigated the possibility of modifying QM while retaining its interpretation as a fundamentally indeterministic wave theory. Motivated by the nonlinear character of classical electromagnetic dynamics, we considered nonlinear corrections to the SE. Within the framework examined there, however, these modifications led to an inconsistency with the second law of thermodynamics, closely related to the possibility of superluminal signalling. The resulting theory therefore proved incompatible with the physical principles underlying our current description of nature.
This motivates the exploration of a conceptually different approach. Rather than modifying the SE directly, in this chapter we introduce the PWT, originally proposed by Louis de Broglie and subsequently developed by David Bohm \textcolor{red}{\cite{Bohm_I,Bohm_II}}. The latter formulation is commonly known as Bohmian Mechanics (BM). In its standard formulation and under the quantum-equilibrium hypothesis, the PWT introduces no corrections to the statistical predictions of QM. Nevertheless, it provides a radically different conceptual framework: particles possess definite configurations and follow continuous trajectories, while their dynamics is determined by the wave function.
The purpose of introducing the PWT is therefore not to propose a different phenomenology at this stage, but to establish a framework in which QM can be related to an underlying deterministic dynamics. In particular, the connection between the PWT, the Hamilton--Jacobi (HJ) formulation of CM, and the dynamics of statistical ensembles will provide the conceptual basis for investigating whether the SE and the associated wave function may be interpreted as emergent, rather than fundamental, structures.

\section{The Hamilton--Jacobi Formalism}

We now introduce the HJ formulation of CM
\textcolor{red}{\cite{Goldstein_classical}}. Although completely equivalent to
Hamiltonian mechanics, this formulation describes the dynamics through a scalar
function defined on configuration space. As we shall see, it also provides a
natural description of an ensemble of trajectories and establishes a direct
connection with the SE and the PWT
\textcolor{red}{\cite{Holland_QTM}}.

\subsection{Canonical transformations}

Consider a classical system with $n$ degrees of freedom, described by the
canonical variables $(\mathbf q,\mathbf p)$ and by the Hamiltonian
$H(\mathbf q,\mathbf p,t)$. Its action is
\begin{equation}
    I[\mathbf q,\mathbf p]
    =
    \int_{t_0}^{t_f}
    \left(
        \sum_i p_i\dot q_i-H
    \right)dt.
\label{eq:classical_phase_space_action}
\end{equation}
The physical trajectories are the curves that make this action stationary,
and therefore satisfy
\begin{equation*}
    \dot q_i=\frac{\partial H}{\partial p_i},
    \qquad
    \dot p_i=-\frac{\partial H}{\partial q_i}.
\end{equation*}

We now introduce an invertible transformation
\begin{equation*}
    Q_i=Q_i(q,p,t),
    \qquad
    P_i=P_i(q,p,t).
\end{equation*}
We seek a canonical transformation: the new variables must
describe the same physical motion while retaining Hamilton's
equations with a transformed Hamiltonian $K$.
In the action formulation, we therefore require the new action
to have the canonical form
\begin{equation*}
    I'[Q,P]
    =
    \int_{t_0}^{t_f}
    \left(\sum_i P_i\dot Q_i-K(Q,P,t)\right)dt,
\end{equation*}
and to satisfy
\begin{equation*}
    \delta I=\delta I'
\end{equation*}
for corresponding variations with vanishing boundary contributions.
\newline The canonical form of $I'$ yields Hamilton's equations for $Q,P$.
Since the transformation is invertible, every curve
$(\mathbf q(t),\mathbf p(t))$ determines a unique curve
$(\mathbf Q(t),\mathbf P(t))$, and vice versa. The equality of the two variations then guarantees that a stationary trajectory in the original variables corresponds to a stationary trajectory in the new variables:
\begin{equation*}
    \delta I=0
    \quad\Longleftrightarrow\quad
    \delta I'=0.
\end{equation*}
Thus, a canonical transformation establishes a one-to-one correspondence
between the physical solutions in the two descriptions.

Varying the transformed action gives
\begin{equation*}
\begin{split}
    \delta I'
    =
    \int_{t_0}^{t_f}
    \sum_i
    \Bigg[
        \delta P_i\,\dot Q_i
        +P_i\,\delta\dot Q_i
        -\frac{\partial K}{\partial Q_i}\delta Q_i
        -\frac{\partial K}{\partial P_i}\delta P_i
    \Bigg]dt.
\end{split}
\end{equation*}
Integrating the term containing $\delta\dot Q_i$ by parts and fixing the
endpoints, we obtain
\begin{equation*}
    \delta I'
    =
    \int_{t_0}^{t_f}
    \sum_i
    \left[
        \left(
            \dot Q_i-\frac{\partial K}{\partial P_i}
        \right)\delta P_i
        -
        \left(
            \dot P_i+\frac{\partial K}{\partial Q_i}
        \right)\delta Q_i
    \right]dt.
\end{equation*}
Since $\delta Q_i$ and $\delta P_i$ are independent, the stationarity
condition $\delta I'=0$ implies
\begin{equation}
    \boxed{
    \dot Q_i=\frac{\partial K}{\partial P_i},
    \qquad
    \dot P_i=-\frac{\partial K}{\partial Q_i}.
    }
    \label{eq:new_hamilton_equations}
\end{equation}
The new variables therefore satisfy Hamilton equations generated by the
transformed Hamiltonian $K(\mathbf Q,\mathbf P,t)$.

The equality $\delta I=\delta I'$ is ensured if the two action integrands
differ only by a total time derivative:
\begin{equation*}
    \sum_i p_i\dot q_i-H
    =
    \sum_iP_i\dot Q_i-K
    +
    \frac{dF_1}{dt},
\end{equation*}
where $F_1(\mathbf q,\mathbf Q,t)$ is called a generating function. Indeed,
after integration, the additional term contributes only at the endpoints:
\begin{equation*}
    I
    =
    I'
    +
    F_1(t_f)-F_1(t_0).
\end{equation*}
For fixed endpoints this term has vanishing variation and consequently does
not change the equations of motion.

The relation between the two action integrands may equivalently be written as
\begin{equation*}
    dF_1
    =
    \sum_i p_i\,dq_i
    -
    \sum_iP_i\,dQ_i
    +(K-H)\,dt.
\end{equation*}
Since $F_1=F_1(\mathbf q,\mathbf Q,t)$, its differential is also
\begin{equation*}
    dF_1
    =
    \sum_i
    \frac{\partial F_1}{\partial q_i}\,dq_i
    +
    \sum_i
    \frac{\partial F_1}{\partial Q_i}\,dQ_i
    +
    \frac{\partial F_1}{\partial t}\,dt.
\end{equation*}
Comparing the coefficients of the independent differentials gives
\begin{equation*}
    p_i=\frac{\partial F_1}{\partial q_i},
    \qquad
    P_i=-\frac{\partial F_1}{\partial Q_i},
    \qquad
    K=H+\frac{\partial F_1}{\partial t}.
\end{equation*}

For the derivation of the HJ equation, it is more convenient to replace the
dependence on $\mathbf Q$ with a dependence on $\mathbf P$. We therefore
introduce the generating function of the second type,
\begin{equation*}
    F_2(\mathbf q,\mathbf P,t)
    =
    F_1(\mathbf q,\mathbf Q,t)
    +
    \sum_iQ_iP_i.
\end{equation*}
Its differential is
\begin{equation*}
\begin{split}
    dF_2
    &=
    dF_1
    +
    \sum_iP_i\,dQ_i
    +
    \sum_iQ_i\,dP_i
    \\
    &=
    \sum_i p_i\,dq_i
    +
    \sum_iQ_i\,dP_i
    +(K-H)\,dt.
\end{split}
\end{equation*}
On the other hand, directly differentiating
$F_2(\mathbf q,\mathbf P,t)$ gives
\begin{equation*}
    dF_2
    =
    \sum_i
    \frac{\partial F_2}{\partial q_i}\,dq_i
    +
    \sum_i
    \frac{\partial F_2}{\partial P_i}\,dP_i
    +
    \frac{\partial F_2}{\partial t}\,dt.
\end{equation*}

Comparing the coefficients, we finally obtain
\begin{equation}
    \boxed{
    p_i=\frac{\partial F_2}{\partial q_i},
    \qquad
    Q_i=\frac{\partial F_2}{\partial P_i},
    \qquad
    K
    =
    H\left(
        \mathbf q,
        \frac{\partial F_2}{\partial\mathbf q},
        t
    \right)
    +
    \frac{\partial F_2}{\partial t}.
    }
    \label{eq:canonical_transformation_F2}
\end{equation}
The first relation is used to replace the old momenta appearing in
$H(\mathbf q,\mathbf p,t)$ by
$\mathbf p=\partial F_2/\partial\mathbf q$. The resulting expression for $K$
is initially written in the mixed variables $(\mathbf q,\mathbf P,t)$ and
therefore does not yet involve only the new canonical variables.

To obtain the transformed Hamiltonian explicitly as
$K(\mathbf Q,\mathbf P,t)$, the second relation,
\begin{equation*}
    \mathbf Q
    =
    \frac{\partial F_2(\mathbf q,\mathbf P,t)}
    {\partial\mathbf P},
\end{equation*}
must be inverted to determine
\begin{equation*}
    \mathbf q=\mathbf q(\mathbf Q,\mathbf P,t).
\end{equation*}
Substituting this result into the last relation of
Eq.~\textcolor{red}{\eqref{eq:canonical_transformation_F2}} expresses $K$ entirely in terms of
the new variables $(\mathbf Q,\mathbf P)$ and time.

\subsection{The Hamilton--Jacobi Equation}

The purpose of the HJ formalism is to determine a canonical transformation for
which the new variables are constants of motion. From Hamilton's equations in the new variables, Eq.~\textcolor{red}{\eqref{eq:new_hamilton_equations}}, the conditions $\dot Q_i=\dot P_i=0$ imply that the transformed Hamiltonian
may depend only on time. Since a function of time alone does not affect
Hamilton's equations and can be absorbed into the generating function, the
transformed Hamiltonian may be chosen, without loss of generality, as
\begin{equation*}
    K=0.
\end{equation*}

Let $S$ denote the generating function of the second type associated with this
transformation. Since a generating function of the second type depends on the
old coordinates and the new momenta, it initially has the form
\begin{equation*}
    S=S(\mathbf q,\mathbf P,t).
\end{equation*}
Using the transformation law
\begin{equation*}
    K
    =
    H\left(
        \mathbf q,
        \frac{\partial S}{\partial\mathbf q},
        t
    \right)
    +
    \frac{\partial S}{\partial t},
\end{equation*}
the choice $K=0$ gives
\begin{equation}
    \boxed{
    \frac{\partial S}{\partial t}
    +
    H\left(
        \mathbf q,
        \frac{\partial S}{\partial\mathbf q},
        t
    \right)
    =0.
    }
    \label{eq:hamilton_jacobi_equation}
\end{equation}
This is the HJ equation, and $S$ is Hamilton's principal function.

For a system with $n$ degrees of freedom, a complete solution of
Eq.~\textcolor{red}{\eqref{eq:hamilton_jacobi_equation}} has the form
\begin{equation*}
    S
    =
    S(q_1,\ldots,q_n;
      \alpha_1,\ldots,\alpha_n;
      t),
\end{equation*}
where the $\alpha_i$ are $n$ independent integration constants, none of which
is purely additive. By definition, a solution is complete only if these
constants are independent, which is expressed by the non-degeneracy condition
\begin{equation*}
    \det\left(
        \frac{\partial^2S}
        {\partial q_i\,\partial\alpha_j}
    \right)
    \neq0.
\end{equation*}
This condition will ensure that the canonical transformation generated by $S$
can be inverted locally.

We may therefore identify the integration constants with the new momenta,
\begin{equation*}
    P_i=\alpha_i.
\end{equation*}
This identification is consistent with the construction because, for $K=0$,
the new momenta are constants of motion.

The canonical transformation generated by $S$ is consequently given by
\begin{equation}
    \boxed{
    p_i
    =
    \frac{\partial S(\mathbf q,\boldsymbol{\alpha},t)}
    {\partial q_i},
    \qquad
    Q_i
    =
    \frac{\partial S(\mathbf q,\boldsymbol{\alpha},t)}
    {\partial\alpha_i}
    =
    \beta_i.
    }
    \label{eq:hj_canonical_transformation}
\end{equation}
The quantities $\beta_i$ are the constant values of the new coordinates,
which are also conserved because $K=0$.

The constants $\boldsymbol{\alpha}$ and $\boldsymbol{\beta}$ are fixed by the
initial conditions of the original system. Let
\begin{equation*}
    \mathbf q(t_0)=\mathbf q_0,
    \qquad
    \mathbf p(t_0)=\mathbf p_0.
\end{equation*}
Evaluating the first set of transformation equations at $t=t_0$ gives
\begin{equation*}
    p_{0i}
    =
    \frac{\partial S(\mathbf q_0,\boldsymbol{\alpha},t_0)}
    {\partial q_i},
    \qquad i=1,\ldots,n.
\end{equation*}
These are $n$ equations for the $n$ constants $\alpha_i$. Owing to the
non-degeneracy condition of the complete solution, they can be inverted
locally to determine
\begin{equation*}
    \boldsymbol{\alpha}
    =
    \boldsymbol{\alpha}(\mathbf q_0,\mathbf p_0).
\end{equation*}

Once these values have been found, the second set of transformation equations
is evaluated at the initial time:
\begin{equation*}
    \beta_i
    =
    \frac{\partial
    S(\mathbf q_0,\boldsymbol{\alpha},t_0)}
    {\partial\alpha_i}.
\end{equation*}
This determines
\begin{equation*}
    \boldsymbol{\beta}
    =
    \boldsymbol{\beta}(\mathbf q_0,\mathbf p_0).
\end{equation*}
Thus, the initial canonical variables determine the constant values of all the
new canonical variables.

Keeping the resulting $\boldsymbol{\alpha}$ and $\boldsymbol{\beta}$ fixed,
the equations
\begin{equation*}
    \beta_i
    =
    \frac{\partial S(\mathbf q,\boldsymbol{\alpha},t)}
    {\partial\alpha_i}
\end{equation*}
can now be inverted with respect to the original coordinates. This gives
\begin{equation*}
    q_i=q_i(\boldsymbol{\alpha},\boldsymbol{\beta},t).
\end{equation*}
Substituting these expressions into the first set of transformation equations
then gives
\begin{equation*}
    p_i
    =
    \left.
    \frac{\partial S(\mathbf q,\boldsymbol{\alpha},t)}
    {\partial q_i}
    \right|_{
        \mathbf q=
        \mathbf q(\boldsymbol{\alpha},\boldsymbol{\beta},t)
    }
    =
    p_i(\boldsymbol{\alpha},\boldsymbol{\beta},t).
\end{equation*}
The complete motion has therefore been reconstructed as
\begin{equation}
    \boxed{
    q_i=q_i(\boldsymbol{\alpha},\boldsymbol{\beta},t),
    \qquad
    p_i=p_i(\boldsymbol{\alpha},\boldsymbol{\beta},t).
    }
    \label{eq:complete_motion_from_hj}
\end{equation}
Since $\boldsymbol{\alpha}$ and $\boldsymbol{\beta}$ are themselves fixed by
$(\mathbf q_0,\mathbf p_0)$, the same solution may equivalently be written as
\begin{equation*}
    q_i=q_i(\mathbf q_0,\mathbf p_0,t),
    \qquad
    p_i=p_i(\mathbf q_0,\mathbf p_0,t).
\end{equation*}
The complete function $S$ determines the family of possible motions, while
the initial conditions select one particular motion from that family.

The fact that the functions in Eq.~\textcolor{red}{\eqref{eq:complete_motion_from_hj}} satisfy
the original equations of motion already follows from the canonical character
of the transformation. It is nevertheless useful to verify explicitly the
relation that will later connect the HJ formulation with the PWT.

Along the reconstructed motion, the quantities $\alpha_a$ and $\beta_a$ are
constant. Taking the total time derivative of
\begin{equation*}
    \beta_a
    =
    \frac{\partial S}{\partial\alpha_a}
\end{equation*}
therefore gives
\begin{equation*}
    0
    =
    \frac{\partial^2S}{\partial\alpha_a\partial t}
    +
    \sum_i
    \frac{\partial^2S}{\partial\alpha_a\partial q_i}
    \dot q_i.
\end{equation*}
On the other hand, differentiating the HJ equation with respect to
$\alpha_a$, while holding $\mathbf q$ and $t$ fixed, gives
\begin{equation*}
    0
    =
    \frac{\partial^2S}{\partial\alpha_a\partial t}
    +
    \sum_i
    \left.
    \frac{\partial H}{\partial p_i}
    \right|_{\mathbf p=\partial S/\partial\mathbf q}
    \frac{\partial^2S}{\partial\alpha_a\partial q_i}.
\end{equation*}
Subtracting these relations, we find
\begin{equation*}
    \sum_i
    \frac{\partial^2S}{\partial\alpha_a\partial q_i}
    \left[
        \dot q_i
        -
        \left.
        \frac{\partial H}{\partial p_i}
        \right|_{\mathbf p=\partial S/\partial\mathbf q}
    \right]
    =0.
\end{equation*}
Since the matrix
$\partial^2S/(\partial\alpha_a\partial q_i)$ is invertible, it follows that
\begin{equation*}
    \dot q_i
    =
    \left.
    \frac{\partial H}{\partial p_i}
    \right|_{
        \substack{
            \mathbf p=\partial S/\partial\mathbf q
        }
    }.
\end{equation*}

The second Hamilton equation follows by taking the total time derivative of
$p_i=\partial S/\partial q_i$:
\begin{equation*}
    \dot p_i
    =
    \frac{\partial^2S}{\partial q_i\partial t}
    +
    \sum_j
    \frac{\partial^2S}{\partial q_i\partial q_j}
    \dot q_j.
\end{equation*}
Differentiating the HJ equation with respect to $q_i$ gives
\begin{equation*}
    \frac{\partial^2S}{\partial q_i\partial t}
    +
    \frac{\partial H}{\partial q_i}
    +
    \sum_j
    \frac{\partial H}{\partial p_j}
    \frac{\partial^2S}{\partial q_i\partial q_j}
    =0.
\end{equation*}
Using the expression obtained above for $\dot q_j$, we obtain
\begin{equation*}
    \dot p_i=-\frac{\partial H}{\partial q_i}.
\end{equation*}
Hence, the curve selected by the initial conditions satisfies
\begin{equation}
    \boxed{
    \dot q_i=\frac{\partial H}{\partial p_i},
    \qquad
    \dot p_i=-\frac{\partial H}{\partial q_i}.
    }
    \label{eq:hamilton_equations_from_S}
\end{equation}
A complete solution of the HJ equation, once the initial conditions have been
specified, therefore determines a trajectory that solves the original
Hamiltonian equations of motion.

Finally, Hamilton's principal function can be related directly to the
classical action.
Since the parameters $\boldsymbol{\alpha}$ are constant along the trajectory,
the total time derivative of $S$ is
\begin{equation*}
\begin{split}
    \frac{dS}{dt}
    &=
    \sum_i
    \frac{\partial S}{\partial q_i}\dot q_i
    +
    \frac{\partial S}{\partial t}
    \\
    &=
    \sum_i p_i\dot q_i-H.
\end{split}
\end{equation*}
Using the Legendre relation
\begin{equation*}
    L=\sum_i p_i\dot q_i-H,
\end{equation*}
we find
\begin{equation*}
    \frac{dS}{dt}=L.
\end{equation*}
Integration along the physical trajectory from $t_0$ to $t$ gives

\begin{equation}
\boxed{
S(\mathbf q(t),\boldsymbol{\alpha},t)
-
S(\mathbf q_0,\boldsymbol{\alpha},t_0)
=
\int_{t_0}^{t}
\left[
\sum_i p_i(t')\dot q_i(t')
-
H(\mathbf q(t'),\mathbf p(t'),t')
\right]dt'.
}
\label{eq:principal_function_as_action}
\end{equation}
Hamilton's principal function is therefore equal, up to an additive constant, to the action evaluated along a physical trajectory, as expressed by Eq.~\textcolor{red}{\eqref{eq:principal_function_as_action}}.

\subsection{From a Single Trajectory to an Ensemble}

A complete solution
$S(\mathbf q,\boldsymbol{\alpha},t)$ of the HJ equation describes a family of
classical motions. Once the constants
$(\boldsymbol{\alpha},\boldsymbol{\beta})$ have been fixed by the initial
conditions, the construction developed above selects one particular
trajectory. We now use the same formalism to describe an ensemble of
trajectories.

We first fix one particular solution of the HJ equation by choosing a definite
value of $\boldsymbol{\alpha}$. Suppressing this fixed parameter from the
notation, we write
\begin{equation*}
    S=S(\mathbf q,t).
\end{equation*}
This function assigns a momentum to every point of configuration space:
\begin{equation}
    \mathbf p(\mathbf q,t)
    =
    \boldsymbol{\nabla}S(\mathbf q,t).
\label{eq:classical_momentum_field}
\end{equation}
This is a momentum field rather than the momentum of one particular
trajectory. The momentum of an individual system is obtained only when this
field is evaluated along its motion.

The first Hamilton equation can be written in vector form as
\begin{equation*}
    \dot{\mathbf q}
    =
    \boldsymbol{\nabla}_{\mathbf p}
    H(\mathbf q,\mathbf p,t).
\end{equation*}
Substituting the momentum field determined by $S$ into its right-hand side defines the velocity field
\begin{equation*}
    \mathbf v(\mathbf q,t)
    \equiv
    \left.
        \boldsymbol{\nabla}_{\mathbf p}
        H(\mathbf q,\mathbf p,t)
    \right|_{\mathbf p=\boldsymbol{\nabla}S(\mathbf q,t)}.
\end{equation*}
The velocity field is therefore not introduced independently of the HJ
formalism. It is the velocity appearing in the first Hamilton equation after
the momentum has been expressed as a function of position and time through
$\mathbf p=\boldsymbol{\nabla}S$.

To distinguish a generic point $\mathbf q$ of configuration space from an
actual trajectory, we denote the latter by
$\mathbf X(t;\mathbf X_0)$. Here $\mathbf X_0$ is its initial position, so
that
\begin{equation*}
    \mathbf X(t_0;\mathbf X_0)=\mathbf X_0.
\end{equation*}
The trajectory is obtained by evaluating the velocity field at the position
occupied by the system:
\begin{equation}
    \boxed{
    \frac{d\mathbf X(t;\mathbf X_0)}{dt}
    =
    \mathbf v\bigl(\mathbf X(t;\mathbf X_0),t\bigr),
    \qquad
    \mathbf X(t_0;\mathbf X_0)=\mathbf X_0.
    }
    \label{eq:ensemble_trajectory}
\end{equation}
Equivalently,
\begin{equation*}
    \frac{d\mathbf X(t;\mathbf X_0)}{dt}
    =
    \left.
        \boldsymbol{\nabla}_{\mathbf p}
        H(\mathbf q,\mathbf p,t)
    \right|_{
        \substack{
            \mathbf q=\mathbf X(t;\mathbf X_0)\\
            \mathbf p=\boldsymbol{\nabla}
            S(\mathbf X(t;\mathbf X_0),t)
        }
    }.
\end{equation*}
Thus, once $S$ has been fixed, it determines the differential
equation governing the motion, while the initial position
$\mathbf X_0$ selects one particular solution.\\
The momentum along the selected trajectory is
\begin{equation*}
    \mathbf p(t;\mathbf X_0)
    =
    \boldsymbol{\nabla}
    S\bigl(\mathbf X(t;\mathbf X_0),t\bigr).
\end{equation*}
As demonstrated in Eq.~\textcolor{red}{\eqref{eq:hamilton_equations_from_S}}, the resulting
pair
$\bigl(\mathbf X(t;\mathbf X_0),\mathbf p(t;\mathbf X_0)\bigr)$ satisfies both
Hamilton equations. Consequently, the integral curves of the velocity field
are physical classical trajectories.\\
An ensemble is obtained by allowing the initial position $\mathbf X_0$ to vary while keeping the same function $S$.The initial positions $\mathbf X_0$ are distributed according
to the density
\begin{equation*}
    \rho_0(\mathbf q):=\rho(\mathbf q,t_0).
\end{equation*}
Thus, $\rho_0(\mathbf q)\,d^nq$ is the probability that
$\mathbf X_0$ lies within the volume element $d^nq$
around $\mathbf q$.
Each initial position is transported by Eq.~\textcolor{red}{\eqref{eq:ensemble_trajectory}}, and
$\rho(\mathbf q,t)$ describes the distribution of the resulting positions at
time $t$.

The flux of systems through configuration space is
\begin{equation*}
    \mathbf J(\mathbf q,t)
    =
    \rho(\mathbf q,t)\mathbf v(\mathbf q,t).
\end{equation*}
Local conservation of the number of systems therefore requires
\begin{equation}
    \boxed{
    \frac{\partial\rho}{\partial t}
    +
    \boldsymbol{\nabla}\cdot
    \left(
        \rho\mathbf v
    \right)
    =0.
    }
    \label{eq:classical_continuity_equation}
\end{equation}
The HJ equation determines the momentum and velocity fields, whereas the
continuity equation determines how the ensemble density is transported by
their flow.

For a non-relativistic spinless particle with Hamiltonian
\begin{equation*}
    H(\mathbf q,\mathbf p,t)
    =
    \frac{\mathbf p^2}{2m}
    +
    V(\mathbf q,t),
\end{equation*}
the velocity field becomes
\begin{equation*}
    \mathbf v(\mathbf q,t)
    =
    \frac{\boldsymbol{\nabla}S(\mathbf q,t)}{m}.
\end{equation*}
Equation~\textcolor{red}{\eqref{eq:ensemble_trajectory}} therefore reduces to
\begin{equation}
    \boxed{
    \frac{d\mathbf X(t;\mathbf X_0)}{dt}
    =
    \frac{1}{m}
    \boldsymbol{\nabla}
    S\bigl(\mathbf X(t;\mathbf X_0),t\bigr),
    \qquad
    \mathbf X(t_0;\mathbf X_0)=\mathbf X_0.
    }
    \label{eq:classical_guidance_equation}
\end{equation}
For the same Hamiltonian, the HJ and continuity equations become
\begin{equation}
    \boxed{
    \begin{aligned}
        &\frac{\partial S}{\partial t}
        +
        \frac{(\boldsymbol{\nabla}S)^2}{2m}
        +
        V
        =0,
        \\[0.4em]
        &\frac{\partial\rho}{\partial t}
        +
        \boldsymbol{\nabla}\cdot
        \left(
            \rho\frac{\boldsymbol{\nabla}S}{m}
        \right)
        =0.
    \end{aligned}
    }
    \label{eq:classical_ensemble_system}
\end{equation}
The first equation is precisely the classical HJ equation associated with
$H=\mathbf p^2/(2m)+V$. For a free particle, $V=0$, and it reduces to
\begin{equation*}
    \frac{\partial S}{\partial t}
    +
    \frac{(\boldsymbol{\nabla}S)^2}{2m}
    =0.
\end{equation*}
Together with the initial density $\rho_0$, the equations in
Eq.~\textcolor{red}{\eqref{eq:classical_ensemble_system}} describe a deterministic ensemble of
classical trajectories.

The pair $(S,\rho)$ represents a particular class of classical ensembles. A
completely general classical ensemble is instead described by a phase-space
density
\begin{equation*}
    f(\mathbf q,\mathbf p,t),
\end{equation*}
which allows positions and momenta to have an arbitrary joint distribution.
In particular, different systems may occupy the same position while having
different momenta.

This possibility is excluded by construction. Once a single-valued function
$S(\mathbf q,t)$ has been fixed, the relation in Eq.~\textcolor{red}{\eqref{eq:classical_momentum_field}} assigns exactly one momentum to each position. Consequently, for a trajectory
starting from $\mathbf X_0$, the initial momentum
\begin{equation*}
    \mathbf p_0
    =
    \boldsymbol{\nabla}S(\mathbf X_0,t_0)
\end{equation*}
is fixed as soon as $\mathbf X_0$ is chosen. The density $\rho_0$ therefore
needs to specify only the distribution of the initial positions, since the
corresponding momenta are already determined by $S$.

The corresponding phase-space distribution can be written as
\begin{equation*}
    f(\mathbf q,\mathbf p,t)
    =
    \rho(\mathbf q,t)
    \delta^{(n)}
    \left(
        \mathbf p-\boldsymbol{\nabla}S(\mathbf q,t)
    \right).
\end{equation*}
The Dirac delta shows that the distribution has support only on the
phase-space submanifold
$\mathbf p=\boldsymbol{\nabla}S$. An arbitrary distribution assigning
different momenta to the same position would require either different
branches of $S$ or a complete phase-space density
$f(\mathbf q,\mathbf p,t)$.

The same restriction can be understood from the complete HJ solution
$S(\mathbf q,\boldsymbol{\alpha},t)$. Fixing
$\boldsymbol{\alpha}$ selects one momentum field,
\begin{equation*}
    \mathbf p_{\boldsymbol{\alpha}}(\mathbf q,t)
    =
    \boldsymbol{\nabla}
    S(\mathbf q,\boldsymbol{\alpha},t).
\end{equation*}
The ensemble considered above is obtained by distributing the initial
positions while keeping this momentum field fixed. A more general classical
ensemble would also require a distribution over different values of
$\boldsymbol{\alpha}$ and hence over different HJ solutions.

The pair $(S,\rho)$ therefore has a precise meaning: $S$ determines the
momentum and velocity associated with each configuration, while $\rho$
determines how the configurations are statistically distributed. This
restricted ensemble structure will be particularly important in the PWT. In
the non-relativistic spinless case, once the wave function and hence its phase
$S$ have been fixed, the particle momentum is not an independent initial
variable but is determined by
$\mathbf p=\boldsymbol{\nabla}S$.

The coupled system in Eq.~\textcolor{red}{\eqref{eq:classical_ensemble_system}} provides the
connection between the classical HJ formalism and QM. This connection becomes
more transparent when the same dynamics is expressed as a Hamiltonian theory
for the fields $(\rho,S)$.

The canonical action of the complete ensemble is obtained by integrating the
action of each trajectory over all possible initial positions:
\begin{equation*}
\begin{split}
    I_{\mathrm E}
    =
    \int dt\,d^nX_0\,
    \rho_0(\mathbf X_0)
    \bigg[
        &\mathbf p(t;\mathbf X_0)\cdot
        \dot{\mathbf X}(t;\mathbf X_0)
        \\
        &-
        H\bigl(
            \mathbf X(t;\mathbf X_0),
            \mathbf p(t;\mathbf X_0),
            t
        \bigr)
    \bigg].
\end{split}
\end{equation*}
Here $\mathbf X_0$ is the integration variable over the initial configuration
space, and $\rho_0(\mathbf X_0)$ assigns the corresponding statistical
weight.

For the ensembles represented by $S$, applying Eq.~\textcolor{red}{\eqref{eq:classical_momentum_field}} along each trajectory and using the chain rule gives
\begin{equation*}
\begin{split}
    \mathbf p(t;\mathbf X_0)\cdot
    \dot{\mathbf X}(t;\mathbf X_0)
    &=
    \frac{d}{dt}
    S\bigl(\mathbf X(t;\mathbf X_0),t\bigr)
    \\
    &\quad
    -
    \left.
        \frac{\partial S(\mathbf q,t)}{\partial t}
    \right|_{\mathbf q=\mathbf X(t;\mathbf X_0)}.
\end{split}
\end{equation*}
The first term is a total time derivative. Its contribution to the action is
\begin{equation*}
\begin{split}
    &\int_{t_0}^{t_f}dt\,d^nX_0\,
    \rho_0(\mathbf X_0)
    \frac{d}{dt}
    S\bigl(\mathbf X(t;\mathbf X_0),t\bigr)
    \\
    &\qquad
    =
    \left[
        \int d^nX_0\,
        \rho_0(\mathbf X_0)
        S\bigl(\mathbf X(t;\mathbf X_0),t\bigr)
    \right]_{t_0}^{t_f}.
\end{split}
\end{equation*}
It therefore depends only on the initial and final configurations and does not
contribute to the equations of motion.

At time $t$, the trajectory flow defines the change of variables
\begin{equation*}
    \mathbf q
    =
    \mathbf X(t;\mathbf X_0).
\end{equation*}
The corresponding volume elements are related by
\begin{equation*}
    d^nq=J(\mathbf X_0,t)\,d^nX_0,
\end{equation*}
where
\begin{equation*}
    J(\mathbf X_0,t)
    =
    \left|
        \det\left(
            \frac{\partial\mathbf X(t;\mathbf X_0)}
            {\partial\mathbf X_0}
        \right)
    \right|.
\end{equation*}
The systems initially contained in the volume element $d^nX_0$
are transported into the corresponding volume element $d^nq$
around $\mathbf X(t;\mathbf X_0)$. Since their statistical weight
is conserved,
\begin{equation*}
    \rho_0(\mathbf X_0)\,d^nX_0
    =
    \rho\bigl(\mathbf X(t;\mathbf X_0),t\bigr)\,d^nq.
\end{equation*}
Substituting $d^nq=J(\mathbf X_0,t)\,d^nX_0$ gives
\begin{equation*}
    \rho_0(\mathbf X_0)\,d^nX_0
    =
    \rho\bigl(\mathbf X(t;\mathbf X_0),t\bigr)
    J(\mathbf X_0,t)\,d^nX_0,
\end{equation*}
and therefore
\begin{equation*}
    \rho\bigl(\mathbf X(t;\mathbf X_0),t\bigr)
    J(\mathbf X_0,t)
    =
    \rho_0(\mathbf X_0).
\end{equation*}
Writing $\mathbf q=\mathbf X(t;\mathbf X_0)$, conservation of
the statistical weight thus takes the form
\begin{equation*}
    \rho_0(\mathbf X_0)\,d^nX_0
    =
    \rho(\mathbf q,t)\,d^nq.
\end{equation*}
Using this change of variables and removing the total time derivative, the
ensemble action becomes
\begin{equation}
    \boxed{
    I_{\mathrm E}[\rho,S]
    =
    -\int dt\,d^nq\,
    \rho
    \left[
        \frac{\partial S}{\partial t}
        +
        H\left(
            \mathbf q,\boldsymbol{\nabla}S,t
        \right)
    \right].
    }
    \label{eq:classical_ensemble_action}
\end{equation}

We define the corresponding ensemble Hamiltonian functional as
\begin{equation}
    \boxed{
    \mathcal H_{\mathrm{HJ}}[\rho,S]
    =
    \int d^nq\,
    \rho(\mathbf q,t)
    H\left(
        \mathbf q,\boldsymbol{\nabla}S,t
    \right).
    }
    \label{eq:classical_ensemble_hamiltonian}
\end{equation}
Equation~\textcolor{red}{\eqref{eq:classical_ensemble_action}} can then be written as
\begin{equation*}
    I_{\mathrm E}
    =
    -\int dt
    \left[
        \int d^nq\,
        \rho\frac{\partial S}{\partial t}
        +
        \mathcal H_{\mathrm{HJ}}[\rho,S]
    \right].
\end{equation*}
Integrating the first term by parts in time gives
\begin{equation*}
    -\int dt\,d^nq\,
    \rho\frac{\partial S}{\partial t}
    =
    -\left[
        \int d^nq\,\rho S
    \right]_{t_0}^{t_f}
    +
    \int dt\,d^nq\,
    S\frac{\partial\rho}{\partial t}.
\end{equation*}
After removing the boundary contribution, the action assumes the equivalent
canonical form
\begin{equation*}
    \boxed{
    I_{\mathrm E}[\rho,S]
    =
    \int dt
    \left[
        \int d^nq\,
        S\frac{\partial\rho}{\partial t}
        -
        \mathcal H_{\mathrm{HJ}}[\rho,S]
    \right].
    }
\end{equation*}
This has the same structure as the ordinary phase-space action in Eq.~\textcolor{red}{\eqref{eq:classical_phase_space_action}}.

At the ensemble level, the role of the canonical coordinate is played by the
field $\rho$, while $S$ is its conjugate momentum. Their functional Hamilton
equations are therefore
\begin{equation*}
    \frac{\partial\rho}{\partial t}
    =
    \frac{\delta\mathcal H_{\mathrm{HJ}}}{\delta S},
    \qquad
    \frac{\partial S}{\partial t}
    =
    -\frac{\delta\mathcal H_{\mathrm{HJ}}}{\delta\rho}.
\end{equation*}
For the functional in
Eq.~\textcolor{red}{\eqref{eq:classical_ensemble_hamiltonian}},
\begin{equation*}
    \frac{\delta\mathcal H_{\mathrm{HJ}}}{\delta\rho}
    =
    H\left(
        \mathbf q,\boldsymbol{\nabla}S,t
    \right),
\end{equation*}
while
\begin{equation*}
    \frac{\delta\mathcal H_{\mathrm{HJ}}}{\delta S}
    =
    -
    \boldsymbol{\nabla}\cdot
    \left[
        \rho
        \left.
            \boldsymbol{\nabla}_{\mathbf p}H
        \right|_{\mathbf p=\boldsymbol{\nabla}S}
    \right].
\end{equation*}
The second relation follows by integrating by parts in configuration space,
with the boundary contribution taken to vanish. The functional Hamilton
equations consequently reproduce the HJ and continuity equations:
\begin{equation}
    \boxed{
    \begin{aligned}
        &\frac{\partial S}{\partial t}
        +
        H\left(
            \mathbf q,\boldsymbol{\nabla}S,t
        \right)
        =0,
        \\[0.4em]
        &\frac{\partial\rho}{\partial t}
        +
        \boldsymbol{\nabla}\cdot
        \left[
            \rho
            \left.
                \boldsymbol{\nabla}_{\mathbf p}H
            \right|_{\mathbf p=\boldsymbol{\nabla}S}
        \right]
        =0.
    \end{aligned}
    }
    \label{eq:functional_classical_ensemble}
\end{equation}
For
$H=\mathbf p^2/(2m)+V$, these are exactly the equations displayed in
Eq.~\textcolor{red}{\eqref{eq:classical_ensemble_system}}. The functional formulation therefore
introduces no new dynamics: it expresses the same classical ensemble through
the canonical fields $(\rho,S)$.

A more general ensemble dynamics can be introduced by adding a functional of
the density:
\begin{equation*}
    \mathcal H_{\mathrm{eff}}[\rho,S]
    =
    \mathcal H_{\mathrm{HJ}}[\rho,S]
    +
    F[\rho].
\end{equation*}
Since $F$ does not depend on $S$,
\begin{equation*}
    \frac{\delta\mathcal H_{\mathrm{eff}}}{\delta S}
    =
    \frac{\delta\mathcal H_{\mathrm{HJ}}}{\delta S},
\end{equation*}
whereas
\begin{equation*}
    \frac{\delta\mathcal H_{\mathrm{eff}}}{\delta\rho}
    =
    H\left(
        \mathbf q,\boldsymbol{\nabla}S,t
    \right)
    +
    \frac{\delta F}{\delta\rho}.
\end{equation*}
The resulting equations are therefore
\begin{equation}
    \boxed{
    \begin{aligned}
        &\frac{\partial S}{\partial t}
        +
        H\left(
            \mathbf q,\boldsymbol{\nabla}S,t
        \right)
        +
        \frac{\delta F}{\delta\rho}
        =0,
        \\[0.4em]
        &\frac{\partial\rho}{\partial t}
        +
        \boldsymbol{\nabla}\cdot
        \left[
            \rho
            \left.
                \boldsymbol{\nabla}_{\mathbf p}H
            \right|_{\mathbf p=\boldsymbol{\nabla}S}
        \right]
        =0.
    \end{aligned}
    }
    \label{eq:effective_ensemble_equations}
\end{equation}
The contribution
\begin{equation*}
    V_{\mathrm{eff}}[\rho]
    =
    \frac{\delta F}{\delta\rho}
\end{equation*}
acts as a state-dependent effective potential. Depending on the physical
meaning of $\rho$, it may represent a collective interaction among underlying
constituents. It may also arise from a statistical description of unresolved
stochastic fluctuations, without representing a direct interaction between
different members of the ensemble.

More generally, the additional functional could also depend on $S$ and its
spatial derivatives, thereby modifying both the HJ equation and the ensemble
current. This possibility will not be considered here. A specific additional
ensemble energy arising from the statistical description of underlying
stochastic fluctuations will be analysed later.

\section{The Hamilton--Jacobi Formulation of Quantum Mechanics}

The classical ensemble introduced in the previous section is described by the
HJ equation and the continuity equation in Eq.~\textcolor{red}{\eqref{eq:classical_ensemble_system}},
\begin{equation*}
    \frac{\partial S}{\partial t}
    +\frac{(\boldsymbol{\nabla}S)^2}{2m}
    +V=0,
    \qquad
    \frac{\partial\rho}{\partial t}
    +
    \boldsymbol{\nabla}\cdot
    \left(
        \rho\frac{\boldsymbol{\nabla}S}{m}
    \right)
    =0.
\end{equation*}
These two real equations can be combined into a single complex equation. In regions where $\rho>0$, we
define
\begin{equation*}
    \psi_{\mathrm{cl}}(\mathbf q,t)
    =
    \sqrt{\rho(\mathbf q,t)}\,
    e^{iS(\mathbf q,t)/\hbar}.
\end{equation*}
where $\hbar$ is introduced here only as a constant with the dimensions of
action, so that the phase is dimensionless.

A direct substitution shows that the classical HJ and continuity equations
are equivalent to
\begin{equation}
    \boxed{
    i\hbar\frac{\partial\psi_{\mathrm{cl}}}{\partial t}
    =
    \left[
        -\frac{\hbar^2}{2m}\nabla^2
        +V
        -Q[\rho]
    \right]\psi_{\mathrm{cl}},
    }
    \label{eq:nonlinear_classical_schrodinger}
\end{equation}
where
\begin{equation*}
    Q[\rho]
    =
    -\frac{\hbar^2}{2m}
    \frac{\nabla^2\sqrt{\rho}}{\sqrt{\rho}}.
\end{equation*}
Since $\rho=|\psi_{\mathrm{cl}}|^2$, the term $Q[\rho]$ depends on the
solution itself, and Eq.~\textcolor{red}{\eqref{eq:nonlinear_classical_schrodinger}} is
therefore nonlinear. The function $\psi_{\mathrm{cl}}$ does not introduce a
new physical field: it is a compact representation of the two classical
ensemble variables $S$ and $\rho$.

The role of the nonlinear term is to cancel exactly the contribution
$Q[\rho]$ generated by the Laplacian when the equation is separated into its
real and imaginary parts. Its real part therefore reproduces the classical HJ
equation, while its imaginary part reproduces the continuity equation.
Because of this nonlinearity, the superposition principle of standard QM
does not hold: a linear combination of two solutions is not, in general,
another solution.

Let us now examine how the same construction changes when applied to QM.

\subsection{Polar Decomposition of the Wave Function}

Consider the SE for a non-relativistic spinless particle,
\begin{equation}
    i\hbar\frac{\partial\psi}{\partial t}
    =
    \left[
        -\frac{\hbar^2}{2m}\nabla^2
        +
        V(\mathbf q,t)
    \right]\psi.
    \label{eq:schrodinger_equation_polar}
\end{equation}
In every region where $\psi\neq0$, the wave function can be written in polar
form as
\begin{equation*}
    \psi(\mathbf q,t)
    =
    \sqrt{\rho(\mathbf q,t)}\,
    e^{iS(\mathbf q,t)/\hbar},
\end{equation*}
where
\begin{equation*}
    \rho(\mathbf q,t)
    =
    |\psi(\mathbf q,t)|^2
\end{equation*}
and $S(\mathbf q,t)$ is a real function representing the phase in units of
action. The phase is locally defined modulo $2\pi\hbar$ and is not defined at
the nodes of the wave function, where $\rho=0$.

The time derivative of $\psi$ is
\begin{equation*}
    \frac{\partial\psi}{\partial t}
    =
    e^{iS/\hbar}
    \left[
        \frac{1}{2\sqrt{\rho}}
        \frac{\partial\rho}{\partial t}
        +
        \frac{i}{\hbar}
        \sqrt{\rho}\,
        \frac{\partial S}{\partial t}
    \right],
\end{equation*}
while its spatial gradient is
\begin{equation*}
    \boldsymbol{\nabla}\psi
    =
    e^{iS/\hbar}
    \left[
        \boldsymbol{\nabla}\sqrt{\rho}
        +
        \frac{i}{\hbar}
        \sqrt{\rho}\,
        \boldsymbol{\nabla}S
    \right].
\end{equation*}
Applying the gradient once more gives
\begin{equation*}
\begin{split}
    \nabla^2\psi
    =
    e^{iS/\hbar}
    \bigg[
        &\nabla^2\sqrt{\rho}
        -
        \frac{\sqrt{\rho}}{\hbar^2}
        (\boldsymbol{\nabla}S)^2
        \\
        &+
        \frac{i}{\hbar}
        \left(
            2\boldsymbol{\nabla}\sqrt{\rho}
            \cdot\boldsymbol{\nabla}S
            +
            \sqrt{\rho}\,\nabla^2S
        \right)
    \bigg].
\end{split}
\end{equation*}

Substituting these expressions into
Eq.~\textcolor{red}{\eqref{eq:schrodinger_equation_polar}} and dividing by
$\sqrt{\rho}\,e^{iS/\hbar}$ yields
\begin{equation*}
\begin{split}
    -\frac{\partial S}{\partial t}
    +
    \frac{i\hbar}{2\rho}
    \frac{\partial\rho}{\partial t}
    =
    &\frac{(\boldsymbol{\nabla}S)^2}{2m}
    +
    V
    -
    \frac{\hbar^2}{2m}
    \frac{\nabla^2\sqrt{\rho}}{\sqrt{\rho}}
    \\
    &-
    \frac{i\hbar}{2m}
    \left[
        \nabla^2S
        +
        \frac{\boldsymbol{\nabla}\rho}{\rho}
        \cdot\boldsymbol{\nabla}S
    \right].
\end{split}
\end{equation*}
Since $\rho$ and $S$ are real, the real and imaginary parts must be satisfied
independently. The real part gives
\begin{equation*}
    \frac{\partial S}{\partial t}
    +
    \frac{(\boldsymbol{\nabla}S)^2}{2m}
    +
    V
    -
    \frac{\hbar^2}{2m}
    \frac{\nabla^2\sqrt{\rho}}{\sqrt{\rho}}
    =0,
\end{equation*}
whereas the imaginary part gives
\begin{equation*}
    \frac{\partial\rho}{\partial t}
    +
    \frac{\rho}{m}\nabla^2S
    +
    \frac{1}{m}
    \boldsymbol{\nabla}\rho
    \cdot\boldsymbol{\nabla}S
    =0.
\end{equation*}
Using
\begin{equation*}
    \boldsymbol{\nabla}\cdot
    \left(
        \rho\boldsymbol{\nabla}S
    \right)
    =
    \rho\nabla^2S
    +
    \boldsymbol{\nabla}\rho
    \cdot\boldsymbol{\nabla}S,
\end{equation*}
the latter assumes the form of a continuity equation. The polar decomposition
of the SE therefore produces the coupled system
\begin{equation}
    \boxed{
    \begin{aligned}
        &\frac{\partial S}{\partial t}
        +
        \frac{(\boldsymbol{\nabla}S)^2}{2m}
        +
        V
        -
        \frac{\hbar^2}{2m}
        \frac{\nabla^2\sqrt{\rho}}{\sqrt{\rho}}
        =0,
        \\[0.4em]
        &\frac{\partial\rho}{\partial t}
        +
        \boldsymbol{\nabla}\cdot
        \left(
            \rho\frac{\boldsymbol{\nabla}S}{m}
        \right)
        =0.
    \end{aligned}
    }
    \label{eq:quantum_hj_continuity_system}
\end{equation}
The second equation has the form of a continuity equation. The first has the structure of the classical HJ equation but contains an additional state-dependent contribution. Together, these equations constitute the Madelung formulation of the SE. Before interpreting that contribution, we must first establish the dynamical and statistical meaning of the fields appearing in the two equations.

\subsection{Particle Trajectories and the Guidance Equation}

We begin with the first equation in
Eq.~\textcolor{red}{\eqref{eq:quantum_hj_continuity_system}}.
For a given wave function, we regard
$\rho(\mathbf q,t)=|\psi(\mathbf q,t)|^2$ as an assigned field
and introduce the effective Hamiltonian
\begin{equation}
    \boxed{
    H_{\mathrm{eff}}(\mathbf q,\mathbf p,t;\rho)
    =
    \frac{\mathbf p^2}{2m}
    +
    V(\mathbf q,t)
    -
    \frac{\hbar^2}{2m}
    \frac{\nabla^2\sqrt{\rho(\mathbf q,t)}}
         {\sqrt{\rho(\mathbf q,t)}}.
    }
    \label{eq:quantum_effective_hamiltonian}
\end{equation}
The first equation then takes the Hamilton--Jacobi form
\begin{equation}
    \boxed{
    \frac{\partial S}{\partial t}
    +
    H_{\mathrm{eff}}
    \bigl(\mathbf q,\boldsymbol{\nabla}S,t;\rho\bigr)
    =0.
    }
    \label{eq:quantum_hamilton_jacobi_equation}
\end{equation}
Here the dependence on $\rho$ specifies a space- and
time-dependent contribution to the Hamiltonian. When taking
partial derivatives with respect to the independent momentum
variable $\mathbf p$, this field is held fixed.

Following the HJ construction developed in the previous section,
$S$ defines the momentum field
\begin{equation*}
    \mathbf p(\mathbf q,t)
    =
    \boldsymbol{\nabla}S(\mathbf q,t).
\end{equation*}
The corresponding velocity field is
\begin{equation}
    \boxed{
    \mathbf v(\mathbf q,t)
    =
    \left.
        \boldsymbol{\nabla}_{\mathbf p}
        H_{\mathrm{eff}}(\mathbf q,\mathbf p,t;\rho)
    \right|_{\mathbf p=\boldsymbol{\nabla}S(\mathbf q,t)}
    =
    \frac{\boldsymbol{\nabla}S(\mathbf q,t)}{m}.
    }
    \label{eq:quantum_velocity_field}
\end{equation}
The last equality follows from the quadratic dependence of
$H_{\mathrm{eff}}$ on $\mathbf p$.

The additional physical postulate of the PWT is that an
individual system possesses an actual particle position
$\mathbf X(t)$ whose motion follows this velocity field:
\begin{equation}
    \boxed{
    \frac{d\mathbf X(t)}{dt}
    =
    \mathbf v\bigl(\mathbf X(t),t\bigr)
    =
    \frac{1}{m}
    \boldsymbol{\nabla}S\bigl(\mathbf X(t),t\bigr),
    \qquad
    \mathbf X(t_0)=\mathbf X_0.
    }
    \label{eq:quantum_guidance_equation}
\end{equation}
This is the guidance equation. The phase $S$ determines
the velocity field, while the initial position $\mathbf X_0$
selects one of its integral curves. Once the wave function
and $\mathbf X_0$ are specified, the trajectory is uniquely
determined wherever the velocity field is regular.

Along this trajectory, the momentum is
\begin{equation*}
    \mathbf p(t)
    =
    \boldsymbol{\nabla}S\bigl(\mathbf X(t),t\bigr).
\end{equation*}
For the Hamiltonian considered here, the guidance equation
therefore implies $\mathbf p(t)=m\dot{\mathbf X}(t)$.
Moreover, since $S$ satisfies the HJ equation, the same
argument used in the classical case gives
\begin{equation*}
    \dot{\mathbf p}(t)
    =
    -\left.
        \boldsymbol{\nabla}_{\mathbf q}
        H_{\mathrm{eff}}(\mathbf q,\mathbf p,t;\rho)
    \right|_{
        \mathbf q=\mathbf X(t),\,
        \mathbf p=\mathbf p(t)
    }.
\end{equation*}
The guided trajectories thus satisfy both Hamilton equations
associated with $H_{\mathrm{eff}}$, with the initial momentum
constrained by
\begin{equation*}
    \mathbf p_0
    =
    \boldsymbol{\nabla}S(\mathbf X_0,t_0).
\end{equation*}
Consequently, once $\psi$ and $\mathbf X_0$ have been fixed,
the initial momentum cannot be chosen independently.

At a node of the wave function, where $\rho=|\psi|^2=0$,
the phase and the effective Hamiltonian may fail to be
regular. The trajectory construction is therefore understood
locally in regions where the corresponding fields are
well defined.

\subsection{Quantum Equilibrium}

We now consider the second equation in
Eq.~\textcolor{red}{\eqref{eq:quantum_hj_continuity_system}}.
Using the velocity field identified in
Eq.~\textcolor{red}{\eqref{eq:quantum_velocity_field}},
it becomes
\begin{equation*}
    \frac{\partial\rho}{\partial t}
    +
    \boldsymbol{\nabla}\cdot(\rho\mathbf v)
    =0.
\end{equation*}
This is a continuity equation with current
\begin{equation*}
    \mathbf j(\mathbf q,t)
    =
    \rho(\mathbf q,t)\mathbf v(\mathbf q,t)
    =
    \rho(\mathbf q,t)
    \frac{\boldsymbol{\nabla}S(\mathbf q,t)}{m}.
\end{equation*}
Thus, the velocity obtained from the HJ construction also
transports the density determined by the wave function.

This density must initially be distinguished from the actual
statistical distribution of particle positions in an ensemble
of systems sharing the same wave function. We denote the
latter by $P(\mathbf q,t)$, so that
$P(\mathbf q,t)d^nq$ is the fraction of systems whose particle
positions lie within the volume element $d^nq$ around
$\mathbf q$ at time $t$.

Every particle follows the guidance
equation~\textcolor{red}{\eqref{eq:quantum_guidance_equation}}.
Conservation of the number of particles therefore requires
$P$ to satisfy the continuity equation associated with the
same velocity field. Consequently,
\begin{equation}
    \boxed{
    \begin{aligned}
        &\frac{\partial P}{\partial t}
        +
        \boldsymbol{\nabla}\cdot(P\mathbf v)
        =0,
        \\[0.4em]
        &\frac{\partial\rho}{\partial t}
        +
        \boldsymbol{\nabla}\cdot(\rho\mathbf v)
        =0,
        \qquad
        \mathbf v
        =
        \frac{\boldsymbol{\nabla}S}{m}.
    \end{aligned}
    }
    \label{eq:particle_and_wave_densities}
\end{equation}

To compare the two distributions, define
\begin{equation*}
    D(\mathbf q,t)=P(\mathbf q,t)-\rho(\mathbf q,t).
\end{equation*}
Subtracting the equations in
Eq.~\textcolor{red}{\eqref{eq:particle_and_wave_densities}}
gives
\begin{equation*}
    \frac{\partial D}{\partial t}
    +
    \boldsymbol{\nabla}\cdot(D\mathbf v)
    =0.
\end{equation*}
If
\begin{equation*}
    P(\mathbf q,t_0)=\rho(\mathbf q,t_0),
\end{equation*}
then $D(\mathbf q,t_0)=0$. Wherever the transport problem
has a unique solution, this implies $D(\mathbf q,t)=0$,
and hence
\begin{equation}
    \boxed{
    P(\mathbf q,t)
    =
    \rho(\mathbf q,t)
    =
    |\psi(\mathbf q,t)|^2.
    }
    \label{eq:quantum_equilibrium}
\end{equation}
The preservation of the $|\psi|^2$ distribution under the
guidance flow is called equivariance.

The initial condition
\begin{equation*}
    P(\mathbf q,t_0)=|\psi(\mathbf q,t_0)|^2
\end{equation*}
is the quantum-equilibrium hypothesis. The guidance equation
does not impose this condition on an arbitrary initial
ensemble; it guarantees that, once satisfied, the condition
is preserved by the evolution.

In quantum equilibrium, the equations obtained from the SE
can therefore be interpreted as an HJ equation coupled to
the transport equation for the corresponding particle ensemble:
\begin{equation}
    \boxed{
    \begin{aligned}
        &\frac{\partial S}{\partial t}
        +
        H_{\mathrm{eff}}
        \bigl(\mathbf q,\boldsymbol{\nabla}S,t;\rho\bigr)
        =0,
        \\[0.4em]
        &\frac{\partial\rho}{\partial t}
        +
        \boldsymbol{\nabla}\cdot
        \left[
            \rho
            \left.
                \boldsymbol{\nabla}_{\mathbf p}
                H_{\mathrm{eff}}
            \right|_{\mathbf p=\boldsymbol{\nabla}S}
        \right]
        =0.
    \end{aligned}
    }
    \label{eq:quantum_hj_ensemble_system}
\end{equation}
The phase determines the velocity field, the actual positions
follow its integral curves, and $\rho=P$ describes their
statistical distribution.

Initial ensembles with
$P(\mathbf q,t_0)\neq|\psi(\mathbf q,t_0)|^2$
describe quantum non-equilibrium. In that case, $\rho$ remains
the density determined by the wave function, while $P$ is
transported by the guidance field without being identified
with $\rho$. In the following, we consider the
quantum-equilibrium regime.

\subsection{The Quantum Potential}

The additional term in
Eq.~\textcolor{red}{\eqref{eq:quantum_effective_hamiltonian}}
is the quantum potential (QP):
\begin{equation}
    \boxed{
    Q[\rho](\mathbf q,t)
    =
    -\frac{\hbar^2}{2m}
    \frac{\nabla^2\sqrt{\rho(\mathbf q,t)}}
         {\sqrt{\rho(\mathbf q,t)}}.
    }
    \label{eq:quantum_potential}
\end{equation}
Thus,
\begin{equation*}
    H_{\mathrm{eff}}
    =
    \frac{\mathbf p^2}{2m}+V+Q[\rho],
    \qquad
    \frac{\partial S}{\partial t}
    +
    \frac{(\boldsymbol{\nabla}S)^2}{2m}
    +
    V+Q[\rho]
    =0.
\end{equation*}

The essential difference from the classical ensemble is that
the effective Hamiltonian depends on the wave-function
density. In the classical case, the HJ equation determines
$S$ independently of the ensemble density, which is then
transported by the resulting velocity field. Here, instead,
$\rho$ enters the HJ equation through $Q[\rho]$.
Although $\rho$ was treated as an assigned field when
constructing the trajectories, its evolution must be
determined together with that of $S$: the phase determines
the velocity in the continuity equation, while the density
contributes to the phase evolution through the QP.

This structure can be expressed through the ensemble
Hamiltonian functional introduced in the previous section.
Consider
\begin{equation}
    \boxed{
    F_Q[\rho]
    =
    \frac{\hbar^2}{8m}
    \int d^nq\,
    \frac{(\boldsymbol{\nabla}\rho)^2}{\rho}.
    }
    \label{eq:quantum_ensemble_energy}
\end{equation}
With boundary conditions that eliminate the surface terms,
its functional derivative is
\begin{equation*}
    \frac{\delta F_Q}{\delta\rho}
    =
    \frac{\hbar^2}{8m}
    \left[
        \frac{(\boldsymbol{\nabla}\rho)^2}{\rho^2}
        -
        2\frac{\nabla^2\rho}{\rho}
    \right].
\end{equation*}
Using
\begin{equation}
    \frac{\nabla^2\sqrt{\rho}}{\sqrt{\rho}}
    =
    \frac{1}{2}\frac{\nabla^2\rho}{\rho}
    -
    \frac{1}{4}
    \frac{(\boldsymbol{\nabla}\rho)^2}{\rho^2},
    \label{eq:sqrt_density_laplacian_identity}
\end{equation}
we obtain
\begin{equation}
    \boxed{
    \frac{\delta F_Q}{\delta\rho}=Q[\rho].
    }
    \label{eq:quantum_potential_functional_derivative}
\end{equation}
The Hamiltonian functional of the quantum ensemble is
therefore
\begin{equation}
    \boxed{
    \mathcal H_Q[\rho,S]
    =
    \mathcal H_{\mathrm{HJ}}[\rho,S]+F_Q[\rho]
    =
    \int d^nq\,\rho
    \left[
        \frac{(\boldsymbol{\nabla}S)^2}{2m}+V
    \right]
    +
    F_Q[\rho].
    }
    \label{eq:quantum_ensemble_hamiltonian}
\end{equation}
Since $F_Q$ depends only on $\rho$, it contributes to the
equation for $S$ while leaving the form of the continuity
equation unchanged. The polar decomposition of the SE
thus realizes the generalized HJ ensemble introduced in
the previous section.

Along a guided trajectory, the second Hamilton equation gives
\begin{equation*}
    \dot{\mathbf p}(t)
    =
    -\left.
        \boldsymbol{\nabla}_{\mathbf q}
        \bigl(V+Q[\rho]\bigr)
    \right|_{\mathbf q=\mathbf X(t)}.
\end{equation*}
Since $\mathbf p=m\dot{\mathbf X}$ for the Hamiltonian
considered here, this becomes
\begin{equation}
    \boxed{
    m\frac{d^2\mathbf X(t)}{dt^2}
    =
    -\left.
        \boldsymbol{\nabla}_{\mathbf q}
        \bigl(V+Q[\rho]\bigr)
    \right|_{\mathbf q=\mathbf X(t)}.
    }
    \label{eq:quantum_newton_equation}
\end{equation}
This Newton-like equation reproduces the guidance dynamics
provided that the initial momentum satisfies
\begin{equation*}
    \mathbf p_0
    =
    \boldsymbol{\nabla}S(\mathbf X_0,t_0).
\end{equation*}
The quantum HJ equation ensures that this initial constraint
is preserved along the resulting regular trajectory, so that
$\mathbf p(t)=\boldsymbol{\nabla}S(\mathbf X(t),t)$
at all subsequent times. Without this constraint, the
second-order equation admits additional trajectories that
do not satisfy the guidance equation.

The QP also explains why the particle position and momentum
alone do not determine the subsequent motion. Two wave
functions may satisfy
\begin{equation*}
    \boldsymbol{\nabla}S_1(\mathbf X_0,t_0)
    =
    \boldsymbol{\nabla}S_2(\mathbf X_0,t_0)
    =
    \mathbf p_0,
\end{equation*}
and hence assign the same initial velocity to a particle
at $\mathbf X_0$. Nevertheless, their densities may produce
different quantum forces,
\begin{equation*}
    \boldsymbol{\nabla}Q[\rho_1](\mathbf X_0,t_0)
    \neq
    \boldsymbol{\nabla}Q[\rho_2](\mathbf X_0,t_0),
\end{equation*}
so that the initial accelerations, and therefore the
trajectories, differ. Even when these forces initially
coincide, their subsequent evolution need not be the same.

For a fixed classical Hamiltonian, by contrast, the same
initial position and momentum determine the same trajectory,
provided the equations of motion admit a unique solution.
In the PWT, the complete wave function is dynamically
relevant, and the state of an individual system is specified
by the pair
\begin{equation*}
    \bigl(\psi,\mathbf X\bigr),
\end{equation*}
rather than by position and momentum alone.

When the QP contribution can be neglected, the coupled
equations reduce to the classical HJ ensemble. A formal
limit $\hbar\to0$ gives this reduction provided the density
and its spatial derivatives remain sufficiently regular
for the quantum contribution to vanish.

\subsection{Many-Particle Dynamics and Nonlocality}

We now briefly examine a fundamental feature of many-particle quantum systems: the wave function is generally defined on the configuration space of the entire system, rather than separately in the physical space of each particle. Within the PWT, this implies that the motion of each particle may depend on the
simultaneous configuration of all the others. To make this structure explicit, we extend the formalism to a system of $N$ spinless particles and denote a point in its $3N$-dimensional configuration space by
\begin{equation*}
    \mathbf q
    =
    (\mathbf q_1,\ldots,\mathbf q_N)
\end{equation*}
and write the wave function in polar form as
\begin{equation*}
    \psi(\mathbf q_1,\ldots,\mathbf q_N,t)
    =
    \sqrt{\rho(\mathbf q_1,\ldots,\mathbf q_N,t)}\,
    e^{iS(\mathbf q_1,\ldots,\mathbf q_N,t)/\hbar}.
\end{equation*}
The polar decomposition of the many-particle SE gives
\begin{equation}
    \boxed{
    \begin{aligned}
        &\frac{\partial S}{\partial t}
        +
        \sum_{k=1}^{N}
        \frac{(\boldsymbol{\nabla}_kS)^2}{2m_k}
        +
        V
        +
        Q[\rho]
        =0,
        \\[0.4em]
        &\frac{\partial\rho}{\partial t}
        +
        \sum_{k=1}^{N}
        \boldsymbol{\nabla}_k\cdot
        \left(
            \rho\frac{\boldsymbol{\nabla}_kS}{m_k}
        \right)
        =0,
    \end{aligned}
    }
    \label{eq:many_particle_quantum_hj_system}
\end{equation}
where
\begin{equation}
    \boxed{
    Q[\rho]
    =
    -\sum_{k=1}^{N}
    \frac{\hbar^2}{2m_k}
    \frac{\nabla_k^2\sqrt{\rho}}
    {\sqrt{\rho}}.
    }
    \label{eq:many_particle_quantum_potential}
\end{equation}
Both $\rho$ and $S$ are functions of the complete configuration, rather than
separate functions defined independently in ordinary three-dimensional
space.

The actual configuration of the system is
\begin{equation*}
    \mathbf X(t)
    =
    \bigl(
        \mathbf X_1(t),\ldots,\mathbf X_N(t)
    \bigr).
\end{equation*}
The current associated with the $k$-th configuration-space coordinate is
\begin{equation*}
    \mathbf j_k
    =
    \rho\frac{\boldsymbol{\nabla}_kS}{m_k},
\end{equation*}
and the corresponding guidance equation is
\begin{equation}
    \boxed{
    \frac{d\mathbf X_k(t)}{dt}
    =
    \left.
        \frac{\boldsymbol{\nabla}_k
        S(\mathbf q_1,\ldots,\mathbf q_N,t)}
        {m_k}
    \right|_{\mathbf q=\mathbf X(t)}.
    }
    \label{eq:many_particle_guidance_equation}
\end{equation}
The notation $\mathbf q=\mathbf X(t)$ means that all the configuration-space
coordinates are simultaneously evaluated at their actual values,
\begin{equation*}
    \mathbf q_1=\mathbf X_1(t),
    \quad\ldots,\quad
    \mathbf q_N=\mathbf X_N(t).
\end{equation*}
Consequently, although
$\boldsymbol{\nabla}_kS$ is a derivative with respect to the coordinate of the
$k$-th particle, its value may depend on the positions of every particle in
the system.

This dependence is absent when the wave function is separable. If, at a given
time,
\begin{equation*}
    \psi(\mathbf q_1,\ldots,\mathbf q_N,t)
    =
    \prod_{k=1}^{N}\psi_k(\mathbf q_k,t),
\end{equation*}
then its phase can be written as
\begin{equation*}
    S(\mathbf q_1,\ldots,\mathbf q_N,t)
    =
    \sum_{k=1}^{N}S_k(\mathbf q_k,t).
\end{equation*}
It follows that
\begin{equation*}
    \boldsymbol{\nabla}_kS
    =
    \boldsymbol{\nabla}_kS_k(\mathbf q_k,t),
\end{equation*}
and the velocity of the $k$-th particle depends only on its own position:
\begin{equation*}
    \frac{d\mathbf X_k(t)}{dt}
    =
    \frac{
        \boldsymbol{\nabla}_k
        S_k(\mathbf X_k(t),t)
    }{m_k}.
\end{equation*}

For an entangled state, no such decomposition is generally possible. The
phase is then a function of the complete configuration, and
Eq.~\textcolor{red}{\eqref{eq:many_particle_guidance_equation}} implies that the instantaneous
velocity of one particle may depend on the simultaneous positions of
spatially separated particles. The same property is visible in the
Newton-like formulation,
\begin{equation*}
    m_k\frac{d^2\mathbf X_k(t)}{dt^2}
    =
    -
    \left.
        \boldsymbol{\nabla}_k
        \left(
            V+Q[\rho]
        \right)
    \right|_{\mathbf q=\mathbf X(t)},
\end{equation*}
because the many-particle QP is defined over the complete
configuration space and need not decompose into independent single-particle
contributions.

The non-local character of the PWT therefore does not arise merely from the
presence of several particles. It is associated with the non-separable
configuration-space dependence of the wave function. In quantum equilibrium,
this dynamics reproduces the non-local correlations predicted by QM. At the
same time, the equilibrium distribution cannot be freely controlled and the
non-local dependence cannot be used to transmit controllable superluminal
signals.

\subsection{The Fundamental or Statistical Nature of the Wave Function}

The preceding analysis shows that, in quantum equilibrium, the SE is
equivalent to a generalized HJ ensemble described by the two real fields
$\rho$ and $S$. Their evolution is generated by Eq.~\textcolor{red}{\eqref{eq:quantum_ensemble_hamiltonian}},
\begin{equation*}
    \boxed{
    \mathcal H_Q[\rho,S]
    =
    \mathcal H_{\mathrm{HJ}}[\rho,S]
    +
    F_Q[\rho],
    }
\end{equation*}
where the functional derivative of $F_Q[\rho]$ produces the QP according to Eq.~\textcolor{red}{\eqref{eq:quantum_potential_functional_derivative}}.

This representation raises the central question considered in the remainder
of this work: whether the wave function is an absolute property of each
individual system or an effective statistical description of an underlying
dynamics. In the standard PWT, the wave function belongs to the complete state
of the individual system, together with its actual configuration, and its
phase determines the corresponding guidance field. The ensemble formulation,
however, shows that the same dynamics can be expressed entirely in terms of a
density $\rho$ and a phase $S$. This leaves open the possibility that these
fields, and hence the wave function constructed from them, may have a
statistical origin.

In particular, the appearance of $\rho$ in the QP does not by
itself imply that $Q[\rho]$ is an ordinary mean-field interaction among the
members of the ensemble. The systems represented by the ensemble are
alternative realizations, rather than simultaneously interacting particles.
Moreover, the QP is invariant under a constant rescaling of
the density:
\begin{equation}
    \boxed{
    Q[C\rho]
    =
    Q[\rho],
    \qquad
    C>0.
    }
    \label{eq:quantum_potential_scale_invariance}
\end{equation}
It therefore depends on the spatial form of the distribution, rather than on
its overall magnitude. This distinguishes it from a conventional mean field
whose strength is generally determined by the amount of matter producing it,
although the scale invariance alone does not establish the microscopic origin
of $Q$.

The functional formulation identifies the precise additional ensemble energy
that an underlying statistical theory must reproduce:
\begin{equation*}
    F_Q[\rho]
    =
    \frac{\hbar^2}{8}
    \sum_{k=1}^{N}
    \frac{1}{m_k}
    \int d^{3N}q\,
    \frac{
        \left(
            \boldsymbol{\nabla}_k\rho
        \right)^2
    }{\rho}.
\end{equation*}
Its functional derivative gives the many-particle QP, while
its combination with the classical ensemble Hamiltonian reproduces the
modified HJ equation and the SE. In the many-particle case, $\rho$ is a
distribution over complete configurations, so any statistical interpretation
must also account for the non-separable correlations described in the
preceding subsection.

The PWT thus provides the required connection between QM and the HJ ensemble
formalism, but it does not determine whether the wave function is fundamental
or statistical. In the following section, we investigate whether
$F_Q[\rho]$ can arise from unresolved stochastic momentum fluctuations. This
will allow us to examine whether the QP and the wave function
may emerge as elements of an effective statistical description, rather than
being introduced as absolute properties of the individual system.

\newpage
\null
\thispagestyle{empty}
\newpage

\chapter{Emergent Quantum Dynamics of Solitons}
\label{chap:3}

The previous chapter showed that QM can be reformulated as a generalized HJ
ensemble described by the density $\rho$ and the phase $S$. Within this
formulation, the additional ensemble energy required to reproduce the QP is known exactly, but its physical origin remains undetermined.
An alternative to the PWT interpretation is to regard the
Madelung equations as describing the mean dynamics of a
stochastic ensemble, with quantum effects represented by
velocity fluctuations and an additional pressure-gradient
force~\textcolor{red}{\cite{HeifetzPlochotnikov2020}}.
In this chapter, we investigate whether this contribution can emerge from an
underlying deterministic field dynamics. In particular, we consider localized
solitonic solutions propagating in a chaotic background and examine whether
the resulting fluctuations of their collective positions and momenta can
generate an effective quantum dynamics at the ensemble level.

\section{The Hall--Reginatto Approach}

To address this question, we first introduce the Hall--Reginatto (HR) exact
uncertainty approach \textcolor{red}{\cite{Hall_Reginatto}}. This construction
starts from the classical HJ ensemble and assumes that the momentum contains
an unresolved random component. By imposing general requirements on the
statistical properties of these fluctuations, their contribution to the
ensemble energy is uniquely determined in terms of the position density. The
resulting variational equations reproduce the quantum HJ equation and the SE.
The approach therefore provides the theoretical conditions that the
fluctuations generated by the solitonic field model must satisfy.

As established in the previous chapter, an ensemble whose local momentum
field is generated by a single-valued HJ function $S(\mathbf x,t)$ is
described by the action
\begin{equation*}
    \mathcal A_{\mathrm C}[\rho,S]
    =
    \int dt\,d^n x\,
    \rho
    \left[
        \frac{\partial S}{\partial t}
        +
        \frac{
            \boldsymbol{\nabla}S\cdot\boldsymbol{\nabla}S
        }{2m}
        +
        V
    \right].
\end{equation*}
Here, $\rho(\mathbf x,t)$ is the normalized position density,
\begin{equation*}
    \int d^n x\,\rho(\mathbf x,t)=1,
\end{equation*}
and
\begin{equation*}
    \mathbf p(\mathbf x,t)
    =
    \boldsymbol{\nabla}S(\mathbf x,t)
\end{equation*}
is the corresponding local momentum field. Independent variations of
$\rho$ and $S$ reproduce the classical HJ equation and the continuity
equation, respectively.

\subsection{Random momentum fluctuations}

The momentum of each member of the ensemble is assumed to contain
a fluctuating contribution,
\begin{equation}
    \mathbf p=\boldsymbol{\nabla}S+\mathbf N,
    \qquad
    \langle\mathbf N\rangle=0,
    \qquad
    \langle\boldsymbol{\nabla}S\cdot\mathbf N\rangle=0.
    \label{eq:momentum_fluctuations}
\end{equation}
Here the brackets denote an average over both the unresolved
fluctuations and the position distribution:
\begin{equation*}
    \langle A\rangle
    :=
    \int d^nx\,\rho(\mathbf x,t)\,
    \overline A(\mathbf x,t),
\end{equation*}
where the bar denotes averaging over fluctuations at fixed position.
The first condition requires the fluctuations to be unbiased globally,
while the second ensures that the cross term in the mean kinetic
energy vanishes.

A stronger sufficient condition is
$\overline{\mathbf N}(\mathbf x,t)=0$ at every position.
This implies both conditions above and identifies
$\boldsymbol{\nabla}S$ with the local mean momentum,
$\overline{\mathbf p}=\boldsymbol{\nabla}S$.
The construction therefore also applies under this stronger assumption.

The mean kinetic energy becomes
\begin{equation*}
    \overline{
        \frac{
            (\boldsymbol{\nabla}S+\mathbf N)^2
        }{2m}
    }
    =
    \frac{
        (\boldsymbol{\nabla}S)^2
    }{2m}
    +
    \frac{
        \overline{\mathbf N\cdot\mathbf N}
    }{2m},
\end{equation*}
because the term linear in $\mathbf N$ vanishes. The ensemble action is
therefore modified according to
\begin{equation}
    \boxed{
    \mathcal A[\rho,S]
    =
    \mathcal A_{\mathrm C}[\rho,S]
    +
    \frac{1}{2m}
    \int dt\,
    (\Delta N)^2,
    }
    \label{eq:HR_modified_action}
\end{equation}
where
\begin{equation}
    \boxed{
    (\Delta N)^2
    =
    \int d^n x\,
    \rho(\mathbf x,t)\,
    \overline{
        \mathbf N(\mathbf x,t)\cdot
        \mathbf N(\mathbf x,t)
    }.
    }
    \label{eq:HR_momentum_fluctuation_strength}
\end{equation}

At this stage, the form of $(\Delta N)^2$ has not yet been specified.
Its independence from $S$ is motivated by the requirement that changing
the momentum contribution $\boldsymbol{\nabla}S$, while keeping the
position density and the fluctuation statistics fixed, should not
change the energy associated with the fluctuations.
The central additional assumption of the exact uncertainty approach
is that the strength of these fluctuations is determined entirely by
the uncertainty in position, and hence by $\rho$ alone. This removes
the need to specify the relevant statistics of $\mathbf N$ independently:
their contribution to the energy is encoded in a functional of $\rho$,
with no additional dependence on $S$ or on the external potential.

\subsection{Locality, causality, and exact uncertainty}

To determine the allowed dependence on $\rho$, the additional contribution
is assumed to preserve the variational formulation and to be local in
configuration space. Following Hall and Reginatto, we further restrict the local contribution to depend on $\rho$ and its first spatial derivatives.
The authors motivate this restriction by requiring that the
evolution be determined by $\rho$ and $S$ on an initial surface. Rotational invariance then allows it to be written as
\begin{equation*}
    (\Delta N)^2
    =
    \int d^n x\,
    \rho\,
    f\left(
        \mathbf x,
        \rho,
        \mathbf x\cdot\boldsymbol{\nabla}\rho,
        \boldsymbol{\nabla}\rho\cdot\boldsymbol{\nabla}\rho
    \right).
\end{equation*}

Consider now the scale transformation
\begin{equation*}
    \rho(\mathbf x)
    \longrightarrow
    \rho_k(\mathbf x)
    =
    k^n\rho(k\mathbf x),
    \qquad k>0.
\end{equation*}
Normalization is preserved by this transformation. We now restrict attention
to direct measures of position uncertainty, defined as quantities that transform
covariantly under a spatial dilation. Thus, when the distribution is contracted
by the factor $k$, an admissible position uncertainty $\delta x$ is required to
satisfy
\begin{equation*}
    \delta x
    \longrightarrow
    \delta x_k
    =
    \frac{\delta x}{k}.
\end{equation*}
This scaling law is not a property of an arbitrary functional of $\rho$, but a
requirement imposed on quantities intended to measure directly the spatial
width of the distribution. The standard deviation is one example of a measure
satisfying this condition.
The exact uncertainty principle requires the momentum fluctuations to scale
inversely,
\begin{equation*}
    \Delta N
    \longrightarrow
    k\Delta N.
\end{equation*}
Consequently,
\begin{equation}
    \boxed{
    f\left(
        k^{-1}\mathbf x,
        k^n u,
        k^n v,
        k^{2n+2}w
    \right)
    =
    k^2 f(\mathbf x,u,v,w),
    }
    \label{eq:HR_homogeneity_condition}
\end{equation}
where, for compactness, we have introduced
\begin{equation*}
    u=\rho,
    \qquad
    v=\mathbf x\cdot\boldsymbol{\nabla}\rho,
    \qquad
    w=\boldsymbol{\nabla}\rho\cdot\boldsymbol{\nabla}\rho.
\end{equation*}

Differentiating Eq.~\textcolor{red}{\eqref{eq:HR_homogeneity_condition}} with respect to $k$
and evaluating the result at $k=1$ gives
\begin{equation*}
    -\sum_{i=1}^{n}
    x_i\frac{\partial f}{\partial x_i}
    +
    nu\frac{\partial f}{\partial u}
    +
    nv\frac{\partial f}{\partial v}
    +
    (2n+2)w\frac{\partial f}{\partial w}
    =
    2f.
\end{equation*}
This first-order partial differential equation can be solved by the method of
characteristics. Introducing a parameter $s$ along each characteristic, the
corresponding system is
\begin{equation*}
    \frac{dx_i}{ds}=-x_i,
    \qquad
    \frac{du}{ds}=nu,
    \qquad
    \frac{dv}{ds}=nv,
    \qquad
    \frac{dw}{ds}=(2n+2)w,
    \qquad
    \frac{df}{ds}=2f.
\end{equation*}
The independent quantities that remain constant along these curves may be
chosen as
\begin{equation*}
    \frac{\sqrt{w}}{u}\mathbf x,
    \qquad
    \frac{v}{u},
    \qquad
    u^{2/n}\mathbf x\cdot\mathbf x,
    \qquad
    \frac{u^2}{w}f.
\end{equation*}
It follows that the general solution of the homogeneity equation has the form
\begin{equation}
    \boxed{
    f
    =
    \frac{w}{u^2}\,
    g\left(
        \frac{\sqrt{w}}{u}\mathbf x,
        \frac{v}{u},
        u^{2/n}\mathbf x\cdot\mathbf x
    \right),
    }
    \label{eq:HR_characteristic_solution}
\end{equation}
where $g$ is an arbitrary function. The scale transformation alone therefore
does not yet determine the fluctuation energy uniquely.

\subsection{Independent subsystems}

The remaining freedom is removed by requiring statistical independence. Let
the complete system consist of two noninteracting and independently prepared
subsystems, with coordinates $\mathbf x_1$ and $\mathbf x_2$. Their joint
density factorizes:
\begin{equation*}
    \rho(\mathbf x_1,\mathbf x_2)
    =
    \rho_1(\mathbf x_1)\rho_2(\mathbf x_2).
\end{equation*}
For independent subsystems, the momentum fluctuations must be uncorrelated,
and their mean-square strengths must add. Equivalently, the fluctuation part
of the ensemble action must decompose into separate subsystem contributions:
\begin{equation}
    \boxed{
    f[\rho_1\rho_2]
    =
    f[\rho_1]+f[\rho_2].
    }
    \label{eq:HR_independence_condition}
\end{equation}

For a factorized density, the combinations entering
Eq.~\textcolor{red}{\eqref{eq:HR_characteristic_solution}} satisfy
\begin{equation*}
    \frac{v}{u}
    =
    \frac{v_1}{u_1}
    +
    \frac{v_2}{u_2},
    \qquad
    \frac{w}{u^2}
    =
    \frac{w_1}{u_1^2}
    +
    \frac{w_2}{u_2^2}.
\end{equation*}
Substitution into Eq.~\textcolor{red}{\eqref{eq:HR_independence_condition}} shows that the
arguments associated with the two subsystems cannot be mixed by the arbitrary
function $g$. Terms depending only on position reduce to additive classical
potentials and can be absorbed into $V$, while the remaining terms would
produce cross-couplings between statistically independent subsystems. The
only nonclassical contribution compatible simultaneously with locality,
scaling, and independence is therefore obtained by taking $g$ to be a
universal positive constant $C$.

Restoring $u=\rho$ and
$w=\boldsymbol{\nabla}\rho\cdot\boldsymbol{\nabla}\rho$, one obtains
\begin{equation}
    \boxed{
    f
    =
    C
    \frac{
        \boldsymbol{\nabla}\rho\cdot
        \boldsymbol{\nabla}\rho
    }{\rho^2},
    \qquad
    C>0.
    }
    \label{eq:HR_unique_fluctuation_density}
\end{equation}
Consequently, the total momentum-fluctuation strength is proportional to the
Fisher information of the position distribution:
\begin{equation}
    \boxed{
    (\Delta N)^2
    =
    C
    \int d^n x\,
    \frac{
        \boldsymbol{\nabla}\rho\cdot
        \boldsymbol{\nabla}\rho
    }{\rho}
    \equiv
    C I[\rho].
    }
    \label{eq:HR_fisher_information}
\end{equation}

This result is not obtained by selecting a particular measure of position
uncertainty in advance. Rather, the Fisher information follows uniquely from
the structural assumptions imposed on the momentum fluctuations.

\subsection{Emergence of the Schrödinger equation}

Using Eq.~\textcolor{red}{\eqref{eq:HR_fisher_information}}, the modified ensemble action
becomes
\begin{equation}
    \boxed{
    \mathcal A_{\mathrm Q}[\rho,S]
    =
    \int dt\,d^n x
    \left\{
        \rho
        \left[
            \frac{\partial S}{\partial t}
            +
            \frac{
                \boldsymbol{\nabla}S\cdot
                \boldsymbol{\nabla}S
            }{2m}
            +
            V
        \right]
        +
        \frac{C}{2m}
        \frac{
            \boldsymbol{\nabla}\rho\cdot
            \boldsymbol{\nabla}\rho
        }{\rho}
    \right\}.
    }
    \label{eq:HR_quantum_ensemble_action}
\end{equation}

Variation with respect to $S$, followed by integration by parts, gives
\begin{equation*}
    \frac{\partial\rho}{\partial t}
    +
    \boldsymbol{\nabla}\cdot
    \left(
        \rho\frac{\boldsymbol{\nabla}S}{m}
    \right)
    =0.
\end{equation*}
The additional Fisher term does not contribute to this equation because it is
independent of $S$.

Variation with respect to $\rho$ requires
\begin{equation*}
    \frac{\delta}{\delta\rho}
    \int d^n x\,
    \frac{
        \boldsymbol{\nabla}\rho\cdot
        \boldsymbol{\nabla}\rho
    }{\rho}
    =
    -2\frac{\nabla^2\rho}{\rho}
    +
    \frac{
        \boldsymbol{\nabla}\rho\cdot
        \boldsymbol{\nabla}\rho
    }{\rho^2}.
\end{equation*}
Using the identity in Eq.~\textcolor{red}{\eqref{eq:sqrt_density_laplacian_identity}}, the two variational equations can be written as
\begin{equation}
    \boxed{
    \begin{array}{@{}l@{}}
        \displaystyle
        \frac{\partial S}{\partial t}
        +
        \frac{
            \boldsymbol{\nabla}S\cdot
            \boldsymbol{\nabla}S
        }{2m}
        +
        V
        -
        \frac{2C}{m}
        \frac{\nabla^2\sqrt{\rho}}{\sqrt{\rho}}
        =0,
        \\[0.6em]
        \displaystyle
        \frac{\partial\rho}{\partial t}
        +
        \boldsymbol{\nabla}\cdot
        \left(
            \rho\frac{\boldsymbol{\nabla}S}{m}
        \right)
        =0.
    \end{array}
    }
    \label{eq:HR_modified_HJ_continuity}
\end{equation}

We now introduce the constant
\begin{equation}
    \boxed{
    \hbar=2\sqrt{C}.
    }
    \label{eq:HR_hbar_definition}
\end{equation}
The first of Eqs.~\textcolor{red}{\eqref{eq:HR_modified_HJ_continuity}} then becomes precisely the quantum HJ equation~\textcolor{red}{\eqref{eq:quantum_hamilton_jacobi_equation}} obtained in the previous chapter.

Finally, defining
\begin{equation*}
    \psi(\mathbf x,t)
    =
    \sqrt{\rho(\mathbf x,t)}\,
    e^{iS(\mathbf x,t)/\hbar}.
\end{equation*}
the modified HJ equation and the continuity equation combine into
\begin{equation}
    \boxed{
    i\hbar\frac{\partial\psi}{\partial t}
    =
    \left[
        -\frac{\hbar^2}{2m}\nabla^2
        +
        V
    \right]\psi.
    }
    \label{eq:HR_schrodinger_equation}
\end{equation}
The SE is therefore recovered from a classical HJ ensemble supplemented by
momentum fluctuations whose strength is fixed by the Fisher information of
the position density.

\subsection{The exact uncertainty relation}

The derivation also identifies the position uncertainty naturally associated
with the momentum fluctuations. In one spatial dimension, define the Fisher
length
\begin{equation}
    \boxed{
    \delta x_{\mathrm F}
    =
    \left[
        \int dx\,
        \rho
        \left(
            \frac{\partial\ln\rho}{\partial x}
        \right)^2
    \right]^{-1/2}
    =
    \frac{1}{\sqrt{I[\rho]}}.
    }
    \label{eq:HR_fisher_length}
\end{equation}
Equation~\textcolor{red}{\eqref{eq:HR_fisher_information}} immediately gives
\begin{equation}
    \boxed{
    \delta x_{\mathrm F}\Delta N
    =
    \sqrt{C}
    =
    \frac{\hbar}{2}.
    }
    \label{eq:HR_exact_uncertainty_relation}
\end{equation}
This is an equality, rather than an inequality, and relates the nonclassical
momentum fluctuations directly to the Fisher length of the position
distribution.

The standard Heisenberg uncertainty relation follows from this result. The
Cramér--Rao inequality \cite{CoverThomas2006} gives
\begin{equation*}
    \Delta x
    \geq
    \delta x_{\mathrm F},
\end{equation*}
where $\Delta x$ is the standard deviation of position. Moreover, from
\begin{equation*}
    p
    =
    \frac{\partial S}{\partial x}+N
\end{equation*}
and the randomness assumptions,
\begin{equation*}
    (\Delta p)^2
    =
    \operatorname{Var}
    \left(
        \frac{\partial S}{\partial x}
    \right)
    +
    (\Delta N)^2
    \geq
    (\Delta N)^2.
\end{equation*}
It follows that
\begin{equation}
    \boxed{
    \Delta x\,\Delta p
    \geq
    \delta x_{\mathrm F}\Delta N
    =
    \frac{\hbar}{2}.
    }
    \label{eq:HR_heisenberg_uncertainty}
\end{equation}
For a Gaussian position density,
$\Delta x=\delta x_{\mathrm F}$. Saturation of the Heisenberg inequality
additionally requires the mean momentum field
$\partial_xS$ to have vanishing variance, as occurs for a phase that is
linear in position.

The HR construction therefore specifies exactly what must be
reproduced by an underlying microscopic theory. Such a theory must provide
zero-mean momentum fluctuations whose ensemble energy is
\begin{equation*}
    (\Delta N)^2
    =
    \frac{\hbar^2}{4}I[\rho].
\end{equation*}
The following sections investigate whether the fluctuations generated by a
soliton interacting with a chaotic field background can realize this
statistical structure and thereby provide a microscopic origin for the
QP.

\section{The Solitonic Field Model}

The HR construction introduced in the previous section, in particular Eq.~\textcolor{red}{\eqref{eq:HR_fisher_information}},
identifies the statistical properties required for momentum fluctuations to
generate the SE. We now introduce the nonlinear field theory whose localized
solutions will provide the underlying particle-like degrees of freedom. The
model is based on the complex sine--Gordon (CSG) theory considered by Bowcock,
Foster, and Sutcliffe
\textcolor{red}{\cite{Bowcock_Qballs}}, and on its Galilean counterpart
constructed by de Melo \textit{et al.}
\textcolor{red}{\cite{deMelo_GCSGE}}.

Consider a complex scalar field $\phi$ with relativistic Lagrangian density
\begin{equation*}
    \mathcal L_{\mathrm R}
    =
    \frac{1}{1-|\phi|^2}
    \left[
        \partial_\mu\phi^*
        \partial^\mu\phi
        +
        |\phi|^2
        -
        |\phi|^4
        +
        V(x)|\phi|^2
    \right].
\end{equation*}
The relative signs of the kinetic and potential contributions depend on the
metric convention; for $V=0$, this is the CSG model studied
in Ref.~\textcolor{red}{\cite{Bowcock_Qballs}}, up to an irrelevant overall
sign of the Lagrangian. The denominator corresponds to a nontrivial
conformally flat target-space metric and plays an essential role in the
existence of the localized solutions considered below.

The theory is invariant under the global transformation
\begin{equation*}
    \phi(x,t)
    \longrightarrow
    e^{i\alpha}\phi(x,t),
\end{equation*}
where $\alpha$ is constant. This symmetry is retained by the Galilean theory
constructed below. Its conserved density and current will be derived
directly from the reduced Galilean Lagrangian, rather than imported from the
relativistic parent theory. The corresponding localized configurations are
non-topological solitons known as Q-balls. Unlike topological solitons, their
existence is not associated with an integer-valued topological invariant,
but with the continuous charge generated by the global $U(1)$ symmetry.

For $V=0$, the relativistic CSG equation is integrable and
admits exact multi-soliton solutions. The present work employs instead the
corresponding Galilean complex sine--Gordon equation (GCSGE), since the
effective dynamics of the soliton centres must be compatible with Galilean
transformations.

\subsection{Bargmann Reduction to the Galilean Field Model}

To construct such a theory systematically, it is useful to
represent Galilean boosts as ordinary isometries of an extended geometry.
In the physical coordinates $(x,t)$, a Galilean boost acts as
\begin{equation*}
    x'=x-vt,
    \qquad
    t'=t.
\end{equation*}
There is no non-degenerate spacetime metric in $(x,t)$ that is invariant
under all these transformations. Galilean spacetime preserves the absolute
time interval $dt$ and spatial distances between simultaneous events, but
these two structures cannot be combined into a single non-degenerate metric
on $(x,t)$. The Bargmann construction resolves this obstruction by adding an
auxiliary null coordinate $s$ and introducing the line element
\begin{equation*}
    d\ell^2
    =
    dx^2-2\,dt\,ds.
\end{equation*}
The boost is lifted to the extended coordinates according to
\begin{equation*}
    x'=x-vt,
    \qquad
    t'=t,
    \qquad
    s'=s-vx+\frac{v^2}{2}t.
\end{equation*}
The transformation of $s$ is fixed by the requirement
\begin{equation*}
    dx'^2-2\,dt'\,ds'
    =
    dx^2-2\,dt\,ds.
\end{equation*}
Thus, the additional coordinate compensates the mixed and quadratic terms
generated by $dx'=dx-v\,dt$, and a Galilean boost becomes an isometry of the
extended space.

In the coordinate order $(x,t,s)$, the corresponding metric and inverse
metric are represented by the same matrix,
\begin{equation*}
    g_{AB}
    =
    g^{AB}
    =
    \begin{pmatrix}
        1 & 0 & 0 \\
        0 & 0 & -1 \\
        0 & -1 & 0
    \end{pmatrix}.
\end{equation*}
Its determinant is nonzero, whereas the directions generated separately by
$\partial_t$ and $\partial_s$ are null. The Bargmann structure additionally
selects
\begin{equation*}
    \xi
    =
    \partial_s
\end{equation*}
as a preferred null direction. Since $\partial_s=\partial_{s'}$ under the
lifted boost, translations along $s$ provide the central generator of the
Galilei group.

Once this geometry has been introduced, an invariant kinetic term can be
formed by contracting field derivatives with the inverse metric. For a
complex scalar field $\widehat\phi(x,t,s)$, the minimal scalar contraction is
\begin{equation*}
    g^{AB}
    \partial_A\widehat\phi^*
    \partial_B\widehat\phi
    =
    \partial_A\widehat\phi^*
    \partial^A\widehat\phi.
\end{equation*}
The kinetic contraction is therefore
\begin{equation*}
    \partial_A\widehat\phi^*
    \partial^A\widehat\phi
    =
    \partial_x\widehat\phi^*
    \partial_x\widehat\phi
    -
    \partial_t\widehat\phi^*
    \partial_s\widehat\phi
    -
    \partial_s\widehat\phi^*
    \partial_t\widehat\phi.
\end{equation*}
We then choose the minimal $U(1)$-invariant Bargmann lift that retains the
nonlinear target-space factor of the CSG model:
\begin{equation*}
    \widehat{\mathcal L}
    =
    \frac{
        \partial_A\widehat\phi^*
        \partial^A\widehat\phi
        +
        V(x)|\widehat\phi|^2
    }{
        1-|\widehat\phi|^2
    }
    +
    |\widehat\phi|^2.
\end{equation*}
This choice is not claimed to be the most general scalar Lagrangian allowed
by the extended geometry. It is the minimal choice, in derivatives and field
content, that reproduces the nonlinear Galilean model required here.

The physical Galilean theory is obtained by selecting an eigensector of
translations along the auxiliary direction:
\begin{equation*}
    i\partial_s\widehat\phi
    =
    m'\widehat\phi.
\end{equation*}
Its general solution is
\begin{equation*}
    \widehat\phi(x,t,s)
    =
    e^{-im's}\phi(x,t).
\end{equation*}
The number $m'$ is therefore the eigenvalue of the central Bargmann
generator $i\partial_s$. After reduction, it appears in the phase of a
Galilean boost and plays the role of mass per unit $U(1)$ charge.

Although the theory is nonlinear, this fixed-$m'$ sector is dynamically
closed. Indeed,
\begin{equation*}
    |\widehat\phi|^2
    =
    |\phi|^2
\end{equation*}
is independent of $s$, and every $U(1)$-covariant term in the field equation carries the same overall factor $e^{-im's}$. A generic $s$-dependent field could contain several Fourier components, but the nonlinearity would in general couple them. The single equivariant component used here is a consistent symmetry reduction to one fixed central-charge sector.
The projection ansatz gives
\begin{equation*}
    \partial_s\widehat\phi
    =
    -im'\widehat\phi,
    \qquad
    \partial_s\widehat\phi^*
    =
    im'\widehat\phi^*.
\end{equation*}
Since the phase factors $e^{\pm im's}$ cancel in every bilinear term, the
kinetic contribution reduces to
\begin{equation*}
    \begin{aligned}
        \partial_A\widehat\phi^*
        \partial^A\widehat\phi
        &=
        \partial_x\phi^*
        \partial_x\phi
        +
        im'
        \left(
            \phi\partial_t\phi^*
            -
            \phi^*\partial_t\phi
        \right).
    \end{aligned}
\end{equation*}
The Galilean Lagrangian density is consequently
\begin{equation}
    \boxed{
    \mathcal L
    =
    \mathcal L_{\mathrm G}
    :=
    \frac{
        \partial_x\phi^*
        \partial_x\phi
        +
        im'
        \left(
            \phi\partial_t\phi^*
            -
            \phi^*\partial_t\phi
        \right)
        +
        V(x)|\phi|^2
    }{
        1-|\phi|^2
    }
    +
    |\phi|^2
    }
    \label{eq:GCSGE_Lagrangian}
\end{equation}
The Lagrangian thus reduces to one for the field $\phi(x,t)$,
with no remaining dependence on the auxiliary coordinate $s$.

Let us derive the field equation generated by this Lagrangian. Define
\begin{equation*}
    D=1-|\phi|^2
\end{equation*}
The derivatives entering the Euler--Lagrange equation for $\phi^*$ are
\begin{equation*}
    \frac{
        \partial\mathcal L_{\mathrm G}
    }{
        \partial(\partial_t\phi^*)
    }
    =
    \frac{
        im'\phi
    }{
        D
    },
    \qquad
    \frac{
        \partial\mathcal L_{\mathrm G}
    }{
        \partial(\partial_x\phi^*)
    }
    =
    \frac{
        \partial_x\phi
    }{
        D
    },
\end{equation*}
and
\begin{equation*}
    \frac{
        \partial\mathcal L_{\mathrm G}
    }{
        \partial\phi^*
    }
    =
    \frac{
        \left(
            -im'\partial_t\phi
            +
            V\phi
        \right)D
        +
        \phi
        \left[
            \partial_x\phi^*
            \partial_x\phi
            +
            im'
            \left(
                \phi\partial_t\phi^*
                -
                \phi^*\partial_t\phi
            \right)
            +
            V|\phi|^2
        \right]
    }{
        D^2
    }
    +
    \phi.
\end{equation*}
Substitution into
\begin{equation*}
    \frac{
        \partial\mathcal L_{\mathrm G}
    }{
        \partial\phi^*
    }
    -
    \partial_t
    \left[
        \frac{
            \partial\mathcal L_{\mathrm G}
        }{
            \partial(\partial_t\phi^*)
        }
    \right]
    -
    \partial_x
    \left[
        \frac{
            \partial\mathcal L_{\mathrm G}
        }{
            \partial(\partial_x\phi^*)
        }
    \right]
    =
    0
\end{equation*}
gives, after multiplication by $D^2$ and simplification,
\begin{equation*}
    D\partial_{xx}\phi
    +
    2im'\partial_t\phi
    +
    \phi^*(\partial_x\phi)^2
    -
    \phi D^2
    -
    V\phi
    =
    0.
\end{equation*}
Solving for the time derivative yields the Galilean complex
sine--Gordon equation (GCSGE),
\begin{equation}
    \boxed{
    2m'\frac{\partial\phi}{\partial t}
    =
    -i\left(1-|\phi|^2\right)^2\phi
    +
    i\left(1-|\phi|^2\right)
    \frac{\partial^2\phi}{\partial x^2}
    +
    i\phi^*
    \left(
        \frac{\partial\phi}{\partial x}
    \right)^2
    -
    iV(x)\phi .
    }
    \label{eq:GCSGE}
\end{equation}
For $V=0$, the extended Lagrangian is invariant under lifted
Galilean boosts: its derivative terms are contracted with the
Bargmann metric, which these transformations preserve, while
the remaining terms depend only on the modulus of the scalar
field. We now show that this symmetry is inherited by the
reduced theory and determine the corresponding transformation
of $\phi(x,t)$.

Let $\mathcal E[\widehat\phi]=0$ denote the field equation
associated with the extended Lagrangian $\mathcal L$, and let
$\mathcal E_{m'}[\phi]=0$ denote the equation associated with
the reduced Lagrangian $\mathcal L_{\mathrm G}$.
Under the equivariance condition, the extended Lagrangian
reduces to $\mathcal L_{\mathrm G}$, and the field equation
factorizes as
\begin{equation*}
    \mathcal E
    \left[\mathrm{e}^{-\mathrm{i}m's}\phi\right]
    =
    \mathrm{e}^{-\mathrm{i}m's}
    \mathcal E_{m'}[\phi].
\end{equation*}
Thus, any solution $\phi(x,t)$ of the reduced equation defines
a solution
$\widehat\phi(x,t,s)=\mathrm{e}^{-\mathrm{i}m's}\phi(x,t)$
of the extended equation.

In the active convention, a lifted Galilean boost acts on
the extended field according to
\begin{equation*}
    \widehat{\phi}_v(x,t,s)
    =
    \widehat{\phi}(
        x-vt,\,
        t,\,
        s-vx+\tfrac12 v^2t
    ).
\end{equation*}
Since this transformation is a symmetry of the extended theory,
the transformed field is also a solution:
\begin{equation*}
    \mathcal E[\widehat\phi_v]=0.
\end{equation*}
Using the equivariance condition, we obtain
\begin{equation*}
    \begin{aligned}
        \widehat{\phi}_v(x,t,s)
        &=
        \mathrm{e}^{-\mathrm{i}m'(s-vx+\tfrac12 v^2t)}
        \phi(x-vt,t)
        \\
        &=
        \mathrm{e}^{-\mathrm{i}m's}
        \mathrm{e}^{\mathrm{i}m'(vx-\tfrac12 v^2t)}
        \phi(x-vt,t).
    \end{aligned}
\end{equation*}
The transformed field therefore retains the same dependence
on $s$ and remains in the sector with fixed $m'$.
Writing
\begin{equation*}
    \widehat\phi_v(x,t,s)
    =
    \mathrm{e}^{-\mathrm{i}m's}\phi_v(x,t),
\end{equation*}
identifies the transformation law of the reduced field:
\begin{equation}
    \boxed{
    \phi_v(x,t)
    =
    \mathrm{e}^{\mathrm{i}m'(vx-\tfrac12 v^2t)}
    \phi(x-vt,t).
    }
    \label{eq:GCSGE_Galilean_boost}
\end{equation}
Applying the extended field equation gives
\begin{equation*}
    0
    =
    \mathcal E[\widehat\phi_v]
    =
    \mathrm{e}^{-\mathrm{i}m's}
    \mathcal E_{m'}[\phi_v].
\end{equation*}
Since the exponential factor never vanishes, it follows that
\begin{equation*}
    \boxed{\mathcal E_{m'}[\phi_v]=0.}
\end{equation*}
Thus, the boost preserves the fixed-$m'$ sector and maps
solutions of the reduced equation into solutions of the same
equation. This establishes the Galilean invariance of the
reduced theory for $V=0$.

We next identify the charge and momentum densities of the
reduced field and derive their local balance equations directly
from Eq.~\textcolor{red}{\eqref{eq:GCSGE}}.
Writing the field in polar form,
\begin{equation}
    \phi(x,t)
    =
    |\phi(x,t)|
    \mathrm{e}^{\mathrm{i}\Theta(x,t)}.
\label{eq:field_polar_decomposition}
\end{equation}
In these variables,
\begin{equation*}
    \left|
        \partial_x\phi
    \right|^2
    =
    \left(
        \partial_x|\phi|
    \right)^2
    +
    |\phi|^2
    \left(
        \partial_x\Theta
    \right)^2
\end{equation*}
and
\begin{equation*}
    \mathrm{i}m'
    \left(
        \phi\partial_t\phi^*
        -
        \phi^*\partial_t\phi
    \right)
    =
    2m'|\phi|^2\partial_t\Theta.
\end{equation*}
The reduced Lagrangian therefore takes the explicit form
\begin{equation*}
    \mathcal L_{\mathrm G}
    =
    \frac{(\partial_x|\phi|)^2}{1-|\phi|^2}
    +
    \frac{|\phi|^2}{1-|\phi|^2}
    \left[
        2m'\partial_t\Theta
        +(\partial_x\Theta)^2
        +V(x)
    \right]
    +|\phi|^2.
\end{equation*}
This expression identifies the normalized density conjugate to the temporal
phase gradient:
\begin{equation}
    \boxed{
    q[\phi]
    =
    \frac{1}{2m'}
    \frac{
        \partial\mathcal L_{\mathrm G}
    }{
        \partial(\partial_t\Theta)
    }
    =
    \frac{
        |\phi|^2
    }{
        1-|\phi|^2
    }.
    }
    \label{eq:GCSGE_charge_density}
\end{equation}
The density conjugate to the spatial phase gradient is
\begin{equation}
    \boxed{
    \begin{aligned}
        p[\phi]
        &=
        \frac{1}{2}
        \frac{
            \partial\mathcal L_{\mathrm G}
        }{
            \partial(\partial_x\Theta)
        }
        \\
        &=
        \frac{
            |\phi|^2
        }{
            1-|\phi|^2
        }
        \partial_x\Theta
        \\
        &=
        \frac{\mathrm{i}}{
            2\left(
                1-|\phi|^2
            \right)
        }
        \left(
            \phi\partial_x\phi^*
            -
            \phi^*\partial_x\phi
        \right).
    \end{aligned}
    }
    \label{eq:GCSGE_momentum_density}
\end{equation}
In the canonical formulation of field theory, these quantities are the
densities of the generators of global phase transformations and spatial
translations, respectively. They can be obtained by applying the general
canonical construction of continuous field transformations, presented in
Ref.~\textcolor{red}{\cite{Srednicki}}, to the polar decomposition of the complex field.
They are therefore identified as the charge density $q[\phi]$ and the
momentum density $p[\phi]$.

Substitution of the polar decomposition~\textcolor{red}{\eqref{eq:field_polar_decomposition}} into Eq.~\textcolor{red}{\eqref{eq:GCSGE}} requires
\begin{equation*}
    \partial_t\phi
    =
    \mathrm{e}^{\mathrm{i}\Theta}
    \left[
        \partial_t|\phi|
        +\mathrm{i}|\phi|\partial_t\Theta
    \right],
\end{equation*}
\begin{equation*}
    \partial_{xx}\phi
    =
    \mathrm{e}^{\mathrm{i}\Theta}
    \left[
        \partial_{xx}|\phi|
        -|\phi|(\partial_x\Theta)^2
        +\mathrm{i}\left(
            2\partial_x|\phi|\partial_x\Theta
            +|\phi|\partial_{xx}\Theta
        \right)
    \right],
\end{equation*}
and
\begin{equation*}
    \phi^*(\partial_x\phi)^2
    =
    \mathrm{e}^{\mathrm{i}\Theta}|\phi|
    \left[
        (\partial_x|\phi|)^2
        -|\phi|^2(\partial_x\Theta)^2
        +2\mathrm{i}|\phi|\partial_x|\phi|\partial_x\Theta
    \right].
\end{equation*}
After canceling the common factor
$\mathrm{e}^{\mathrm{i}\Theta}$, the real and imaginary parts of the field equation give
\begin{empheq}[box=\fbox]{align}
    2m'\partial_t|\phi|
    &=
    -2(\partial_x|\phi|)(\partial_x\Theta)
    -|\phi|(1-|\phi|^2)\partial_{xx}\Theta
    \label{eq:GCSGE_polar_amplitude_equation}
    \\[0.5em]
    2m'\partial_t\Theta
    &=
    \frac{(1-|\phi|^2)\partial_{xx}|\phi|}{|\phi|}
    +(\partial_x|\phi|)^2
    -(1-|\phi|^2)^2
    -(\partial_x\Theta)^2
    -V(x).
    \label{eq:GCSGE_polar_phase_equation}
\end{empheq}
These are the two real equations contained in the complex field equation.

We first derive the charge balance. From the explicit definition of
$q[\phi]$,
\begin{equation*}
    \partial_t q[\phi]
    =
    \frac{2|\phi|\partial_t|\phi|}{(1-|\phi|^2)^2}.
\end{equation*}
Substitution of
Eq.~\textcolor{red}{\eqref{eq:GCSGE_polar_amplitude_equation}} gives
\begin{equation*}
    \partial_t q[\phi]
    =
    -\frac{1}{m'}
    \left[
        \frac{2|\phi|\,\partial_x|\phi|}{(1-|\phi|^2)^2}
        \,\partial_x\Theta
        +
        \frac{|\phi|^2}{1-|\phi|^2}
        \,\partial_{xx}\Theta
    \right].
\end{equation*}
The expression inside the square brackets is exactly
\begin{equation*}
    \partial_x
    \left[
        \frac{
            |\phi|^2
        }{
            1-|\phi|^2
        }
        \partial_x\Theta
    \right]
    =
    \partial_xp[\phi].
\end{equation*}
Consequently,
\begin{equation}
    \boxed{
    m'\partial_tq[\phi]
    +
    \partial_xp[\phi]
    =
    0.
    }
    \label{eq:GCSGE_charge_continuity}
\end{equation}

The local momentum balance follows from the phase equation. Since
\begin{equation*}
    p[\phi]
    =
    q[\phi]\partial_x\Theta,
\end{equation*}
we have
\begin{equation*}
    \partial_tp[\phi]
    =
    \partial_x\Theta\,
    \partial_tq[\phi]
    +
    q[\phi]\,
    \partial_x\partial_t\Theta.
\end{equation*}
Differentiating Eq.~\textcolor{red}{\eqref{eq:GCSGE_polar_phase_equation}} with respect to
$x$ gives
\begin{equation*}
    2m'\partial_x\partial_t\Theta
    =
    \partial_x
    \left[
        \frac{(1-|\phi|^2)\partial_{xx}|\phi|}{|\phi|}
        +(\partial_x|\phi|)^2
        -(1-|\phi|^2)^2
        -(\partial_x\Theta)^2
        -V(x)
    \right].
\end{equation*}
Using the charge-continuity equation to replace
$\partial_tq[\phi]$, we obtain
\begin{equation*}
    \begin{aligned}
        \partial_t p[\phi]
        ={}&
        -\frac{\partial_x\Theta}{m'}
        \partial_x\left[q[\phi]\partial_x\Theta\right]
        \\
        &+\frac{q[\phi]}{2m'}
        \partial_x\left[
            \frac{(1-|\phi|^2)\partial_{xx}|\phi|}{|\phi|}
            +(\partial_x|\phi|)^2
            -(1-|\phi|^2)^2
            -(\partial_x\Theta)^2
            -V(x)
        \right].
    \end{aligned}
\end{equation*}
The terms containing the phase gradient combine according to
\begin{equation*}
    -\frac{\partial_x\Theta}{m'}
    \partial_x\left[q[\phi]\partial_x\Theta\right]
    -\frac{q[\phi]}{2m'}
    \partial_x\left[(\partial_x\Theta)^2\right]
    =
    -\frac{1}{m'}
    \partial_x\left[q[\phi](\partial_x\Theta)^2\right].
\end{equation*}
The remaining amplitude contribution satisfies the identity
\begin{equation*}
    \begin{aligned}
        &\frac{|\phi|^2}{1-|\phi|^2}
        \partial_x\left[
            \frac{(1-|\phi|^2)\partial_{xx}|\phi|}{|\phi|}
            +(\partial_x|\phi|)^2
            -(1-|\phi|^2)^2
        \right]
        \\
        &\qquad=
        \partial_x\left[
            |\phi|\partial_{xx}|\phi|
            -(\partial_x|\phi|)^2
            +|\phi|^4
        \right].
    \end{aligned}
\end{equation*}
Therefore, the momentum flux is
\begin{equation}
    \boxed{
    J_p[\phi]
    =
    \frac{1}{2m'}
    \left[
        |\phi|\partial_{xx}|\phi|
        -(\partial_x|\phi|)^2
        +|\phi|^4
        -2\frac{|\phi|^2}{1-|\phi|^2}(\partial_x\Theta)^2
    \right].
    }
    \label{eq:GCSGE_momentum_flux}
\end{equation}
The exact local momentum equation becomes
\begin{equation}
    \boxed{
    \partial_tp[\phi]
    =
    \partial_xJ_p[\phi]
    -
    \frac{1}{2m'}
    q[\phi]\partial_xV.
    }
    \label{eq:GCSGE_momentum_balance}
\end{equation}
Both balance equations have been obtained directly from the GCSGE. No
soliton profile, weak-potential approximation, or rigidity assumption has
been introduced.

For a localized configuration, integration of the charge equation gives
\begin{equation*}
    \begin{aligned}
        \frac{dQ}{dt}
        &=
        \frac{d}{dt}
        \int_{-\infty}^{+\infty}
        q[\phi]\,dx
        \\
        &=
        -\frac{1}{m'}
        \left[
            p[\phi]
        \right]_{-\infty}^{+\infty}.
    \end{aligned}
\end{equation*}
If the momentum density has the same asymptotic value at the two boundaries,
and in particular if it vanishes there, then
\begin{equation}
    \boxed{
    Q
    =
    \int_{-\infty}^{+\infty}
    q[\phi]\,dx,
    \qquad
    \frac{dQ}{dt}
    =
    0.
    }
    \label{eq:GCSGE_conserved_charge}
\end{equation}

We now consider the quantity associated with spatial translations. The corresponding momentum density is given in Eq.~\textcolor{red}{\eqref{eq:GCSGE_momentum_density}}, and its local balance law is Eq.~\textcolor{red}{\eqref{eq:GCSGE_momentum_balance}}, where $J_p[\phi]$ is the momentum flux. Integrating this equation over the real line gives
\begin{equation*}
    \frac{d}{dt}
    \int_{-\infty}^{+\infty}
    p[\phi]\,dx
    =
    \left[
        J_p[\phi]
    \right]_{-\infty}^{+\infty}
    -
    \frac{1}{2m'}
    \int_{-\infty}^{+\infty}
    q[\phi]\partial_xV\,dx.
\end{equation*}
For a localized field configuration, or more generally whenever the momentum
flux has the same value at the two spatial boundaries, the boundary
contribution vanishes. We can therefore define the total field momentum as
\begin{equation}
    \boxed{
    P
    =
    \int_{-\infty}^{+\infty}
    p[\phi]\,dx,
    \qquad
    \frac{dP}{dt}
    =
    -\frac{1}{2m'}
    \int_{-\infty}^{+\infty}
    q[\phi]\partial_xV\,dx
    \equiv
    F_V[\phi].
    }
    \label{eq:GCSGE_total_momentum_balance}
\end{equation}
Thus, in the absence of an external potential gradient,
\begin{equation*}
    \partial_xV=0,
\end{equation*}
the total momentum is conserved:
\begin{equation*}
    \frac{dP}{dt}=0.
\end{equation*}
In the presence of a nonuniform external potential, the same quantity is no
longer conserved, but its time derivative is exactly equal to the total
external force acting on the field.

Finally, we determine the mass carried by this charge. From
Eq.~\textcolor{red}{\eqref{eq:GCSGE_Galilean_boost}}, the modulus and phase transform as
\begin{equation*}
    |\phi_v(x,t)|
    =
    |\phi(x-vt,t)|
\end{equation*}
and
\begin{equation*}
    \Theta_v(x,t)
    =
    \Theta(x-vt,t)
    +m'(vx-\tfrac12 v^2t).
\end{equation*}
It follows that
\begin{equation*}
    q[\phi_v](x,t)
    =
    q[\phi](x-vt,t)
\end{equation*}
and
\begin{equation*}
    \begin{aligned}
        p[\phi_v](x,t)
        &=
        q[\phi_v](x,t)
        \partial_x\Theta_v(x,t)
        \\
        &=
        p[\phi](x-vt,t)
        +
        m'v\,q[\phi](x-vt,t).
    \end{aligned}
\end{equation*}
Using the total momentum defined in Eq.~\textcolor{red}{\eqref{eq:GCSGE_total_momentum_balance}} and changing the integration variable from $x$ to $x-vt$, we obtain
\begin{equation*}
    P_v
    =
    P+m'Qv.
\end{equation*}
The Galilean transformation law for the momentum of an object with total
mass $\mathcal M$ is
\begin{equation*}
    P_v
    =
    P+\mathcal Mv.
\end{equation*}
Comparison of the two expressions gives
\begin{equation}
    \boxed{
    \mathcal M
    =
    m'Q.
    }
    \label{eq:GCSGE_mass_charge_relation}
\end{equation}
The parameter $m'$ is fixed by the selected eigensector of
$i\partial_s$, whereas $Q$ is conserved by the field equation. Hence the
total Galilean mass $\mathcal M=m'Q$ remains constant during the evolution.
The construction therefore provides a Galilean counterpart of the
relativistic theory while preserving its characteristic nonlinear field
structure; it is not a direct low-velocity expansion of the original
relativistic equation.

\subsection{One-Soliton Solutions and Effective Dynamics}

We now focus on the one-soliton sector of the Galilean field theory. This
choice is motivated by the fact that an explicit localized one-soliton
solution can be constructed analytically. By contrast, a general exact
multi-soliton family and its complete scattering dynamics have not been
established within the present Galilean construction. Configurations
containing several solitons can therefore be considered through
well-separated-profile approximations and direct numerical evolution, rather than being assumed to belong to a known exact multi-soliton family.

For a vanishing external potential, we first search for a stationary
localized solution of Eq.~\textcolor{red}{\eqref{eq:GCSGE}} in the form
\begin{equation*}
    \phi(x,t)
    =
    f(x-X_0)\,
    \mathrm{e}^{\mathrm{i}\Omega t},
\end{equation*}
where $f$ is real and satisfies
\begin{equation*}
    f(x-X_0)
    \longrightarrow
    0
    \qquad
    \text{as}
    \qquad
    |x|\longrightarrow\infty.
\end{equation*}
Smooth localization also requires
\begin{equation*}
    \frac{df}{dx}
    \longrightarrow
    0
    \qquad
    \text{as}
    \qquad
    |x|\longrightarrow\infty.
\end{equation*}

Since
\begin{equation*}
    \partial_t\phi
    =
    \mathrm{i}\Omega\phi,
\end{equation*}
and
\begin{equation*}
    \partial_x\phi
    =
    \frac{df}{dx}
    \mathrm{e}^{\mathrm{i}\Omega t},
    \qquad
    \partial_{xx}\phi
    =
    \frac{d^2f}{dx^2}
    \mathrm{e}^{\mathrm{i}\Omega t},
\end{equation*}
substitution into Eq.~\textcolor{red}{\eqref{eq:GCSGE}}, with $V=0$, gives
\begin{equation*}
    2m'\Omega f
    =
    -
    \left(
        1-f^2
    \right)^2f
    +
    \left(
        1-f^2
    \right)
    \frac{d^2f}{dx^2}
    +
    f
    \left(
        \frac{df}{dx}
    \right)^2.
\end{equation*}
Equivalently,
\begin{equation*}
    \left(
        1-f^2
    \right)
    \frac{d^2f}{dx^2}
    +
    f
    \left(
        \frac{df}{dx}
    \right)^2
    =
    \left(
        1-f^2
    \right)^2f
    +
    2m'\Omega f.
\end{equation*}

Multiplying both sides by
\begin{equation*}
    \frac{
        2\,df/dx
    }{
        \left(
            1-f^2
        \right)^2
    },
\end{equation*}
the left-hand side becomes
\begin{equation*}
    \frac{d}{dx}
    \left[
        \frac{
            \left(
                df/dx
            \right)^2
        }{
            1-f^2
        }
    \right].
\end{equation*}
The resulting equation can therefore be written as
\begin{equation*}
    \frac{d}{dx}
    \left[
        \frac{
            \left(
                df/dx
            \right)^2
        }{
            1-f^2
        }
    \right]
    =
    \frac{d}{dx}
    \left[
        f^2
        +
        \frac{
            2m'\Omega
        }{
            1-f^2
        }
    \right].
\end{equation*}
Integration with respect to $x$ gives
\begin{equation*}
    \frac{
        \left(
            df/dx
        \right)^2
    }{
        1-f^2
    }
    =
    f^2
    +
    \frac{
        2m'\Omega
    }{
        1-f^2
    }
    +
    C.
\end{equation*}
The localization conditions at spatial infinity imply
\begin{equation*}
    C=-2m'\Omega.
\end{equation*}
Consequently,
\begin{equation*}
    \left(
        \frac{df}{dx}
    \right)^2
    =
    f^2
    \left(
        1+2m'\Omega-f^2
    \right).
\end{equation*}

We introduce the soliton parameter $w'$ through
\begin{equation*}
    \boxed{
    w^{\prime 2}
    =
    1+2m'\Omega.
    }
\end{equation*}
The first integral then takes the form
\begin{equation*}
    \left(
        \frac{df}{dx}
    \right)^2
    =
    f^2
    \left(
        w^{\prime 2}-f^2
    \right).
\end{equation*}
Its positive, localized and reflection-symmetric solution is
\begin{equation*}
    f(x-X_0)
    =
    w'\operatorname{sech}
    \left[
        w'(x-X_0)
    \right].
\end{equation*}
Indeed,
\begin{equation*}
    \frac{df}{dx}
    =
    -
    w^{\prime 2}
    \operatorname{sech}
    \left[
        w'(x-X_0)
    \right]
    \operatorname{tanh}
    \left[
        w'(x-X_0)
    \right],
\end{equation*}
and direct substitution verifies the first-order equation.

Localization requires $w^{\prime 2}>0$, while regularity of the target-space
metric requires $|\phi|<1$. Since the maximum field amplitude is $w'$, the
allowed interval is
\begin{equation*}
    0<w'<1.
\end{equation*}
For $m'>0$, the corresponding frequency lies in the interval
\begin{equation*}
    -\frac{1}{2m'}
    <
    \Omega
    <
    0.
\end{equation*}

The stationary solution can now be set into uniform motion by applying the
Galilean boost derived above. Starting from the solution centered at $X_s(0)$,
we obtain
\begin{equation}
    \boxed{
    \begin{aligned}
        &\phi_s(x,t)
        =
        w'\operatorname{sech}[w'(x-X_s(t))]
        \,\mathrm{e}^{\mathrm{i}m'(v_0x-\tfrac12 v_0^2t)}
        \mathrm{e}^{\mathrm{i}\Omega t},
        \\
        &X_s(t)=X_s(0)+v_0t.
    \end{aligned}
    }
    \label{eq:GCSGE_single_soliton}
\end{equation}
This expression is an exact solution of the field equation for $V=0$. The
envelope is centered at $X_s(t)$ and moves with the constant velocity $v_0$.

The spatial phase gradient of this solution is
\begin{equation*}
    \partial_x\Theta_s
    =
    m'v_0.
\end{equation*}
Using
\begin{equation*}
    p[\phi_s]
    =
    q[\phi_s]\partial_x\Theta_s,
\end{equation*}
we find
\begin{equation*}
    p[\phi_s]
    =
    m'v_0
    \frac{
        |\phi_s|^2
    }{
        1-|\phi_s|^2
    }.
\end{equation*}
The soliton charge is
\begin{equation*}
    Q_s
    =
    \int_{-\infty}^{+\infty}
    \frac{|\phi_s|^2}{1-|\phi_s|^2}\,dx
    =
    \frac{2\operatorname{arcsin}(w')}{\sqrt{1-w'^2}}.
\end{equation*}
Since the Galilean mass carried by a field configuration is
$\mathcal M=m'Q$, the mass of the one-soliton solution is
\begin{equation}
    \boxed{
    M_s
    =
    m'Q_s
    =
    \frac{2m'\operatorname{arcsin}(w')}{\sqrt{1-w'^2}}.
    }
    \label{eq:reference_soliton_mass}
\end{equation}
Integrating the momentum density therefore gives
\begin{equation}
    \boxed{
    P_s
    =
    \int_{-\infty}^{+\infty}
    p[\phi_s]\,dx
    =
    m'Q_sv_0
    =
    M_sv_0.
    }
    \label{eq:GCSGE_soliton_momentum}
\end{equation}
Equation~\textcolor{red}{\eqref{eq:GCSGE_soliton_momentum}} is an exact relation for the
uniformly moving free soliton. It shows that the parameter $v_0$ introduced
by the Galilean boost is exactly the velocity of the centre of the localized
field configuration.

We now introduce an external potential. In this case the uniformly moving
solution in Eq.~\textcolor{red}{\eqref{eq:GCSGE_single_soliton}} is no longer an exact solution:
the potential can accelerate the centre and, in general, can also deform the
profile and generate radiation. We restrict the following treatment to a
potential that varies slowly over the soliton width and is sufficiently weak
that changes of the internal profile can be neglected.

Under this rigid-profile approximation, we use the free one-soliton family
as a time-dependent collective-coordinate ansatz:
\begin{equation*}
    \phi_s(x,t)
    \simeq
    w'\operatorname{sech}[w'(x-X_s(t))]
    \,\mathrm{e}^{\mathrm{i}m'\dot X_s(t)(x-X_s(t))}
    \mathrm{e}^{\mathrm{i}\Gamma_s(t)}.
\end{equation*}
Here $X_s(t)$ is not prescribed in advance as a classical trajectory. It is
an unknown collective coordinate that must be determined from the field
equation. In a numerical evolution, it can equivalently be extracted by
fitting the localized profile, or by evaluating an appropriate centre of its
localized density. The purely time-dependent phase $\Gamma_s(t)$ is fixed by
the remaining phase equation but does not enter the spatial momentum.

Pointwise substitution of this rigid ansatz into the complete field equation
does not produce an exact solution for a generic potential, because the
potential also excites deformation and radiation modes. The evolution
equation for $X_s(t)$ is instead obtained by projecting the exact field
dynamics onto the translational collective mode. The integrated microscopic
momentum balance provides precisely this projection.

The spatial phase gradient of the collective-coordinate ansatz is
\begin{equation*}
    \partial_x\Theta_s
    =
    m'\dot X_s(t).
\end{equation*}
Since the rigid profile carries the constant charge $Q_s$, its collective
momentum is
\begin{equation}
    \begin{aligned}
        P_s(t)
        &\simeq
        \int_{-\infty}^{+\infty}
        q_s
        \left(
            x-X_s(t)
        \right)
        m'\dot X_s(t)\,dx
        \\
        &=
        m'Q_s\dot X_s(t)
        \\
        &=
        M_s\dot X_s(t),
    \end{aligned}
\label{eq:rigid_soliton_momentum}
\end{equation}
where
\begin{equation*}
    q_s
    \left(
        x-X_s
    \right)
    =
    \frac{
        w^{\prime 2}
        \operatorname{sech}^{2}
        \left[
            w'
            \left(
                x-X_s
            \right)
        \right]
    }{
        1-
        w^{\prime 2}
        \operatorname{sech}^{2}
        \left[
            w'
            \left(
                x-X_s
            \right)
        \right]
    }.
\end{equation*}
Unlike Eq.~\textcolor{red}{\eqref{eq:GCSGE_soliton_momentum}}, this relation is not being
claimed as an exact identity for an arbitrarily deformed field. It is the
momentum--velocity relation obtained after restricting the dynamics to the
rigid translational one-soliton manifold.

We now apply the exact local momentum balance,
Eq.~\textcolor{red}{\eqref{eq:GCSGE_momentum_balance}}, to the field and then evaluate it in
the rigid-profile approximation. Integration over the real line gives
\begin{equation*}
    \begin{aligned}
        \frac{dP_s}{dt}
        &\simeq
        \int_{-\infty}^{+\infty}
        \partial_t p[\phi_s]\,dx
        \\
        &=
        \int_{-\infty}^{+\infty}
        \partial_xJ_p[\phi_s]\,dx
        -
        \frac{1}{2m'}
        \int_{-\infty}^{+\infty}
        q_s
        \left(
            x-X_s
        \right)
        \partial_xV(x)\,dx
        \\
        &=
        \left[
            J_p[\phi_s]
        \right]_{-\infty}^{+\infty}
        -
        \frac{1}{2m'}
        \int_{-\infty}^{+\infty}
        q_s
        \left(
            x-X_s
        \right)
        \partial_xV(x)\,dx.
    \end{aligned}
\end{equation*}
Because the localized configuration has the same asymptotic momentum flux at
the two spatial boundaries,
\begin{equation*}
    \left[
        J_p[\phi_s]
    \right]_{-\infty}^{+\infty}
    =
    0.
\end{equation*}
The projected microscopic equation is therefore
\begin{equation*}
    \frac{dP_s}{dt}
    \simeq
    -
    \frac{1}{2m'}
    \int_{-\infty}^{+\infty}
    q_s
    \left(
        x-X_s
    \right)
    \partial_xV(x)\,dx.
\end{equation*}

The potential energy experienced by the rigid soliton is defined by
\begin{equation*}
    V_s(X_s)
    =
    \frac{1}{2m'}
    \int_{-\infty}^{+\infty}
    q_s
    \left(
        x-X_s
    \right)
    V(x)\,dx.
\end{equation*}
Its derivative with respect to the unknown collective coordinate is
\begin{equation*}
    \begin{aligned}
        \frac{\partial V_s(X_s)}{\partial X_s}
        &=
        \frac{1}{2m'}
        \int_{-\infty}^{+\infty}
        V(x)
        \frac{\partial}{\partial X_s}
        q_s
        \left(
            x-X_s
        \right)
        \,dx
        \\
        &=
        -
        \frac{1}{2m'}
        \int_{-\infty}^{+\infty}
        V(x)
        \frac{\partial}{\partial x}
        q_s
        \left(
            x-X_s
        \right)
        \,dx.
    \end{aligned}
\end{equation*}
Integration by parts gives
\begin{equation*}
    \begin{aligned}
        \frac{\partial V_s(X_s)}{\partial X_s}
        &=
        -
        \frac{1}{2m'}
        \left[
            V(x)
            q_s
            \left(
                x-X_s
            \right)
        \right]_{-\infty}^{+\infty}
        \\
        &\quad+
        \frac{1}{2m'}
        \int_{-\infty}^{+\infty}
        q_s
        \left(
            x-X_s
        \right)
        \partial_xV(x)\,dx.
    \end{aligned}
\end{equation*}
The boundary term vanishes because $q_s(x-X_s)$ is localized. Hence
\begin{equation*}
    \frac{\partial V_s(X_s)}{\partial X_s}
    =
    \frac{1}{2m'}
    \int_{-\infty}^{+\infty}
    q_s
    \left(
        x-X_s
    \right)
    \partial_xV(x)\,dx.
\end{equation*}
Comparison with the projected microscopic momentum balance gives
\begin{equation*}
    \frac{dP_s}{dt}
    \simeq
    -
    \frac{\partial V_s(X_s)}{\partial X_s}.
\end{equation*}

Finally, within the same rigid-profile approximation, Eq.~\textcolor{red}{\eqref{eq:rigid_soliton_momentum}} holds and $M_s=m' Q_s$ remains constant because the external potential preserves the
global phase symmetry and therefore conserves $Q_s$. We consequently obtain
the collective-coordinate equation
\begin{equation}
    \boxed{
    M_s\ddot X_s
    =
    -
    \frac{\partial V_s(X_s)}{\partial X_s}.
    }
    \label{eq:GCSGE_Newton_equation}
\end{equation}
Thus, $X_s(t)$ has not been assumed to follow a Newtonian trajectory.
Instead, it was introduced as an undetermined centre of the instantaneous
one-soliton profile, and its Newtonian path emerges from the microscopic
field equation after projection onto the translational collective mode.

The validity of Eq.~\textcolor{red}{\eqref{eq:GCSGE_Newton_equation}} is limited by the
rigid-profile approximation. If the potential varies significantly across
the soliton, or if it excites internal modes and radiation, the field cannot
be represented by the one-soliton family with $X_s(t)$ as its only dynamical
collective coordinate. Additional collective variables or the complete field
evolution must then be retained.

\section{Solitons in a Fluctuating Background}

We now examine the statistical results obtained by Gomila
\textcolor{red}{\cite{Gomila_solitons}}, with the aim of clarifying their mathematical basis
and the collective dynamics underlying them. This provides a concrete
setting in which to investigate the origin of the momentum fluctuations
introduced in the HR construction discussed above.

The possibility of relating quantum dynamics to an underlying stochastic
motion has a long history, including Nelson's formulation in terms of
diffusive particle trajectories \textcolor{red}{\cite{Nelson_1966}}. A related perspective
is developed by Tsekov, Heifetz, and Cohen \textcolor{red}{\cite{Tsekov_2017}}, who determine
statistical properties of a stochastic force by comparing ensemble dynamics
with the Madelung equations, while leaving its microscopic mechanism
unspecified. Their analysis provides useful motivation for investigating
how such fluctuations may originate from an underlying field theory.

In the solitonic model, the localized excitation and the fluctuating
background belong to the same nonlinear field. Randomness enters through
the preparation of the initial background, after which the complete field
evolves deterministically. We formulate the collective variables using
background-subtracted conserved densities, recover the initial position
and momentum fluctuations in Gomila's regime of approximation, and derive
the subsequent centre dynamics from the microscopic conservation laws.
The connection with the SE will be developed in a later section.

\subsection{Background-Subtracted Collective Variables}

In the original construction, the position fluctuations are evaluated locally
around a prescribed soliton centre $X_s(t)$, which follows the classical
trajectory in the absence of the background. This provides a suitable
description as long as all realizations remain close to the same reference
trajectory. To define the actual collective position of each realization
without introducing a finite integration interval, we now formulate the
construction in terms of background-subtracted conserved densities.

For each realization $r$, let $\phi_r(x,t)$ denote the complete field
containing the soliton and the chaotic background. We introduce a reference
field $\eta_r^{(0)}(x,t)$ obtained by evolving the same initial realization of
the background, with the same external potential, but without inserting the
soliton. At the preparation time $t_0$, the two initial conditions are
therefore

\begin{equation*}
    \phi_r(x,t_0)
    =
    \phi_s(x-X_s)+\zeta_r(x),
    \qquad
    \eta_r^{(0)}(x,t_0)
    =
    \zeta_r(x).
\end{equation*}

The additive form above specifies only the initial preparation. At later
times, $\phi_r(x,t)$ is the full solution of the GCSGE evolving from this
initial condition; no soliton-plus-background ansatz is imposed on its
subsequent evolution. The reference field $\eta_r^{(0)}(x,t)$ evolves
separately from the background-only initial condition. All quantities
defined below at generic times are evaluated on these dynamically evolved
fields.

The charge and momentum densities of the field, defined in Eqs.~\textcolor{red}{\eqref{eq:GCSGE_charge_density}} and~\textcolor{red}{\eqref{eq:GCSGE_momentum_density}}, are

\begin{equation*}
    q[\phi]
    =
    \frac{|\phi|^2}{1-|\phi|^2},
    \qquad
    p[\phi]
    =
    \frac{i}{
        2\left(1-|\phi|^2\right)
    }
    \left(
        \phi\,\partial_x\phi^*
        -
        \phi^*\partial_x\phi
    \right).
\end{equation*}

We define the charge and momentum densities in excess of the corresponding
background realization as

\begin{equation}
    \boxed{
    \Delta q_r
    =
    q[\phi_r]
    -
    q[\eta_r^{(0)}],
    \qquad
    \Delta p_r
    =
    p[\phi_r]
    -
    p[\eta_r^{(0)}].
    }
    \label{eq:background_subtracted_densities}
\end{equation}

These differences do not remove the influence of the background on the
soliton. They remove only the charge and momentum that would have been
present in the same realization even in the absence of the soliton.
Interference terms, nonlinear corrections, deformations, and radiation
generated by the interaction between the soliton and the background remain
included in $\Delta q_r$ and $\Delta p_r$.

Assuming that these excess densities vanish sufficiently rapidly away from
the localized excitation, we define

\begin{equation}
    \boxed{
    \begin{aligned}
        Q_r^{\mathrm{ex}}
        &=
        \int_{\mathbb R}dx\,\Delta q_r(x,t),
        \\
        P_r^{\mathrm{ex}}
        &=
        \int_{\mathbb R}dx\,\Delta p_r(x,t),
        \\
        X_r(t)
        &=
        \frac{
            \displaystyle
            \int_{\mathbb R}dx\,x\,\Delta q_r(x,t)
        }{
            \displaystyle
            Q_r^{\mathrm{ex}}
        }.
    \end{aligned}
    }
    \label{eq:background_subtracted_collective_variables}
\end{equation}

The quantity $X_r(t)$ is the centre of the complete localized excitation in
realization $r$, including the dressing induced by the background. Unlike the
local definition centred on $X_s(t)$, it does not require the introduction of
an arbitrary finite interval and remains meaningful when different
realizations cease to be concentrated around a single classical trajectory.

We first determine the joint initial statistics of $X_r(t_0)$ and
$P_r^{\mathrm{ex}}(t_0)$. These will supply the initial condition for the
ensemble dynamics derived in the next subsection, after expressing the
momentum fluctuations in terms of the effective translational momentum
$\Pi_r$ introduced there. Throughout the following statistical calculation,
all quantities are evaluated at $t_0$, even when the time argument is
omitted, and $X_s$ denotes $X_s(t_0)$.

At this preparation time, the charge density in excess of the background is

\begin{equation*}
    \Delta q_r(x,t_0)
    =
    q\left[
        \phi_s(x-X_s)+\zeta_r(x)
    \right]
    -
    q[\zeta_r(x)].
\end{equation*}

This subtraction does not reduce $\Delta q_r$ to the unperturbed soliton
density. To see this explicitly, let

\begin{equation*}
    u=|\phi_s|^2,
    \qquad
    v=|\zeta_r|^2,
    \qquad
    c=
    \phi_s^*\zeta_r
    +
    \phi_s\zeta_r^*.
\end{equation*}

A direct calculation gives

\begin{equation*}
    \Delta q_r
    =
    \frac{
        u+c
    }{
        (1-v)(1-u-v-c)
    }.
\end{equation*}

The pure background contribution cancels from the numerator, but the
interference term $c$ and the nonlinear dependence on $v$ remain. Thus,
$\Delta q_r$ describes the charge density of a background-dressed soliton,
rather than that of the bare profile.

We now verify that this definition reproduces the position fluctuations
obtained in the original calculation. Introducing the coordinate

\begin{equation*}
    y=x-X_s,
\end{equation*}

the exact displacement of the excess centre is

\begin{equation*}
    X_r-X_s
    =
    \frac{
        \displaystyle
        \int_{\mathbb R}dy\,
        y\,\Delta q_r(X_s+y,t_0)
    }{
        \displaystyle
        Q_r^{\mathrm{ex}}
    }.
\end{equation*}

For a weak background, $\zeta_r=O(\epsilon)$, the excess density can be
expanded as

\begin{equation*}
    \Delta q_r
    =
    q_s
    +
    \delta q_r^{(1)}
    +
    O(\epsilon^2),
\end{equation*}

where

\begin{equation*}
    q_s(y)
    =
    \frac{
        |\phi_s(y)|^2
    }{
        1-|\phi_s(y)|^2
    }
\end{equation*}

and

\begin{equation*}
    \delta q_r^{(1)}(y)
    =
    \frac{
        \phi_s^*(y)\zeta_r(X_s+y)
        +
        \phi_s(y)\zeta_r^*(X_s+y)
    }{
        \left[
            1-|\phi_s(y)|^2
        \right]^2
    }.
\end{equation*}

Terms beyond first order in $\epsilon$ have been neglected in this expansion. This truncation remains consistent after spatial integration because the background subtraction removes all contributions involving the background alone: every remaining term contains at least one factor of the localized soliton profile and therefore produces no volume-enhanced contribution. Hence, provided that the denominators remain bounded away from zero, spatial integration does not alter the perturbative ordering in $\epsilon$.

Since $q_s(y)$ is even,

\begin{equation*}
    \int_{\mathbb R}dy\,yq_s(y)=0.
\end{equation*}

Moreover,

\begin{equation*}
    Q_r^{\mathrm{ex}}
    =
    Q_s+\delta Q_r^{(1)}+O(\epsilon^2),
    \qquad
    Q_s
    =
    \int_{\mathbb R}dy\,q_s(y).
\end{equation*}

The numerator of $X_r-X_s$ is already of order $O(\epsilon)$. Replacing
$Q_r^{\mathrm{ex}}$ by $Q_s$ therefore changes the displacement only at
second order, and hence

\begin{equation*}
    \delta X_r
    \equiv
    X_r-X_s
    =
    \frac{1}{Q_s}
    \int_{\mathbb R}dy\,
    y\,\delta q_r^{(1)}(y)
    +
    O(\epsilon^2).
\end{equation*}

For the homogeneous complex Gaussian background,

\begin{equation*}
    \left\langle\zeta_r(x)\right\rangle=0,
    \qquad
    \left\langle
        \zeta_r(x)\zeta_r(x')
    \right\rangle=0,
    \qquad
    \left\langle
        \zeta_r(x)\zeta_r^*(x')
    \right\rangle
    =
    \epsilon^2\delta(x-x'),
\end{equation*}

we obtain

\begin{equation*}
    \left\langle
        \delta q_r^{(1)}(y)
        \delta q_r^{(1)}(y')
    \right\rangle
    =
    2\epsilon^2
    \frac{
        |\phi_s(y)|^2
    }{
        \left[
            1-|\phi_s(y)|^2
        \right]^4
    }
    \delta(y-y').
\end{equation*}

The reflection symmetry of the soliton profile and of the background
statistics gives

\begin{equation*}
    \left\langle X_r(t_0)\right\rangle=X_s.
\end{equation*}

The position variance is

\begin{equation*}
    \sigma_X^2
    =
    \frac{
        2\epsilon^2
    }{
        Q_s^2
    }
    \int_{\mathbb R}dy\,
    y^2
    \frac{
        |\phi_s(y)|^2
    }{
        \left[
            1-|\phi_s(y)|^2
        \right]^4
    }
    +
    O(\epsilon^4).
\end{equation*}

For

\begin{equation*}
    |\phi_s(y)|^2
    =
    w'^2\operatorname{sech}^2(w'y),
\end{equation*}

the unperturbed charge is

\begin{equation*}
    Q_s
    =
    \frac{
        2\arcsin w'
    }{
        \sqrt{1-w'^2}
    }
    =
    2w'+O(w'^3).
\end{equation*}

Using

\begin{equation*}
    \int_{-\infty}^{+\infty}
    dz\,z^2\operatorname{sech}^2z
    =
    \frac{\pi^2}{6},
\end{equation*}

the small-soliton approximation $w'\ll1$ gives

\begin{equation}
    \boxed{
    \left\langle X_r(t_0)\right\rangle
    =
    X_s,
    \qquad
    \sigma_X^2
    =
    \frac{
        \pi^2\epsilon^2
    }{
        12w'^3
    }
    +
    O(\epsilon^2w'^{-1},\epsilon^4).
    }
    \label{eq:background_subtracted_position_statistics}
\end{equation}

We next calculate the initial excess momentum in the inertial frame in
which the reference soliton is instantaneously at rest at $t_0$. Thus its
velocity is $v=0$ and its unperturbed momentum is $P_s=0$, while the
individual realizations retain their background-induced momentum
fluctuations. In this frame,

\begin{equation*}
    P_r^{\mathrm{ex}}(t_0)
    =
    \int_{\mathbb R}dx\,
    \left\{
        p[\phi_s+\zeta_r]
        -
        p[\zeta_r]
    \right\}.
\end{equation*}

Since $p[\zeta_r]$ is at least quadratic in the background amplitude, its
subtraction does not modify the linear momentum fluctuation. Writing the
phase fluctuation around the soliton as $\theta_r$, the leading contribution
in this rest frame is therefore the same as in the original calculation:

The first-order momentum fluctuation in the soliton rest frame is
\begin{equation*}
    \delta P_r^{(1)}
    =
    -2\int_{\mathbb R}dx\,
    \frac{\partial_x\phi_s(x-X_s)}
    {[1-\phi_s^2(x-X_s)]^2}
    \zeta_{r,\mathrm I}(x).
\end{equation*}
Using the zero mean and spatial covariance of the imaginary
noise component gives
\begin{equation*}
    \left\langle(\delta P_r^{(1)})^2\right\rangle
    =
    2\epsilon^2\int_{\mathbb R}dx\,
    \frac{[\partial_x\phi_s(x-X_s)]^2}
    {[1-\phi_s^2(x-X_s)]^4}.
\end{equation*}
In the small-amplitude regime $w'\ll1$, this reduces to
\begin{equation*}
    \begin{aligned}
        \left\langle(\delta P_r^{(1)})^2\right\rangle
        &\simeq
        2\epsilon^2 w'^4
        \int_{\mathbb R}dx\,
        \operatorname{sech}^2[w'(x-X_s)]
        \operatorname{tanh}^2[w'(x-X_s)]
        \\
        &=\frac{4}{3}w'^3\epsilon^2.
    \end{aligned}
\end{equation*}
Since the phase fluctuation has zero mean, the momentum statistics at the preparation time are
\begin{equation}
    \boxed{
    \left\langle
        P_r^{\mathrm{ex}}(t_0)
    \right\rangle
    =
    P_s=0,
    \qquad
    \sigma_P^2
    =
    \frac{4}{3}w'^3\epsilon^2
    +
    O(\epsilon^4).
    }
    \label{eq:background_subtracted_momentum_statistics}
\end{equation}

Consequently,

\begin{equation}
    \boxed{
    \sigma_X\sigma_P
    =
    \frac{\pi\epsilon^2}{3}
    =
    \frac{\hbar_{\mathrm{eff}}}{2},
    \qquad
    \hbar_{\mathrm{eff}}
    =
    \frac{2\pi\epsilon^2}{3}.
    }
    \label{eq:background_subtracted_exact_uncertainty}
\end{equation}

The detailed perturbative calculation of the initial position and momentum
distributions is given in Section~\ref{app:initial-statistics} of
Appendix~\ref{app:field}.

The background-subtracted variables therefore reproduce, at the preparation
time and within the same weak-background and small-soliton approximations,
the mean values and leading fluctuations obtained in the original local
construction. The difference is conceptual and dynamical: the original
definition evaluates fluctuations inside a finite neighbourhood of the
prescribed classical position $X_s$, whereas
Eq.~\textcolor{red}{\eqref{eq:background_subtracted_collective_variables}} defines the actual
collective position and momentum of every realization relative to the same
background evolved without the soliton.

\subsection{Dynamics of the Background-Subtracted Center}

We now return to generic times and use the full dynamically evolved fields.
The relation between the collective momentum and the motion of the new center
follows from the local conservation law
associated with the global $U(1)$ symmetry. In the conventions used here, the
charge and momentum densities satisfy Eq.~\textcolor{red}{\eqref{eq:GCSGE_charge_continuity}},

\begin{equation*}
    \partial_tq[\phi]
    +
    \frac{1}{m'}\partial_xp[\phi]
    =
    0.
\end{equation*}

Subtracting the equation for the reference background from that for the complete field and using Eq.~\textcolor{red}{\eqref{eq:background_subtracted_densities}} gives

\begin{equation*}
    \partial_t\Delta q_r
    +
    \frac{1}{m'}\partial_x\Delta p_r
    =
    0.
\end{equation*}

Integration over the real line shows that the excess charge is conserved.
Indeed, from its definition,

\begin{equation*}
    Q_r^{\mathrm{ex}}
    =
    \int_{\mathbb R}dx\,
    \Delta q_r,
\end{equation*}

we obtain

\begin{equation*}
    \begin{aligned}
        \frac{dQ_r^{\mathrm{ex}}}{dt}
        &=
        \int_{\mathbb R}dx\,
        \partial_t\Delta q_r
        \\
        &=
        -\frac{1}{m'}
        \int_{\mathbb R}dx\,
        \partial_x\Delta p_r
        \\
        &=
        -\frac{1}{m'}
        \left[
            \Delta p_r
        \right]_{-\infty}^{+\infty}.
    \end{aligned}
\end{equation*}

The complete momentum density and the background momentum density do not need
to vanish separately at infinity. Since the excitation produced by the
soliton is localized, it is only their difference that must satisfy

\begin{equation*}
    \Delta p_r(x,t)
    =
    p[\phi_r(x,t)]
    -
    p[\eta_r^{(0)}(x,t)]
    \longrightarrow0
\end{equation*}

as $|x|\rightarrow\infty$. Consequently,

\begin{equation*}
    \frac{dQ_r^{\mathrm{ex}}}{dt}
    =
    0.
\end{equation*}

We now use the definition of the background-subtracted center,

\begin{equation*}
    X_r(t)
    =
    \frac{
        \displaystyle
        \int_{\mathbb R}dx\,
        x\,\Delta q_r(x,t)
    }{
        \displaystyle
        Q_r^{\mathrm{ex}}
    }.
\end{equation*}

Multiplying this relation by $Q_r^{\mathrm{ex}}$ gives

\begin{equation*}
    Q_r^{\mathrm{ex}}X_r
    =
    \int_{\mathbb R}dx\,
    x\,\Delta q_r.
\end{equation*}

Taking its time derivative and using the continuity equation,

\begin{equation*}
    \begin{aligned}
        \frac{d}{dt}
        \left(
            Q_r^{\mathrm{ex}}X_r
        \right)
        &=
        \int_{\mathbb R}dx\,
        x\,\partial_t\Delta q_r
        \\
        &=
        -\frac{1}{m'}
        \int_{\mathbb R}dx\,
        x\,\partial_x\Delta p_r.
    \end{aligned}
\end{equation*}

The last integral is evaluated by parts:

\begin{equation*}
    \int_{\mathbb R}dx\,
    x\,\partial_x\Delta p_r
    =
    \left[
        x\Delta p_r
    \right]_{-\infty}^{+\infty}
    -
    \int_{\mathbb R}dx\,
    \Delta p_r.
\end{equation*}

The existence of the first moment requires the stronger localization
condition

\begin{equation*}
    x\Delta p_r(x,t)
    \longrightarrow0
\end{equation*}

as $|x|\rightarrow\infty$. Therefore,

\begin{equation*}
    \begin{aligned}
        \frac{d}{dt}
        \left(
            Q_r^{\mathrm{ex}}X_r
        \right)
        &=
        \frac{1}{m'}
        \int_{\mathbb R}dx\,
        \Delta p_r
        \\
        &=
        \frac{1}{m'}P_r^{\mathrm{ex}},
    \end{aligned}
\end{equation*}

where the last equality follows from the definition in Eq.~\textcolor{red}{\eqref{eq:background_subtracted_collective_variables}}.

Since $Q_r^{\mathrm{ex}}$ is constant, the product rule gives

\begin{equation*}
    \begin{aligned}
        \frac{d}{dt}
        \left(
            Q_r^{\mathrm{ex}}X_r
        \right)
        &=
        Q_r^{\mathrm{ex}}\dot X_r
        +
        X_r\frac{dQ_r^{\mathrm{ex}}}{dt}
        \\
        &=
        Q_r^{\mathrm{ex}}\dot X_r.
    \end{aligned}
\end{equation*}

Comparing the two expressions for the same derivative, we obtain

\begin{equation*}
    Q_r^{\mathrm{ex}}\dot X_r
    =
    \frac{1}{m'}P_r^{\mathrm{ex}}.
\end{equation*}

It follows that

\begin{equation}
    \boxed{
    P_r^{\mathrm{ex}}
    =
    M_r^{\mathrm{ex}}\dot X_r,
    \qquad
    M_r^{\mathrm{ex}}
    =
    m' Q_r^{\mathrm{ex}}.
    }
    \label{eq:exact_excess_momentum_velocity_relation}
\end{equation}

Equation~\textcolor{red}{\eqref{eq:exact_excess_momentum_velocity_relation}} is exact for the
background-subtracted collective variables. The excess mass
$M_r^{\mathrm{ex}}$ is constant in time within each realization, because the
corresponding excess charge is conserved, but it may fluctuate from one
realization to another. We therefore do not replace it by a common mass at
this stage. We denote by $M_s=m' Q_s$ the mass of the unperturbed soliton, which
coincides with the ensemble-averaged excess mass to the order considered
here.

We now derive the force acting on the center directly from the microscopic
momentum balance. In the presence of an external potential $V(x)$, an
arbitrary field configuration satisfies

\begin{equation*}
    \partial_t p[\phi]
    =
    \partial_xJ_p[\phi]
    -
    \frac{1}{2m'}q[\phi]V_x.
\end{equation*}

Applying this equation separately to $\phi_r$ and $\eta_r^{(0)}$, and defining

\begin{equation*}
    \Delta J_{p,r}
    =
    J_p[\phi_r]
    -
    J_p[\eta_r^{(0)}],
\end{equation*}

we obtain

\begin{equation*}
    \partial_t\Delta p_r
    =
    \partial_x\Delta J_{p,r}
    -
    \frac{1}{2m'}\Delta q_rV_x.
\end{equation*}

Since only the excess current enters, the homogeneous background does not
need to vanish separately at infinity. It is sufficient that

\begin{equation*}
    \Delta J_{p,r}(+\infty,t)
    -
    \Delta J_{p,r}(-\infty,t)
    =
    0.
\end{equation*}

Integration over space then gives

\begin{equation}
    \boxed{
    \frac{dP_r^{\mathrm{ex}}}{dt}
    =
    -
    \frac{1}{2m'}
    \int_{\mathbb R}dx\,
    \Delta q_r(x,t)V_x(x).
    }
    \label{eq:microscopic_excess_momentum_equation}
\end{equation}

Combining Eqs.~\textcolor{red}{\eqref{eq:exact_excess_momentum_velocity_relation}} and
\textcolor{red}{\eqref{eq:microscopic_excess_momentum_equation}}, and using the conservation of
$M_r^{\mathrm{ex}}$, gives the exact equation

\begin{equation}
    \boxed{
    M_r^{\mathrm{ex}}\ddot X_r
    =
    -
    \frac{1}{2m'}
    \int_{\mathbb R}dx\,
    \Delta q_r(x,t)V_x(x).
    }
    \label{eq:exact_background_subtracted_centre_dynamics}
\end{equation}

If the microscopic external potential is at most quadratic, it can be written
as

\begin{equation*}
    V(x)
    =
    V_0
    +
    ax
    +
    \frac{b}{2}x^2,
\end{equation*}

where $V_0$ is dynamically irrelevant. Its gradient is therefore

\begin{equation*}
    V_x(x)
    =
    a+bx.
\end{equation*}

Substitution into
Eq.~\textcolor{red}{\eqref{eq:exact_background_subtracted_centre_dynamics}} gives

\begin{equation*}
    \begin{aligned}
        M_r^{\mathrm{ex}}\ddot X_r
        &=
        -\frac{a}{2m'}
        \int_{\mathbb R}dx\,
        \Delta q_r
        -
        \frac{b}{2m'}
        \int_{\mathbb R}dx\,
        x\Delta q_r
        \\
        &=
        -\frac{a}{2m'}Q_r^{\mathrm{ex}}
        -
        \frac{b}{2m'}Q_r^{\mathrm{ex}}X_r.
    \end{aligned}
\end{equation*}

Using $Q_r^{\mathrm{ex}}=M_r^{\mathrm{ex}}/m'$, we find

\begin{equation*}
    M_r^{\mathrm{ex}}\ddot X_r
    =
    -
    \frac{M_r^{\mathrm{ex}}}{2m^{\prime 2}}
    \left(
        a+bX_r
    \right).
\end{equation*}

The realization-dependent mass appears on both sides and cancels exactly:

\begin{equation}
    \boxed{
    \ddot X_r
    =
    -
    \frac{1}{2m^{\prime 2}}
    \left(
        a+bX_r
    \right).
    }
    \label{eq:exact_quadratic_center_dynamics}
\end{equation}

This cancellation has an important physical consequence. Although the total
amount of excess field and therefore $M_r^{\mathrm{ex}}$ may fluctuate among
realizations, every center follows the same deterministic equation of motion.
For a linear potential, the acceleration is identical in every realization.
For a quadratic potential, the acceleration depends on $X_r$ and therefore
varies among realizations, but no additional stochastic force appears in the
center dynamics. The realization-dependent normalization of the localized
excitation does not enter explicitly in either case. The center can therefore
be described by a common effective mass $M_s$, equal to the mass of the unperturbed soliton in Eq.~\textcolor{red}{\eqref{eq:reference_soliton_mass}} or, equivalently to the order considered here, to the
ensemble-averaged excess mass.

We consequently define the effective kinematic momentum

\begin{equation}
    \boxed{
    \Pi_r
    \equiv
    M_s\dot X_r
    =
    \frac{M_s}{M_r^{\mathrm{ex}}}
    P_r^{\mathrm{ex}}.
    }
    \label{eq:effective_kinematic_momentum}
\end{equation}

The microscopic momentum $P_r^{\mathrm{ex}}$ measures the complete momentum
carried by the excess field, including fluctuations of its normalization.
The variable $\Pi_r$, instead, measures the motion of the center with the
common effective mass $M_s$. Under a change of inertial frame by a constant
velocity $v$, it transforms as

\begin{equation*}
    \Pi_r
    \longrightarrow
    \Pi_r+M_sv,
\end{equation*}

so the shift is the same in every realization and its variance is independent
of the inertial frame.

Multiplying
Eq.~\textcolor{red}{\eqref{eq:exact_quadratic_center_dynamics}}
by $M_s$ gives
\begin{equation*}
    \dot\Pi_r
    =
    -\frac{M_s}{2m'^2}(a+bX_r).
\end{equation*}
This force is generated by an effective potential of the form
\begin{equation}
    U(X)=U_0+U_1X+U_2X^2,
    \qquad
    U_1=\frac{M_sa}{2m'^2},
    \qquad
    U_2=\frac{M_sb}{4m'^2}.
    \label{eq:effective_potential_matching}
\end{equation}
Here $U_0$ is an arbitrary additive constant. The coefficients
$U_1$ and $U_2$ relate the potential energy of the center to
the microscopic potential $V(x)=V_0+ax+bx^2/2$.
For these quadratic potentials, $U$ differs from the rigid-soliton
potential energy $V_s$ introduced previously only by an
$X$-independent constant.

The exact center equations therefore become
\begin{equation}
    \boxed{
    \dot X_r=\frac{\Pi_r}{M_s},
    \qquad
    \dot\Pi_r=-U'(X_r)=-U_1-2U_2X_r.
    }
    \label{eq:effective_center_hamilton_equations}
\end{equation}
They are generated by the Hamiltonian
\begin{equation}
    \boxed{
    H_{\mathrm{eff}}(X,\Pi)
    =
    \frac{\Pi^2}{2M_s}+U(X).
    }
    \label{eq:effective_center_hamiltonian}
\end{equation}

This single expression includes free evolution
($U_1=U_2=0$), motion under a uniform force
($U_2=0$, with $\mathcal F_0=-U_1$), and harmonic motion
about the origin
($U_1=0$, with $U_2=M_s\omega^2/2>0$).
The constant $U_0$ does not affect the center dynamics.

The statistics calculated in Eqs.~\textcolor{red}{\eqref{eq:background_subtracted_position_statistics}} and~\textcolor{red}{\eqref{eq:background_subtracted_momentum_statistics}} now provide the initial condition for this ensemble dynamics. In the rest frame of the
reference soliton at $t_0$, $P_s=0$, so that
$P_r^{\mathrm{ex}}(t_0)=\delta P_r^{(1)}+O(\epsilon^2)$, while
$M_r^{\mathrm{ex}}=M_s+O(\epsilon)$. Consequently,

\begin{equation*}
    \Pi_r(t_0)
    =
    \frac{M_s}{M_r^{\mathrm{ex}}}P_r^{\mathrm{ex}}(t_0)
    =
    P_r^{\mathrm{ex}}(t_0)+O(\epsilon^2)
    =
    \delta P_r^{(1)}+O(\epsilon^2).
\end{equation*}

The initial excess-momentum calculation, together with Eq.~\textcolor{red}{\eqref{eq:effective_kinematic_momentum}}, therefore gives the leading
fluctuation of the translational momentum $\Pi_r$. Its Gaussian statistics,
together with those of $X_r(t_0)$, determine the initial ensemble
distribution. The frame transformation given above shifts the mean of
$\Pi_r$ by the common amount $M_sv$, so these fluctuations also specify the
initial distribution in a frame with nonzero mean momentum. We may therefore
write

\begin{equation*}
    X_r(t_0)
    =
    X_0+\delta X_r,
    \qquad
    \Pi_r(t_0)
    =
    \Pi_0+\delta\Pi_r.
\end{equation*}

with

\begin{equation*}
    \left\langle\delta X_r\right\rangle
    =
    \left\langle\delta\Pi_r\right\rangle
    =
    0
\end{equation*}

and

\begin{equation}
    \sigma_X^2
    =
    \frac{\pi^2\epsilon^2}{12w^{\prime3}},
    \qquad
    \sigma_\Pi^2
    =
    \frac{4}{3}
    w^{\prime3}\epsilon^2.
\label{eq:initial_collective_variances}
\end{equation}

The two fluctuations depend on independent quadratures of the Gaussian
background and are therefore independent at this order. Their product is

\begin{equation*}
    \sigma_X\sigma_\Pi
    =
    \frac{\pi\epsilon^2}{3}.
\end{equation*}

Introducing

\begin{equation}
    \hbar_{\mathrm{eff}}
    =
    \frac{2\pi\epsilon^2}{3},
\label{eq:effective_hbar_definition}
\end{equation}

we obtain

\begin{equation}
    \boxed{
    \sigma_X\sigma_\Pi
    =
    \frac{\hbar_{\mathrm{eff}}}{2}.
    }
    \label{eq:effective_center_exact_uncertainty}
\end{equation}

The initial phase-space density is consequently

\begin{equation}
    \boxed{
        f_0(X,\Pi)
        =
        \frac{1}{2\pi\sigma_X\sigma_\Pi}
        \operatorname{exp}\left[
            -\frac{(X-X_0)^2}{2\sigma_X^2}
            -\frac{(\Pi-\Pi_0)^2}{2\sigma_\Pi^2}
        \right].
    }
    \label{eq:effective_center_initial_distribution}
\end{equation}

We now determine how this density evolves. The phase-space distribution at a
generic time is defined by

\begin{equation*}
    f(X,\Pi,t)
    =
    \left\langle
        \delta\left[X-X_r(t)\right]
        \delta\left[\Pi-\Pi_r(t)\right]
    \right\rangle.
\end{equation*}

Conservation of probability along the Hamiltonian flow gives

\begin{equation}
    \boxed{
    \frac{\partial f}{\partial t}
    +
    \frac{\Pi}{M_s}
    \frac{\partial f}{\partial X}
    -
    U'(X)
    \frac{\partial f}{\partial\Pi}
    =
    0.
    }
    \label{eq:effective_center_liouville_equation}
\end{equation}

The meaning of this equation becomes transparent by following its
characteristics. Let

\begin{equation*}
    X=X_r(t),
    \qquad
    \Pi=\Pi_r(t)
\end{equation*}

be a solution of
Eq.~\textcolor{red}{\eqref{eq:effective_center_hamilton_equations}}. Along this trajectory,

\begin{equation*}
    \begin{aligned}
        \frac{df}{dt}
        &=
        \frac{\partial f}{\partial t}
        +
        \dot X_r
        \frac{\partial f}{\partial X}
        +
        \dot\Pi_r
        \frac{\partial f}{\partial\Pi}
        \\
        &=
        \frac{\partial f}{\partial t}
        +
        \frac{\Pi}{M_s}
        \frac{\partial f}{\partial X}
        -
        U'(X)
        \frac{\partial f}{\partial\Pi}
        \\
        &=
        0.
    \end{aligned}
\end{equation*}

Thus, $f$ remains constant along every phase-space trajectory. If
$(X_r(t_0),\Pi_r(t_0))$ is the initial point of the trajectory that reaches
$(X,\Pi)$ at time $t$, then

\begin{equation}
    \boxed{
    f(X,\Pi,t)
    =
    f_0
    \left(
        X_r(t_0),
        \Pi_r(t_0)
    \right).
    }
    \label{eq:effective_center_characteristic_solution}
\end{equation}

The initial coordinates $X_r(t_0)$ and $\Pi_r(t_0)$ of the trajectory
reaching $(X,\Pi)$ at time $t$ are obtained by solving the equations of
motion and then inverting them.

For free evolution, setting $\tau=t-t_0$, the trajectories are

\begin{equation*}
    X
    =
    X_r(t_0)+\frac{\Pi_r(t_0)}{M_s}\tau,
    \qquad
    \Pi
    =
    \Pi_r(t_0).
\end{equation*}

Therefore,

\begin{equation*}
    X_r(t_0)
    =
    X-\frac{\Pi}{M_s}\tau,
    \qquad
    \Pi_r(t_0)
    =
    \Pi,
\end{equation*}

and

\begin{equation*}
    f(X,\Pi,t)
    =
    f_0
    \left(
        X-\frac{\Pi}{M_s}\tau,
        \Pi
    \right).
\end{equation*}

For a uniform force $\mathcal F_0$, corresponding to

\begin{equation*}
    U(X)
    =
    -\mathcal F_0X,
\end{equation*}

the trajectories are

\begin{equation*}
    \Pi
    =
    \Pi_r(t_0)+\mathcal F_0\tau
\end{equation*}

and

\begin{equation*}
    X
    =
    X_r(t_0)
    +
    \frac{\Pi_r(t_0)}{M_s}\tau
    +
    \frac{\mathcal F_0}{2M_s}\tau^2.
\end{equation*}

Their inverse is

\begin{equation*}
    \Pi_r(t_0)
    =
    \Pi-\mathcal F_0\tau
\end{equation*}

and

\begin{equation*}
    X_r(t_0)
    =
    X
    -
    \frac{\Pi}{M_s}\tau
    +
    \frac{\mathcal F_0}{2M_s}\tau^2.
\end{equation*}

Consequently,

\begin{equation*}
    f(X,\Pi,t)
    =
    f_0
    \left(
        X-\frac{\Pi}{M_s}\tau
        +\frac{\mathcal F_0}{2M_s}\tau^2,
        \Pi-\mathcal F_0\tau
    \right).
\end{equation*}

For the harmonic potential,
$U(X)=M_s\omega^2X^2/2$, the trajectories are

\begin{equation*}
    X
    =
    X_r(t_0)\cos{\omega\tau}
    +
    \frac{\Pi_r(t_0)}{M_s\omega}
    \sin{\omega\tau}
\end{equation*}

and

\begin{equation*}
    \Pi
    =
    \Pi_r(t_0)\cos{\omega\tau}
    -
    M_s\omega X_r(t_0)\sin{\omega\tau}.
\end{equation*}

Inverting this rotation gives

\begin{equation*}
    X_r(t_0)
    =
    X\cos{\omega\tau}
    -
    \frac{\Pi}{M_s\omega}
    \sin{\omega\tau}
\end{equation*}

and

\begin{equation*}
    \Pi_r(t_0)
    =
    \Pi\cos{\omega\tau}
    +
    M_s\omega X\sin{\omega\tau}.
\end{equation*}

Therefore,

\begin{equation*}
    \begin{aligned}
        f(X,\Pi,t)
        =
        f_0\bigg(
            &X\cos{\omega\tau}
            -
            \frac{\Pi}{M_s\omega}
            \sin{\omega\tau},
            \\
            &\Pi\cos{\omega\tau}
            +
            M_s\omega X\sin{\omega\tau}
        \bigg).
    \end{aligned}
\end{equation*}

These explicit solutions show directly why the Gaussian form is preserved.
In all three cases, the initial variables $(X_r(t_0),\Pi_r(t_0))$ are affine functions of the current variables $(X,\Pi)$. Substitution into the Gaussian $f_0(X_r(t_0),\Pi_r(t_0))$ therefore produces another Gaussian in $(X,\Pi)$.

\subsection{From Ensemble Dynamics to the Schr\"odinger Equation}

The detailed calculations underlying this subsection are collected in
Appendix~\ref{app:gaussian}.

We now express the phase-space dynamics in terms of the position density and
the local mean momentum. Define

\begin{equation*}
    \rho(X,t)
    =
    \int_{\mathbb R}d\Pi\,
    f(X,\Pi,t)
\end{equation*}

and

\begin{equation*}
    \overline\Pi(X,t)
    =
    \frac{
        \displaystyle
        \int_{\mathbb R}d\Pi\,
        \Pi f(X,\Pi,t)
    }{
        \displaystyle
        \rho(X,t)
    }.
\end{equation*}

The corresponding velocity field is

\begin{equation*}
    u(X,t)
    =
    \frac{\overline\Pi(X,t)}{M_s}.
\end{equation*}

Integration of
Eq.~\textcolor{red}{\eqref{eq:effective_center_liouville_equation}} over momentum gives

\begin{equation}
    \boxed{
    \frac{\partial\rho}{\partial t}
    +
    \frac{\partial}{\partial X}
    \left(
        \rho u
    \right)
    =
    0.
    }
    \label{eq:effective_center_continuity_equation}
\end{equation}

To obtain the equation for the local momentum, we multiply the Liouville
equation by $\Pi$ and integrate over momentum. Assuming that $f$ vanishes
sufficiently rapidly for $|\Pi|\rightarrow\infty$, integration by parts gives

\begin{equation*}
    \frac{\partial}{\partial t}
    \left(
        \rho\overline\Pi
    \right)
    +
    \frac{\partial}{\partial X}
    \left[
        \frac{\rho}{M_s}
        \left(
            \overline\Pi^2
            +
            \sigma_{\Pi|X}^2
        \right)
    \right]
    +
    \rho U'(X)
    =
    0,
\end{equation*}

where

\begin{equation*}
    \sigma_{\Pi|X}^2
    =
    \frac{
        \displaystyle
        \int_{\mathbb R}d\Pi\,
        \left[
            \Pi-\overline\Pi(X,t)
        \right]^2
        f(X,\Pi,t)
    }{
        \displaystyle
        \rho(X,t)
    }
\end{equation*}

is the momentum variance conditioned on the position. Combining this equation
with the continuity equation yields

\begin{equation}
    \boxed{
    M_s
    \left(
        \frac{\partial u}{\partial t}
        +
        u\frac{\partial u}{\partial X}
    \right)
    =
    -
    U'(X)
    -
    \frac{1}{M_s\rho}
    \frac{\partial}{\partial X}
    \left(
        \rho\sigma_{\Pi|X}^2
    \right).
    }
    \label{eq:effective_center_euler_equation}
\end{equation}

The last term is the force generated by the local dispersion of momentum
inside the ensemble. Its form is determined by the Gaussian phase-space
distribution. For a bivariate Gaussian,

\begin{equation}
    \overline\Pi(X,t)
    =
    \left\langle\Pi\right\rangle
    +
    \frac{C_{X\Pi}}{\Sigma_X^2}
    \left(
        X-\left\langle X\right\rangle
    \right)
\label{eq:gaussian_local_mean_momentum}
\end{equation}

and

\begin{equation*}
    \sigma_{\Pi|X}^2
    =
    \Sigma_\Pi^2
    -
    \frac{C_{X\Pi}^2}{\Sigma_X^2}.
\end{equation*}

As shown in Appendix~\ref{app:gaussian-density}, the covariance
determinant is conserved:
\begin{equation*}
    \Sigma_X^2\Sigma_\Pi^2-C_{X\Pi}^2
    =
    \sigma_X^2\sigma_\Pi^2
    =
    \frac{\hbar_{\mathrm{eff}}^2}{4}.
\end{equation*}
The conditional momentum variance therefore becomes
\begin{equation}
    \boxed{
        \sigma_{\Pi|X}^2
        =
        \frac{\hbar_{\mathrm{eff}}^2}{4\Sigma_X^2}.
    }
    \label{eq:effective_center_conditional_variance}
\end{equation}

The position density is

\begin{equation}
    \rho(X,t)
    =
    \frac{1}{
        \sqrt{2\pi\Sigma_X^2(t)}
    }
    e^{-\left[X-\langle X\rangle\right]^2/
    \left[2\Sigma_X^2(t)\right]}.
\label{eq:gaussian_position_marginal}
\end{equation}

For this density, define

\begin{equation*}
    Q[\rho]
    =
    -
    \frac{\hbar_{\mathrm{eff}}^2}{2M_s}
    \frac{
        \partial_X^2\sqrt\rho
    }{
        \sqrt\rho
    }.
\end{equation*}

Direct evaluation gives

\begin{equation*}
    Q[\rho]
    =
    \frac{
        \hbar_{\mathrm{eff}}^2
    }{
        4M_s\Sigma_X^2
    }
    -
    \frac{
        \hbar_{\mathrm{eff}}^2
        \left[
            X-\langle X\rangle
        \right]^2
    }{
        8M_s\Sigma_X^4
    }.
\end{equation*}

On the other hand, using
Eq.~\textcolor{red}{\eqref{eq:effective_center_conditional_variance}},

\begin{equation*}
    \begin{aligned}
        \frac{1}{M_s\rho}
        \frac{\partial}{\partial X}
        \left(
            \rho\sigma_{\Pi|X}^2
        \right)
        &=
        \frac{
            \hbar_{\mathrm{eff}}^2
        }{
            4M_s\Sigma_X^2
        }
        \frac{\partial_X\rho}{\rho}
        \\
        &=
        -
        \frac{
            \hbar_{\mathrm{eff}}^2
            \left[
                X-\langle X\rangle
            \right]
        }{
            4M_s\Sigma_X^4
        }
        \\
        &=
        \partial_XQ[\rho].
    \end{aligned}
\end{equation*}

The momentum-dispersion term derived from the Liouville equation therefore
coincides exactly with the gradient of the QP. Equation
\textcolor{red}{\eqref{eq:effective_center_euler_equation}} becomes

\begin{equation*}
    M_s
    \left(
        \frac{\partial u}{\partial t}
        +
        u\frac{\partial u}{\partial X}
    \right)
    =
    -
    \frac{\partial}{\partial X}
    \left(
        U+Q[\rho]
    \right).
\end{equation*}

In one spatial dimension, the velocity field can be represented by a scalar
momentum potential $S(X,t)$:

\begin{equation*}
    \partial_XS
    =
    M_su.
\end{equation*}

It follows that

\begin{equation*}
    M_s
    \left(
        \frac{\partial u}{\partial t}
        +
        u\frac{\partial u}{\partial X}
    \right)
    =
    \frac{\partial}{\partial X}
    \left[
        \frac{\partial S}{\partial t}
        +
        \frac{
            (\partial_XS)^2
        }{
            2M_s
        }
    \right].
\end{equation*}

Therefore,

\begin{equation*}
    \frac{\partial}{\partial X}
    \left[
        \frac{\partial S}{\partial t}
        +
        \frac{
            (\partial_XS)^2
        }{
            2M_s
        }
        +
        U
        +
        Q[\rho]
    \right]
    =
    0.
\end{equation*}

The expression inside brackets may consequently depend only on time:

\begin{equation*}
    \frac{\partial S}{\partial t}
    +
    \frac{
        (\partial_XS)^2
    }{
        2M_s
    }
    +
    U
    +
    Q[\rho]
    =
    c(t).
\end{equation*}

The phase-space distribution determines $\partial_XS$, but leaves a purely
time-dependent integration term in $S$ undetermined. Fixing this term does
not change either $\rho$ or the local mean momentum. Define

\begin{equation}
    \widetilde S(X,t)
    =
    S(X,t)
    -
    \int_{t_0}^{t}c(t')\,dt'
\label{eq:collective_temporal_phase_choice}
\end{equation}

so that $\widetilde S(X,t_0)=S(X,t_0)$ and
$\partial_X\widetilde S=\partial_XS$. This choice fixes the temporal
integration term while preserving the initial phase. We then obtain

\begin{equation}
    \boxed{
    \frac{\partial\widetilde S}{\partial t}
    +
    \frac{
        (\partial_X\widetilde S)^2
    }{
        2M_s
    }
    +
    U
    -
    \frac{\hbar_{\mathrm{eff}}^2}{2M_s}
    \frac{
        \partial_X^2\sqrt\rho
    }{
        \sqrt\rho
    }
    =
    0.
    }
    \label{eq:effective_center_quantum_HJ_equation}
\end{equation}

This equation has not been imposed. It follows from the Liouville evolution
of the soliton ensemble and from the conditional momentum variance generated
by the initial Gaussian fluctuations.

Finally, define

\begin{equation*}
    \psi(X,t)
    =
    \sqrt{\rho(X,t)}\,
    e^{i\widetilde S(X,t)/\hbar_{\mathrm{eff}}}.
\end{equation*}

Equation~\textcolor{red}{\eqref{eq:effective_center_continuity_equation}} and
Eq.~\textcolor{red}{\eqref{eq:effective_center_quantum_HJ_equation}} are respectively the
imaginary and real parts of

\begin{equation}
    \boxed{
    i\hbar_{\mathrm{eff}}
    \frac{\partial\psi}{\partial t}
    =
    \left[
        -
        \frac{\hbar_{\mathrm{eff}}^2}{2M_s}
        \frac{\partial^2}{\partial X^2}
        +
        U(X)
    \right]
    \psi.
    }
    \label{eq:effective_center_Schrodinger_equation}
\end{equation}

Thus, within the weak-background and small-soliton approximations used to
obtain the initial Gaussian minimum-uncertainty ensemble, the subsequent
center dynamics reproduces the SE for free, linear and harmonic potentials.

The same wave function can also be checked by direct substitution into the
SE. This check uses the full phase $\widetilde S$ already constructed in the
HJ treatment; its temporal integration term must therefore be retained from
the outset.

For the Gaussian ensemble, let $\xi=X-\langle X\rangle$. The local mean momentum in Eq.~\textcolor{red}{\eqref{eq:gaussian_local_mean_momentum}} is

\begin{equation*}
    \partial_XS(X,t)
    =\overline\Pi(X,t)
    =\langle\Pi\rangle
    +\frac{C_{X\Pi}}{\Sigma_X^2}\xi.
\end{equation*}

Integrating with respect to $X$ at fixed $t$ gives

\begin{equation*}
    S(X,t)=S_{\mathrm{sp}}(X,t)+C(t),
    \qquad
    S_{\mathrm{sp}}(X,t)
    =\langle\Pi\rangle\xi
    +\frac{C_{X\Pi}}{2\Sigma_X^2}\xi^2.
\end{equation*}

Here $C(t)$ is the integration function left undetermined by the momentum
field. Its presence follows directly from spatial integration. To connect
this expression with the preceding HJ construction, define

\begin{equation*}
    K(t)
    =\frac{\langle\Pi\rangle^2}{2M_s}
    -U(\langle X\rangle)
    -\frac{\hbar_{\mathrm{eff}}^2}{4M_s\Sigma_X^2}.
\end{equation*}

The HJ expression for
$S_{\mathrm{sp}}$ is $-K(t)$, as shown explicitly in
~\ref{app:hj-residual}. Retaining the integration function therefore gives

\begin{equation*}
    \partial_tS+\frac{(\partial_XS)^2}{2M_s}+U+Q[\rho]
    =\dot C(t)-K(t)=c(t).
\end{equation*}

The function $c(t)$ consequently depends on the chosen $C(t)$. Substituting both into the definition of $\widetilde S$ in Eq.~\textcolor{red}{\eqref{eq:collective_temporal_phase_choice}} yields

\begin{equation*}
    \begin{aligned}
        \widetilde S(X,t)
        &=S(X,t)-\int_{t_0}^{t}c(t')\,dt'
        \\
        &=S_{\mathrm{sp}}(X,t)+C(t)
        -\int_{t_0}^{t}\bigl[\dot C(t')-K(t')\bigr]\,dt'
        \\
        &=S_{\mathrm{sp}}(X,t)+C(t_0)
        +\int_{t_0}^{t}K(t')\,dt'.
    \end{aligned}
\end{equation*}

Thus, the arbitrary time dependence of $C(t)$ cancels in the corrected
phase, leaving only its initial value. No particular evolution has to be
imposed on the original integration function $C(t)$. The temporal term in
$\widetilde S$ is already fixed by the preceding HJ construction.

For compactness, denote this existing temporal term by

\begin{equation*}
    \hbar_{\mathrm{eff}}\gamma(t)
    :=C(t_0)+\int_{t_0}^{t}K(t')\,dt',
    \qquad
    \hbar_{\mathrm{eff}}\dot\gamma(t)=K(t).
\end{equation*}

The wave function $\psi=\sqrt\rho\,
\operatorname{exp}[i\widetilde S/\hbar_{\mathrm{eff}}]$ then takes the explicit
form

\begin{equation*}
    \psi(X,t)
    =\frac{1}{(2\pi\Sigma_X^2)^{1/4}}
    \operatorname{exp}\Biggl[
        -\frac{\xi^2}{4\Sigma_X^2}
        +\frac{i}{\hbar_{\mathrm{eff}}}
        \biggl(
            \langle\Pi\rangle\xi
            +\frac{C_{X\Pi}}{2\Sigma_X^2}\xi^2
        \biggr)
        +i\gamma(t)
    \Biggr].
\end{equation*}

The symbol $\gamma$ introduces no additional phase freedom at this stage:
it abbreviates the temporal part of $\widetilde S$. Gomila's notation
$\psi=\sqrt\rho\,\operatorname{exp}(iS/\hbar)$ keeps the full phase in a
single function~\textcolor{red}{\cite{Gomila_solitons}}; the separate temporal term here
appears only because we have written that function explicitly.

For an at-most-quadratic potential, the Liouville equation gives the moment equations in Eq.~\eqref{app:closed-moments}

\begin{equation}
    \begin{aligned}
        \frac{d\langle X\rangle}{dt}
        &=\frac{\langle\Pi\rangle}{M_s},
        &
        \frac{d\langle\Pi\rangle}{dt}
        &=-U'(\langle X\rangle),
        \\
        \frac{d\Sigma_X^2}{dt}
        &=\frac{2C_{X\Pi}}{M_s},
        &
        \frac{d\Sigma_\Pi^2}{dt}
        &=-2U''C_{X\Pi},
        \\
        \frac{dC_{X\Pi}}{dt}
        &=\frac{\Sigma_\Pi^2}{M_s}-U''\Sigma_X^2.
    \end{aligned}
\label{eq:collective_gaussian_moment_evolution}
\end{equation}

These relations follow from the first and second moments of $f$.
Differentiating the wave function with its already determined phase and
using these moment equations gives

\begin{equation*}
    \begin{aligned}
        \frac{1}{\psi}
        \biggl[
            i\hbar_{\mathrm{eff}}\partial_t
            +\frac{\hbar_{\mathrm{eff}}^2}{2M_s}\partial_X^2
            -U(X)
        \biggr]\psi
        ={}&
        \frac{
            \hbar_{\mathrm{eff}}^2/4
            -\bigl(\Sigma_X^2\Sigma_\Pi^2-C_{X\Pi}^2\bigr)
        }{2M_s\Sigma_X^4}\,\xi^2
        \\
        &+K(t)-\hbar_{\mathrm{eff}}\dot\gamma(t),
    \end{aligned}
\end{equation*}

The coefficient of $\xi^2$ vanishes by the conserved covariance determinant.
The spatially uniform term also vanishes, since
$\hbar_{\mathrm{eff}}\dot\gamma=K$ already follows from the definition of
$\widetilde S$. The substitution therefore verifies the SE without imposing
any further condition on the phase. Choosing $C(t_0)=\Pi_0X_0$ gives the initial wave function
in the phase convention used above:
\begin{equation*}
    \psi(X,t_0)
    =
    \frac{1}{(2\pi\sigma_X^2)^{1/4}}
    \operatorname{exp}\biggl[
        -\frac{(X-X_0)^2}{4\sigma_X^2}
    \biggr]
    \operatorname{exp}\biggl[
        \frac{i\Pi_0X}{\hbar_{\mathrm{eff}}}
    \biggr].
\end{equation*}

The reconstruction from $f$ fixes the density and the spatial phase
differences. With the temporal phase term already fixed by the HJ construction, uniqueness
of the SE evolution for this initial state establishes equality with the
solution of the SE with effective potential $U$. A different constant initial
phase multiplies both solutions by the same constant phase factor.

For free and uniform-force evolution, the position variance is

\begin{equation*}
    \Sigma_X^2(t)
    =
    \sigma_X^2
    +
    \frac{\sigma_\Pi^2}{M_s^2}
    (t-t_0)^2.
\end{equation*}

Using the exact uncertainty relation, this becomes

\begin{equation*}
    \Sigma_X^2(t)
    =
    \sigma_X^2
    \left[
        1+
        \left(
            \frac{
                \hbar_{\mathrm{eff}}(t-t_0)
            }{
                2M_s\sigma_X^2
            }
        \right)^2
    \right],
\end{equation*}

which is the quantum spreading of a free Gaussian. The uniform force changes
the mean trajectory but not the covariance.

For the harmonic potential,

\begin{equation*}
    \Sigma_X^2(t)
    =
    \sigma_X^2
    \left[
        \operatorname{cos}
        \left(
            \omega(t-t_0)
        \right)
    \right]^2
    +
    \frac{\sigma_\Pi^2}{M_s^2\omega^2}
    \left[
        \operatorname{sin}
        \left(
            \omega(t-t_0)
        \right)
    \right]^2.
\end{equation*}

If

\begin{equation*}
    \sigma_X^2
    =
    \frac{
        \hbar_{\mathrm{eff}}
    }{
        2M_s\omega
    },
\end{equation*}

the width remains constant and the ensemble corresponds to a coherent state.
For any other minimum-uncertainty width, it corresponds to a squeezed
Gaussian state.

For a nonquadratic microscopic potential, the exact force depends on higher
moments of the dressed excess density and need not close on the center and
its momentum alone. Even if a closed classical Liouville description is
adopted, its flow need not preserve Gaussianity, and the conditional
momentum variance need not produce the gradient of $Q[\rho]$. The
correspondence established above therefore applies to the Gaussian sector
preserved by potentials at most quadratic. An extension to a barrier or a
generic potential requires further information from the field dynamics;
the closure used here cannot simply be assumed to remain valid.

\subsection{Exact uncertainty relation in the Gaussian sector}

We can now establish a precise connection between the collective fluctuations
derived above and the exact uncertainty relation of Hall and Reginatto. Let

\begin{equation*}
    f(X,\Pi,t)
\end{equation*}

denote the joint distribution of the effective center and momentum. We define
the local mean momentum by

\begin{equation*}
    \overline{\Pi}(X,t)
    =
    \mathbb E
    \left[
        \Pi_r
        \mid
        X_r=X
    \right]
    =
    \frac{\partial S(X,t)}{\partial X},
\end{equation*}

and introduce the residual momentum fluctuation

\begin{equation}
    \boxed{
        N_r
        =
        \Pi_r
        -
        \frac{\partial S(X_r,t)}{\partial X}.
    }
    \label{eq:HR_residual_collective_momentum}
\end{equation}

By construction,

\begin{equation*}
    \mathbb E
    \left[
        N_r
        \mid
        X_r=X
    \right]
    =
    0.
\end{equation*}

For the free, linear, and quadratic cases considered above, an initially
Gaussian distribution remains Gaussian under the effective Hamiltonian
evolution. Its covariance matrix can be written as

\begin{equation*}
    \Sigma(t)
    =
    \begin{pmatrix}
        \Sigma_X^2(t)
        &
        C_{X\Pi}(t)
        \\
        C_{X\Pi}(t)
        &
        \Sigma_\Pi^2(t)
    \end{pmatrix},
\end{equation*}

where

\begin{equation*}
    C_{X\Pi}(t)
    =
    \operatorname{Cov}
    \left(
        X_r,\Pi_r
    \right).
\end{equation*}

For a bivariate Gaussian distribution, the conditional momentum variance is

\begin{equation*}
    \begin{aligned}
        \mathbb E
        \left[
            N_r^2
            \mid
            X_r=X
        \right]
        &=
        \operatorname{Var}
        \left(
            \Pi_r
            \mid
            X_r=X
        \right)
        \\
        &=
        \Sigma_\Pi^2(t)
        -
        \frac{
            C_{X\Pi}^2(t)
        }{
            \Sigma_X^2(t)
        }
        \\
        &=
        \frac{
            \det\Sigma(t)
        }{
            \Sigma_X^2(t)
        }.
    \end{aligned}
\end{equation*}

This quantity is independent of \(X\), although it may depend on time. For a
Hamiltonian containing a potential of at most second order in \(X\), the
phase-space evolution is affine and symplectic. Consequently,

\begin{equation*}
    \Sigma(t)=A(t)\Sigma(t_0)A^T(t),
    \qquad
    \operatorname{det}A(t)=1.
\end{equation*}

and therefore

\begin{equation*}
    \det\Sigma(t)
    =
    \det\Sigma(t_0).
\end{equation*}

At the preparation time, the position and momentum fluctuations calculated
from the background are uncorrelated. Hence,

\begin{equation*}
    \det\Sigma(t_0)
    =
    \sigma_X^2\sigma_\Pi^2.
\end{equation*}

Using the variances obtained from the solitonic field in Eq.~\textcolor{red}{\eqref{eq:initial_collective_variances}}, we find

\begin{equation*}
    \sigma_X^2\sigma_\Pi^2
    =
    \frac{\pi^2\epsilon^4}{9}.
\end{equation*}

Using $\hbar_{\mathrm{eff}}$ as defined in Eq.~\textcolor{red}{\eqref{eq:effective_hbar_definition}}, this relation becomes

\begin{equation}
    \boxed{
        \det\Sigma(t)
        =
        \frac{
            \hbar_{\mathrm{eff}}^2
        }{
            4
        }.
    }
    \label{eq:preserved_collective_covariance_determinant}
\end{equation}

The conditional fluctuation variance is consequently

\begin{equation}
    \boxed{
        \mathbb E
        \left[
            N_r^2
            \mid
            X_r=X
        \right]
        =
        \frac{
            \hbar_{\mathrm{eff}}^2
        }{
            4\Sigma_X^2(t)
        }.
    }
    \label{eq:conditional_collective_momentum_variance}
\end{equation}

The conditional second moment is precisely the local fluctuation average
that enters the HR formulation. In the present one-dimensional
setting, we identify

\begin{equation*}
    \overline{N^2}(X,t)
    \equiv
    \mathbb E\bigl[N_r^2\mid X_r=X\bigr].
\end{equation*}

For the Gaussian position marginal, the conditional variance already has
the value of the full Fisher-information functional, multiplied by
$\hbar_{\mathrm{eff}}^2/4$. To make this connection explicit, recall the marginal in Eq.~\textcolor{red}{\eqref{eq:gaussian_position_marginal}},

\begin{equation*}
    \rho(X,t)
    =
    \frac{1}{\sqrt{2\pi\Sigma_X^2(t)}}
    \operatorname{exp}\Biggl[
        -\frac{[X-\langle X\rangle]^2}{2\Sigma_X^2(t)}
    \Biggr].
\end{equation*}

The Fisher information is

\begin{equation*}
    \begin{aligned}
        I[\rho](t)
        &=
        \int_{\mathbb R}dY\,
        \frac{[\partial_Y\rho(Y,t)]^2}{\rho(Y,t)}
        \\
        &=
        \frac{1}{\Sigma_X^4(t)}
        \int_{\mathbb R}dY\,
        \rho(Y,t)[Y-\langle X\rangle]^2
        \\
        &=
        \frac{1}{\Sigma_X^2(t)}.
    \end{aligned}
\end{equation*}

Consequently, Eq.~\textcolor{red}{\eqref{eq:conditional_collective_momentum_variance}} can
already be written as

\begin{equation}
    \boxed{
        \mathbb E\bigl[N_r^2\mid X_r=X\bigr]
        =
        \frac{\hbar_{\mathrm{eff}}^2}{4}I[\rho](t).
    }
    \label{eq:conditional_variance_Fisher_functional}
\end{equation}

The right-hand side is independent of the conditioning position $X$, but it
depends on the density through its Fisher information and can therefore
change in time. Spatial uniformity does not imply temporal constancy.
It is the integrated
functional $I[\rho]$, rather than the local quantity
$[\partial_X\ln\rho(X,t)]^2$.

We now insert this result into the ensemble action used in the HJ
formulation. At a fixed position $X$, the momentum decomposes as
$\Pi_r=\partial_XS(X,t)+N_r$. Expanding the square before taking the
conditional average gives

\begin{equation*}
    \frac{\Pi_r^2}{2M_s}
    =
    \frac{(\partial_XS)^2}{2M_s}
    +\frac{\partial_XS}{M_s}N_r
    +\frac{N_r^2}{2M_s}.
\end{equation*}

Since $\partial_XS(X,t)$ is fixed under this conditional average and
$\mathbb E[N_r\mid X_r=X]=0$, the mixed term vanishes:

\begin{equation*}
    \begin{aligned}
        \mathbb E\biggl[
            \frac{\Pi_r^2}{2M_s}\,\bigg|\,X_r=X
        \biggr]
        &=
        \frac{(\partial_XS)^2}{2M_s}
        +\frac{\partial_XS}{M_s}
        \mathbb E\bigl[N_r\mid X_r=X\bigr]
        \\
        &\quad+
        \frac{1}{2M_s}
        \mathbb E\bigl[N_r^2\mid X_r=X\bigr]
        \\
        &=
        \frac{(\partial_XS)^2}{2M_s}
        +\frac{1}{2M_s}
        \mathbb E\bigl[N_r^2\mid X_r=X\bigr].
    \end{aligned}
\end{equation*}

Averaging over position therefore gives the ensemble Hamiltonian

\begin{equation*}
    \begin{aligned}
        \mathcal H[\rho,S]
        &=
        \int_{\mathbb R}dX\int_{\mathbb R}d\Pi\,
        f(X,\Pi,t)
        \biggl[\frac{\Pi^2}{2M_s}+U(X)\biggr]
        \\
        &=
        \int_{\mathbb R}dX\,\rho(X,t)
        \biggl[\frac{(\partial_XS)^2}{2M_s}+U(X)\biggr]
        +\Delta\mathcal H[\rho],
    \end{aligned}
\end{equation*}

where

\begin{equation*}
    \Delta\mathcal H[\rho]
    =
    \frac{1}{2M_s}
    \int_{\mathbb R}dX\,\rho(X,t)
    \mathbb E\bigl[N_r^2\mid X_r=X\bigr].
\end{equation*}

In the HJ variational formulation, with the overall sign convention adopted
previously, the ensemble action is

\begin{equation*}
    \mathcal A[\rho,S]
    =
    \int dt\,
    \biggl[
        \int_{\mathbb R}dX\,\rho\,\partial_tS
        +\mathcal H[\rho,S]
    \biggr].
\end{equation*}

Thus, the added kinetic energy contributes its time integral to the action:

\begin{equation*}
    \Delta\mathcal A[\rho]
    =
    \int dt\,\Delta\mathcal H[\rho]
    =
    \frac{1}{2M_s}
    \int dt\int_{\mathbb R}dX\,
    \rho(X,t)\,
    \mathbb E\bigl[N_r^2\mid X_r=X\bigr].
\end{equation*}

Substituting Eq.~\textcolor{red}{\eqref{eq:conditional_variance_Fisher_functional}} and using
normalization at each time gives

\begin{equation*}
    \begin{aligned}
        \Delta\mathcal A[\rho]
        &=
        \frac{\hbar_{\mathrm{eff}}^2}{8M_s}
        \int dt\,I[\rho](t)
        \int_{\mathbb R}dX\,\rho(X,t)
        \\
        &=
        \frac{\hbar_{\mathrm{eff}}^2}{8M_s}
        \int dt\,I[\rho](t)
        \\
        &=
        \frac{\hbar_{\mathrm{eff}}^2}{8M_s}
        \int dt\int_{\mathbb R}dX\,
        \frac{[\partial_X\rho(X,t)]^2}{\rho(X,t)}.
    \end{aligned}
\end{equation*}

Thus, the spatial average in the action introduces no additional factor:
its weight is the normalized probability density. The conditional variance
already supplies the Fisher functional, and this functional remains after
taking the ensemble average.

Finally, writing the Fisher integrand with an explicit probability weight,

\begin{equation*}
    \frac{(\partial_X\rho)^2}{\rho}
    =
    \rho\biggl(\frac{\partial_X\rho}{\rho}\biggr)^2
    =
    \rho(\partial_X\ln\rho)^2,
\end{equation*}

puts all contributions into the same ensemble-average form:

\begin{equation}
    \boxed{
        \mathcal A[\rho,S]
        =
        \int dt\int_{\mathbb R}dX\,\rho
        \biggl[
            \partial_tS
            +\frac{(\partial_XS)^2}{2M_s}
            +U
            +\frac{\hbar_{\mathrm{eff}}^2}{8M_s}
            (\partial_X\ln\rho)^2
        \biggr].
    }
    \label{eq:Gaussian_collective_Hall_Reginatto_action}
\end{equation}

Thus, within the Gaussian sector considered here, the collective momentum fluctuations reproduce the Fisher-information contribution required by the HR action in Eq.~\textcolor{red}{\eqref{eq:HR_quantum_ensemble_action}}. This establishes the correspondence with that
formulation for our ensemble. The connection between soliton-ensemble
dynamics and the SE also has numerical support in Gomila's
work~\textcolor{red}{\cite{Gomila_solitons}}, where the position density and local mean
velocity obtained from soliton trajectories were compared with the
predictions of the time-dependent SE in a potential-barrier scattering
problem.

\newpage
\null
\thispagestyle{empty}
\newpage

\chapter{Conclusions}

The starting point of this thesis was the possibility of introducing nonlinearity into QM and understanding whether it could provide a deeper description of quantum phenomena. Weinberg's formulation offered a systematic framework for exploring this possibility through nonlinear extensions of the standard theory~\cite{Weinberg1989}. Its analysis, however, also exposed the difficulties that arise when the dynamics is modified while the usual statistical interpretation is retained. In particular, the objections discussed in connection with Peres' argument concern the evolution of statistical mixtures and its compatibility with thermodynamics~\cite{Peres1989}. These difficulties indicate that introducing nonlinear corrections requires reconsidering the relation between the dynamics and the statistical meaning assigned to quantum states.

This observation motivated the change of perspective developed throughout the thesis. Drawing on the PWT and its description of quantum phenomena in terms of trajectories and ensembles, we investigated the possibility that nonlinearity belongs to a more fundamental dynamical level, from which QM emerges as an effective statistical theory. In this setting, the nonlinear evolution governs the underlying physical field, while the wave function describes the collective behavior of its localized excitations. The role of nonlinearity is therefore more fundamental than that of a perturbative correction to the SE: it supports the structures whose ensemble dynamics is to be described by the effective theory.

Solitons provide a concrete setting in which to explore this possibility. Their localization and persistence allow collective positions and momenta to be associated with extended field configurations. When these structures evolve in a fluctuating background, their collective variables acquire statistical distributions even though each realization of the complete field evolves deterministically. This offers a possible connection between nonlinear microscopic dynamics and the probabilistic description of quantum phenomena. It also motivates the question of whether some quantum correlations could originate from interactions between localized structures and their common background.

The main contribution of this thesis has been the construction of a mathematical framework for investigating this connection and clarifying the statistical results reported by Gomila~\cite{Gomila_solitons}. By defining collective quantities through background subtraction, we distinguished the localized excitation from the fluctuating field surrounding it and connected its motion to the microscopic conservation laws. Within the approximations considered, we recovered the initial position and momentum fluctuations and used them as the initial conditions for the subsequent ensemble dynamics. This made it possible to identify explicitly the assumptions under which a quantum description of the collective motion is obtained.

For the Gaussian sector preserved by effective potentials at most quadratic, the resulting ensemble evolution admits a wave-function description satisfying the SE. The collective momentum fluctuations also reproduce the Fisher-information contribution required by the HR action~\cite{Hall_Reginatto}. The effective Planck constant is then related to the amplitude of the background fluctuations. These results provide an analytical basis for part of the correspondence observed in Gomila's work and identify a concrete origin for the additional ensemble contribution associated with quantum dynamics. Their scope is determined by the approximations and closure conditions used in the collective description.

A useful comparison is provided by the work of Folli and Conti on the random walk of solitary and shock waves in nonlocal disordered media~\cite{FolliConti_RandomWalk_2013}. Their study examines how disorder and nonlocality influence nonlinear wave propagation, illustrating the importance of the fluctuating environment for the motion of localized structures. A complementary perspective is developed by Conti in the study of the second-quantized nonlocal Schr\"odinger--Newton soliton~\cite{Conti_RandomWalk_2023}, where quantum diffusion of the soliton parameters and the emergence of non-Gaussian statistics are investigated through a stochastic representation of the quantum field dynamics.

The origin and dynamical role of the fluctuations differ across these settings. In the model considered here, randomness is introduced through the initial field configuration, after which the soliton and its background evolve together according to deterministic nonlinear equations. In Conti's second-quantized treatment, stochastic terms represent fluctuations of an already quantized field. The Gaussianity obtained in our collective description also relies on the initial statistics and their preservation under the restricted effective dynamics. A quantitative comparison would therefore need to examine the same observables and account for the different field equations, noise correlations, and couplings between soliton parameters. This comparison offers a useful direction for understanding when Gaussian collective statistics persist and when more general distributions develop.

The first extension of the present work concerns more general potentials. For potentials beyond quadratic order, the force acting on the localized excitation can depend on its deformation and on higher moments of the field distribution. The collective dynamics may then fail to close on position and momentum alone, and Gaussianity need not be preserved. Establishing a correspondence with the SE in this regime requires returning to the field dynamics and determining how these additional degrees of freedom affect the ensemble evolution. Potential barriers provide a particularly relevant test, allowing the analytical framework developed here to be compared more directly with the scattering problem investigated by Gomila. Such an analysis should identify both the conditions under which the quantum description remains valid and the mechanisms responsible for possible departures from it.

A second direction is the extension to ensembles containing several solitons. Their nonlinear interactions and their coupling to a common fluctuating background may generate correlations between collective coordinates. The central question is whether these correlations can support an effective description in terms of nonseparable quantum states. Addressing it requires deriving the joint ensemble dynamics and examining whether the reconstructed density and phase satisfy the appropriate many-particle SE. This would make it possible to distinguish ordinary statistical correlations from quantum entanglement and to determine whether the mechanism identified for a single soliton extends consistently to composite systems.

The results obtained support the investigation of nonlinear field dynamics as a possible foundation for an effective quantum description. Within the sector studied, the statistical motion of solitons provides the structures required by the SE and the HR formulation. Extending this correspondence to more general potentials and interacting solitons will be essential for assessing how far this mechanism can account for quantum phenomena.

\addtocontents{toc}{\protect\setcounter{tocdepth}{0}}
\appendix

\chapter{Technical Details of Nonlinear Quantum Mechanics}

This appendix presents supplementary calculations related to
Chapter~\textcolor{red}{\ref{chap:1}}. We first derive the commutator of infinitesimal transformations and explain how closure of the generators determines the Lie algebra of these transformations.
We then calculate the eigenvalues of a statistical mixture of
two pure states.

\section{Commutators of Infinitesimal Transformations}
\label{app:commutators}

We now consider a family of infinitesimal symmetry transformations generated by observables $\{\lambda_i\}$. By definition, each generator $\lambda_i$ induces
\begin{equation*}
\delta_{\lambda_i}\psi_k=-i\frac{\partial\lambda_i}{\partial\psi_k^*},
\qquad
\delta_{\lambda_i}\psi_k^*=i\frac{\partial\lambda_i}{\partial\psi_k}.
\end{equation*}
Let us compute the commutator of two such transformations acting on $\psi_k$:
\begin{align*}
\delta_{\lambda_i}(\delta_{\lambda_j}\psi_k)
&=\delta_{\lambda_i}\!\left(-i\frac{\partial\lambda_j}{\partial\psi_k^*}\right)\\
&=-i\sum_m\left(
\frac{\partial^2\lambda_j}{\partial\psi_k^*\partial\psi_m}\,\delta_{\lambda_i}\psi_m
+
\frac{\partial^2\lambda_j}{\partial\psi_k^*\partial\psi_m^*}\,\delta_{\lambda_i}\psi_m^*
\right).
\end{align*}
Substituting the transformation laws,
\begin{align*}
\delta_{\lambda_i}(\delta_{\lambda_j}\psi_k)
&=-i\sum_m\left[
\frac{\partial^2\lambda_j}{\partial\psi_k^*\partial\psi_m}
\left(-i\frac{\partial\lambda_i}{\partial\psi_m^*}\right)
+
\frac{\partial^2\lambda_j}{\partial\psi_k^*\partial\psi_m^*}
\left(i\frac{\partial\lambda_i}{\partial\psi_m}\right)
\right].
\end{align*}
Exchanging $i\leftrightarrow j$ and taking the difference, we obtain
\begin{equation*}
(\delta_{\lambda_i}\delta_{\lambda_j}-\delta_{\lambda_j}\delta_{\lambda_i})\psi_k
=\frac{\partial}{\partial\psi_k^*}
\left(
\sum_m
\frac{\partial\lambda_i}{\partial\psi_m}\frac{\partial\lambda_j}{\partial\psi_m^*}
-
\frac{\partial\lambda_j}{\partial\psi_m}\frac{\partial\lambda_i}{\partial\psi_m^*}
\right).
\end{equation*}
Using the definition of the product (cf. $a*b$) and of the bracket, we can write
\begin{equation*}
[\lambda_i,\lambda_j]
:=\sum_m\left(
\frac{\partial\lambda_i}{\partial\psi_m}\frac{\partial\lambda_j}{\partial\psi_m^*}
-
\frac{\partial\lambda_j}{\partial\psi_m}\frac{\partial\lambda_i}{\partial\psi_m^*}
\right),
\end{equation*}
so that
\begin{equation*}
    (\delta_{\lambda_i}\delta_{\lambda_j}
    -\delta_{\lambda_j}\delta_{\lambda_i})\psi_k
    =
    \frac{\partial[\lambda_i,\lambda_j]}{\partial\psi_k^*}.
\end{equation*}
We define the usual commutator of the variation operators,
acting on any differentiable function $b$ of the state, by
\begin{equation*}
    [\delta_{\lambda_i},\delta_{\lambda_j}]_{\mathrm C}\,b
    :=
    \delta_{\lambda_i}(\delta_{\lambda_j}b)
    -
    \delta_{\lambda_j}(\delta_{\lambda_i}b).
\end{equation*}
Recalling the definition of the variation generated by an
observable, we have
\begin{equation*}
    \delta_{\,i[\lambda_i,\lambda_j]}\psi_k
    =
    -i\frac{\partial\bigl(i[\lambda_i,\lambda_j]\bigr)}
    {\partial\psi_k^*}
    =
    \frac{\partial[\lambda_i,\lambda_j]}{\partial\psi_k^*}.
\end{equation*}
The corresponding identity holds on the conjugate coordinates
and therefore, by the chain rule, on any differentiable function
of the state. Hence,
\begin{equation}
    \boxed{
        [\delta_{\lambda_i},\delta_{\lambda_j}]_{\mathrm C}
        =
        \delta_{\,i[\lambda_i,\lambda_j]}.
    }
    \label{eq:generator_transformation_bracket}
\end{equation}
This identity connects the Weinberg bracket of the generators
with the commutator of the transformations they induce: the
commutator is itself a transformation generated by
$i[\lambda_i,\lambda_j]$. For real generators, this function
is also real.

However, the resulting generator need not belong to the linear
span of the chosen family $\{\lambda_i\}$. This closure property
must be verified or required in both the linear and nonlinear
theories. To ensure that the corresponding transformations form
a Lie algebra, we require $i[\lambda_i,\lambda_j]$ to be a real
linear combination of the chosen generators. With our convention,
we write this condition as
\begin{equation*}
    [\lambda_i,\lambda_j]
    =
    i\sum_h C_{ijh}\lambda_h,
\end{equation*}
where $C_{ijh}$ are real constants.
This condition selects a subspace of observables which is stable under the bracket and therefore forms a Lie algebra.
Using this closure relation, we obtain
\begin{align*}
(\delta_{\lambda_i}\delta_{\lambda_j}-\delta_{\lambda_j}\delta_{\lambda_i})\psi_k
&=\frac{\partial}{\partial\psi_k^*}\left(i\sum_h C_{ijh}\lambda_h\right)\\
&=-\sum_h C_{ijh}\,\delta_{\lambda_h}\psi_k,
\end{align*}
Therefore,
\begin{equation*}
    [\delta_{\lambda_i},\delta_{\lambda_j}]_{\mathrm{C}}
    =
    -\sum_h C_{ijh}\,\delta_{\lambda_h}.
\end{equation*}
Hence, the infinitesimal transformations close under commutation
and form a Lie algebra acting on the space of states. Their
commutation relations are determined by the Weinberg brackets
of the generators.

\section{Eigenvalues of a Mixture of Two Pure States}
\label{app:entropy}

Consider the density operator
\begin{equation*}
    \rho
    =
    \lambda|\phi\rangle\langle\phi|
    +(1-\lambda)|\psi\rangle\langle\psi|,
    \qquad 0<\lambda<1,
\end{equation*}
where $|\phi\rangle$ and $|\psi\rangle$ are normalized states.
To determine its eigenvalues, we restrict $\rho$ to the subspace
spanned by these two states, since it vanishes on the orthogonal
complement.

For $|\langle\phi|\psi\rangle|<1$, an orthonormal basis of this
subspace is formed by $|\phi\rangle$ and
\begin{equation*}
    |\phi_\perp\rangle
    =
    \frac{
        |\psi\rangle-\langle\phi|\psi\rangle|\phi\rangle
    }{
        \sqrt{1-|\langle\phi|\psi\rangle|^2}
    }.
\end{equation*}
Consequently,
\begin{equation*}
    |\psi\rangle
    =
    \langle\phi|\psi\rangle|\phi\rangle
    +
    \sqrt{1-|\langle\phi|\psi\rangle|^2}\,
    |\phi_\perp\rangle.
\end{equation*}
Substituting this decomposition into the density operator gives
its matrix in the ordered basis
$\{|\phi\rangle,|\phi_\perp\rangle\}$:
\begin{equation*}
    \rho
    =
    \begin{pmatrix}
        \lambda+(1-\lambda)|\langle\phi|\psi\rangle|^2
        &
        (1-\lambda)\langle\phi|\psi\rangle
        \sqrt{1-|\langle\phi|\psi\rangle|^2}
        \\
        (1-\lambda)\langle\psi|\phi\rangle
        \sqrt{1-|\langle\phi|\psi\rangle|^2}
        &
        (1-\lambda)(1-|\langle\phi|\psi\rangle|^2)
    \end{pmatrix}.
\end{equation*}

The trace of this matrix is one. Its determinant is
\begin{equation*}
    \begin{aligned}
        \det\rho
        ={}&
        \bigl[
            \lambda+(1-\lambda)|\langle\phi|\psi\rangle|^2
        \bigr]
        (1-\lambda)
        \bigl[
            1-|\langle\phi|\psi\rangle|^2
        \bigr]
        \\
        &-
        (1-\lambda)^2|\langle\phi|\psi\rangle|^2
        \bigl[
            1-|\langle\phi|\psi\rangle|^2
        \bigr]
        \\
        ={}&
        \lambda(1-\lambda)
        \bigl[
            1-|\langle\phi|\psi\rangle|^2
        \bigr].
    \end{aligned}
\end{equation*}
The terms proportional to
$(1-\lambda)^2|\langle\phi|\psi\rangle|^2$ cancel.
Therefore, the characteristic equation is
\begin{equation*}
    w^2-w
    +\lambda(1-\lambda)
    \bigl[
        1-|\langle\phi|\psi\rangle|^2
    \bigr]
    =0.
\end{equation*}
Solving this quadratic equation gives
\begin{equation}
    \boxed{
        w_\pm
        =
        \frac{
            1\pm
            \sqrt{
                1-4\lambda(1-\lambda)
                \bigl[
                    1-|\langle\phi|\psi\rangle|^2
                \bigr]
            }
        }{2}.
    }
    \label{app:mixture-spectrum}
\end{equation}

For orthogonal states, these eigenvalues reduce to $\lambda$ and
$1-\lambda$. If $|\langle\phi|\psi\rangle|=1$, the two states
represent the same ray, so $\rho$ is a rank-one projector with
eigenvalues $1$ and $0$. This case is also recovered continuously
from Eq.~\eqref{app:mixture-spectrum}.

\chapter{Field Identities and Initial Soliton Statistics}
\label{app:field}

We now derive the field identities used in the third chapter and then
calculate the initial collective distribution. The parameter $m'>0$
is kept generic. No small-$w'$ approximation is used in the field
balances, the soliton profile or the charge integral. That approximation
will enter only when evaluating the leading statistical coefficients.

\section{Integrals of the One-Soliton Profile}
\label{app:soliton}

The one-soliton solution derived in the main text has modulus
\begin{equation}
    |\phi_s(x,t)|
    =
    w'\operatorname{sech}
    \bigl[w'(x-X_s(t))\bigr],
    \qquad 0<w'<1.
    \label{app:soliton-profile}
\end{equation}
All spatial integrals below are evaluated at fixed time, so $X_s(t)$
is constant during integration. Introducing the integration variable
\begin{equation*}
    z=w'(x-X_s(t)),
    \qquad
    dx=\frac{dz}{w'},
\end{equation*}
removes the position of the center from the integrals.

We first evaluate the soliton charge:
\begin{equation*}
    \begin{aligned}
        Q_s
        &=
        \int_{\mathbb R}dx\,
        \frac{|\phi_s(x,t)|^2}{1-|\phi_s(x,t)|^2}
        \\
        &=
        w'\int_{\mathbb R}dz\,
        \frac{\operatorname{sech}^2z}
        {1-w^{\prime2}\operatorname{sech}^2z}.
    \end{aligned}
\end{equation*}
Using
\begin{equation*}
    u=\operatorname{tanh}z,
    \qquad
    du=\operatorname{sech}^2z\,dz,
    \qquad
    \operatorname{sech}^2z=1-u^2,
\end{equation*}
the integration limits become $u=-1$ and $u=1$, and hence
\begin{equation*}
    Q_s
    =
    w'\int_{-1}^{1}
    \frac{du}{1-w^{\prime2}+w^{\prime2}u^2}.
\end{equation*}
An antiderivative of this integrand is
\begin{equation*}
    \frac{1}{\sqrt{1-w^{\prime2}}}
    \operatorname{arctan}
    \biggl(
        \frac{w'u}{\sqrt{1-w^{\prime2}}}
    \biggr).
\end{equation*}
Evaluating it at the two endpoints gives
\begin{equation*}
    Q_s
    =
    \frac{2}{\sqrt{1-w^{\prime2}}}
    \operatorname{arctan}
    \biggl(
        \frac{w'}{\sqrt{1-w^{\prime2}}}
    \biggr).
\end{equation*}
For $0<w'<1$, the angle in this expression is
$\operatorname{arcsin}w'$. Therefore,
\begin{equation}
    \boxed{
        Q_s
        =
        \frac{2\operatorname{arcsin}w'}
        {\sqrt{1-w^{\prime2}}},
        \qquad
        M_s=m'Q_s.
    }
    \label{app:soliton-charge}
\end{equation}
This result is exact throughout the allowed range of $w'$.
For the small-amplitude regime used in the statistical calculation,
its expansion is
\begin{equation*}
    Q_s
    =
    2w'
    +\frac{4}{3}w^{\prime3}
    +\frac{16}{15}w^{\prime5}
    +O(w^{\prime7}).
\end{equation*}

The leading fluctuation variances require three further integrals.
The first follows directly from the derivative of
$\operatorname{tanh}z$:
\begin{equation*}
    \int_{\mathbb R}\operatorname{sech}^2z\,dz
    =
    \bigl[\operatorname{tanh}z\bigr]_{-\infty}^{+\infty}
    =2.
\end{equation*}
For the second, the same substitution
$u=\operatorname{tanh}z$ gives
\begin{equation*}
    \begin{aligned}
        \int_{\mathbb R}
        \operatorname{sech}^2z
        \operatorname{tanh}^2z\,dz
        &=
        \int_{-1}^{1}u^2\,du
        \\
        &=
        \biggl[\frac{u^3}{3}\biggr]_{-1}^{1}
        =\frac{2}{3}.
    \end{aligned}
\end{equation*}

For the third integral, evenness gives
\begin{equation*}
    \int_{\mathbb R}z^2\operatorname{sech}^2z\,dz
    =
    2\int_0^\infty z^2\operatorname{sech}^2z\,dz.
\end{equation*}
Since
\begin{equation*}
    \operatorname{sech}^2z
    =
    \frac{d}{dz}\bigl(\operatorname{tanh}z-1\bigr),
\end{equation*}
integration by parts yields
\begin{equation*}
    \begin{aligned}
        \int_{\mathbb R}z^2\operatorname{sech}^2z\,dz
        &=
        2\bigl[z^2(\operatorname{tanh}z-1)\bigr]_0^\infty
        +4\int_0^\infty z(1-\operatorname{tanh}z)\,dz
        \\
        &=
        8\int_0^\infty\frac{z}{e^{2z}+1}\,dz.
    \end{aligned}
\end{equation*}
The boundary term vanishes because
$\operatorname{tanh}z-1$ decays exponentially at infinity.
With $u=2z$, the remaining integral becomes
\begin{equation*}
    \int_{\mathbb R}z^2\operatorname{sech}^2z\,dz
    =
    2\int_0^\infty\frac{u}{e^u+1}\,du
    =
    \frac{\pi^2}{6},
\end{equation*}
where we used the standard definite integral
$\int_0^\infty u/(e^u+1)\,du=\pi^2/12$.

The three required identities are therefore
\begin{equation}
    \begin{aligned}
        \int_{\mathbb R}\operatorname{sech}^2z\,dz
        &=2,
        \\
        \int_{\mathbb R}
        \operatorname{sech}^2z\operatorname{tanh}^2z\,dz
        &=\frac{2}{3},
        \\
        \int_{\mathbb R}z^2\operatorname{sech}^2z\,dz
        &=\frac{\pi^2}{6}.
    \end{aligned}
    \label{app:sech-integrals}
\end{equation}

Returning to the original spatial coordinate makes their role in
the fluctuation calculation explicit. The squared-profile integral is
\begin{equation*}
    \int_{\mathbb R}dx\,|\phi_s(x,t)|^2
    =
    w'\int_{\mathbb R}dz\,\operatorname{sech}^2z
    =
    2w'.
\end{equation*}
The position-weighted integral is
\begin{equation*}
    \begin{aligned}
        \int_{\mathbb R}dx\,
        (x-X_s(t))^2|\phi_s(x,t)|^2
        &=
        \frac{1}{w'}
        \int_{\mathbb R}dz\,z^2\operatorname{sech}^2z
        \\
        &=\frac{\pi^2}{6w'}.
    \end{aligned}
\end{equation*}
Finally, differentiating the modulus of the profile gives
\begin{equation*}
    \partial_x|\phi_s(x,t)|
    =
    -w^{\prime2}\operatorname{sech}z\operatorname{tanh}z,
\end{equation*}
so that
\begin{equation*}
    \begin{aligned}
        \int_{\mathbb R}dx\,
        \bigl(\partial_x|\phi_s(x,t)|\bigr)^2
        &=
        w^{\prime3}\int_{\mathbb R}dz\,
        \operatorname{sech}^2z\operatorname{tanh}^2z
        \\
        &=\frac{2}{3}w^{\prime3}.
    \end{aligned}
\end{equation*}
The derivative here acts on the modulus. In the initial rest frame,
where the soliton profile is chosen real, it coincides with the
profile derivative entering the momentum-fluctuation calculation.
These integrals supply the leading statistical coefficients when
the nonlinear denominators are expanded for $w'\ll1$.

\section[Initial Collective Statistics]{Perturbative Initial Position and Momentum Distributions}
\label{app:initial-statistics}

\subsection{Preparation and the noise convention}

The superposition used in this calculation is an initial condition:
\begin{equation}
 \phi_r(x,t_0)=\phi_s(x-X_s)+\zeta_r(x),\qquad
 \eta_r^{(0)}(x,t_0)=\zeta_r(x).
 \label{app:preparation}
\end{equation}

For a soliton moving with velocity $v$, the field has the form
\begin{equation*}
    \phi_s(x,t)
    =
    |\phi_s(x,t)|\,
    \operatorname{exp}\bigl[i(m'vx+\theta(t))\bigr].
\end{equation*}
A global phase rotation can remove the spatially uniform phase
$\theta(t_0)$, but not the position-dependent term $m'vx$.
The soliton field can therefore be made real everywhere at $t_0$
by such a rotation only in its rest frame.

We consequently perform the initial statistical calculation at
$v=0$ and remove the phase $\theta(t_0)$, so that the unperturbed
profile $\phi_s$ is real. The same rotation is applied to the
background, whose circular Gaussian statistics are unchanged.

This phase choice leaves the collective position unchanged.
Indeed, the charge density
\begin{equation*}
    q[\phi]=\frac{|\phi|^2}{1-|\phi|^2}
\end{equation*}
depends only on the field modulus. The excess charge density
and its normalized first spatial moment are therefore invariant
under the common phase rotation.

The fields $\phi_r(x,t)$ and
$\eta_r^{(0)}(x,t)$ subsequently solve the nonlinear field equation
separately. No decomposition of the evolved $\phi_r$ as a fixed
soliton plus an independently specified noise field is assumed.

Write the two physical quadratures of the preparation noise as
$\zeta_r=\zeta_{r,\mathrm R}+i\zeta_{r,\mathrm I}$. The circular
Gaussian covariance convention used here is
\begin{align*}
 \mathbb E[\zeta_r(x)]&=0,&
 \mathbb E[\zeta_r(x)\zeta_r(x')]&=0,\\
 \mathbb E[\zeta_r(x)\zeta_r^*(x')]&=\epsilon^2\delta(x-x'),&
 \mathbb E[\zeta_{r,\mathrm R}(x)\zeta_{r,\mathrm I}(x')]&=0,\\
 \mathbb E[\zeta_{r,\mathrm R}(x)\zeta_{r,\mathrm R}(x')]
 &=\frac{\epsilon^2}{2}\delta(x-x'),&
 \mathbb E[\zeta_{r,\mathrm I}(x)\zeta_{r,\mathrm I}(x')]
 &=\frac{\epsilon^2}{2}\delta(x-x').
\end{align*}
The delta-correlated covariance is used to evaluate the integrated
linear fluctuations. At the field level, the background is understood
to have a finite spatial resolution, and the expansion is restricted
to weak perturbations within $|\phi|<1$.
\subsection{Expansion of the excess charge and its normalization}

At $t_0$, the exact algebraic background subtraction reads
\begin{equation*}
 \begin{split}
 \Delta q_r
 &=\frac{|\phi_s+\zeta_r|^2}{1-|\phi_s+\zeta_r|^2}
 -\frac{|\zeta_r|^2}{1-|\zeta_r|^2}\\
 &=\frac{|\phi_s|^2+\phi_s^*\zeta_r+\phi_s\zeta_r^*}
 {(1-|\zeta_r|^2)(1-|\phi_s+\zeta_r|^2)}.
 \end{split}
\end{equation*}
Thus the background cancels as a separate contribution, but its
interference with the soliton remains. A direct Taylor expansion gives
\begin{equation}
 \Delta q_r=q[\phi_s]
 +\frac{2\phi_s\zeta_{r,\mathrm R}}{(1-\phi_s^2)^2}
 +O(\epsilon^2).
 \label{app:charge-linear}
\end{equation}

The linear charge fluctuation is
\begin{equation*}
 Q_r^{\mathrm{ex}}=Q_s+\delta Q_r^{(1)}+O(\epsilon^2),\qquad
 \delta Q_r^{(1)}=2\int dx\,
 \frac{\phi_s(x-X_s)\zeta_{r,\mathrm R}(x)}
 {[1-\phi_s^2(x-X_s)]^2}.
\end{equation*}
The center is a ratio. To expand it correctly, first subtract its
unperturbed center before expanding the denominator:
\begin{equation*}
 X_r(t_0)-X_s
 =\frac{\int dx\,(x-X_s)\Delta q_r(x,t_0)}{Q_r^{\mathrm{ex}}}.
\end{equation*}
The zeroth-order numerator is zero by reflection symmetry. Since
\begin{equation*}
 \frac{1}{Q_r^{\mathrm{ex}}}
 =\frac{1}{Q_s}-\frac{\delta Q_r^{(1)}}{Q_s^2}
 +O(\epsilon^2),
\end{equation*}
the fluctuating denominator contributes only at second order to the
center displacement. The first-order result is
\begin{equation}
 \delta X_r^{(1)}=\frac{2}{Q_s}\int dx\,
 \frac{(x-X_s)\phi_s(x-X_s)}{[1-\phi_s^2(x-X_s)]^2}
 \zeta_{r,\mathrm R}(x).
 \label{app:position-linear}
\end{equation}
At the next order the ratio includes
$-\delta Q_r^{(1)}\delta X_r^{(1)}/Q_s$, as well as the
first moment of the second-order density. Thus the normalization
fluctuation was retained in the expansion rather than discarded at
the outset.

The mean of Eq.~\eqref{app:position-linear} vanishes. Contracting its
two noise factors and using the delta covariance gives
\begin{equation}
 \operatorname{Var}(\delta X_r^{(1)})
 =\frac{2\epsilon^2}{Q_s^2}\int dx\,
 \frac{(x-X_s)^2\phi_s^2(x-X_s)}{[1-\phi_s^2(x-X_s)]^4}.
 \label{app:position-variance-finite-width}
\end{equation}
The delta-covariance approximation is valid when the noise
correlation length is much smaller than the soliton width.

\subsection{Expansion of the excess momentum in the rest frame}

The momentum density of a generic complex field is
\begin{equation*}
    p[\phi]
    =
    \frac{i}{2(1-|\phi|^2)}
    \left(
        \phi\,\partial_x\phi^*
        -\phi^*\,\partial_x\phi
    \right).
\end{equation*}
At the preparation time, the complete field and the reference
background satisfy
\begin{equation*}
    \phi_r(x,t_0)=\phi_s(x-X_s)+\zeta_r(x),
    \qquad
    \eta_r^{(0)}(x,t_0)=\zeta_r(x).
\end{equation*}
The corresponding excess momentum is therefore
\begin{equation*}
    P_r^{\mathrm{ex}}(t_0)
    =
    \int_{\mathbb R}dx\,
    \bigl[
        p[\phi_s+\zeta_r]-p[\zeta_r]
    \bigr].
\end{equation*}

Since the unperturbed profile $\phi_s$ is real at $t_0$, we have
\begin{equation*}
    p[\phi_s]
    =
    \frac{i}{2(1-\phi_s^2)}
    \left(
        \phi_s\partial_x\phi_s
        -\phi_s\partial_x\phi_s
    \right)
    =0.
\end{equation*}
Thus, in the rest frame, the momentum density
vanishes everywhere.

Substituting the initial field into the full momentum density gives
\begin{equation*}
    p[\phi_s+\zeta_r]
    =
    \frac{i}{
        2\bigl(1-|\phi_s+\zeta_r|^2\bigr)
    }
    \bigl[
        (\phi_s+\zeta_r)\partial_x(\phi_s+\zeta_r^*)
        -(\phi_s+\zeta_r^*)\partial_x(\phi_s+\zeta_r)
    \bigr].
\end{equation*}
Expanding the products and cancelling the unperturbed terms yields
\begin{equation*}
    \begin{aligned}
        p[\phi_s+\zeta_r]
        =
        \frac{i}{
            2\bigl(1-|\phi_s+\zeta_r|^2\bigr)
        }
        \Big[
            &\phi_s(\partial_x\zeta_r^*-\partial_x\zeta_r)
            +(\zeta_r-\zeta_r^*)\partial_x\phi_s
            \\
            &+\zeta_r\partial_x\zeta_r^*
            -\zeta_r^*\partial_x\zeta_r
        \Big].
    \end{aligned}
\end{equation*}
At fixed spatial regularization, $\zeta_r$ and its spatial derivative
are of order $\epsilon$. The last two terms in the numerator are
therefore of order $\epsilon^2$ and do not contribute to the
first-order momentum.

The denominator must be expanded consistently. Since
\begin{equation*}
    |\phi_s+\zeta_r|^2
    =
    \phi_s^2+\phi_s(\zeta_r+\zeta_r^*)+|\zeta_r|^2,
\end{equation*}
we have
\begin{equation*}
    \frac{1}{1-|\phi_s+\zeta_r|^2}
    =
    \frac{1}{1-\phi_s^2}
    +
    \frac{\phi_s(\zeta_r+\zeta_r^*)}{(1-\phi_s^2)^2}
    +O(\epsilon^2).
\end{equation*}
The numerator has no zeroth-order contribution in the rest frame.
Its product with the linear correction to the denominator is
therefore already of order $\epsilon^2$. Retaining only first-order
terms gives
\begin{equation*}
    p[\phi_s+\zeta_r]
    =
    \frac{i}{2(1-\phi_s^2)}
    \left[
        \phi_s(\partial_x\zeta_r^*-\partial_x\zeta_r)
        +(\zeta_r-\zeta_r^*)\partial_x\phi_s
    \right]
    +O(\epsilon^2).
\end{equation*}
Using $\zeta_r=\zeta_{r,\mathrm R}+i\zeta_{r,\mathrm I}$, this becomes
\begin{equation*}
    p[\phi_s+\zeta_r]
    =
    \frac{
        \phi_s\partial_x\zeta_{r,\mathrm I}
        -(\partial_x\phi_s)\zeta_{r,\mathrm I}
    }{1-\phi_s^2}
    +O(\epsilon^2).
\end{equation*}
The reference-background momentum is itself quadratic in the noise:
\begin{equation*}
    p[\zeta_r]
    =
    \frac{i}{
        2(1-|\zeta_r|^2)
    }
    \left(
        \zeta_r\partial_x\zeta_r^*
        -\zeta_r^*\partial_x\zeta_r
    \right)
    =O(\epsilon^2).
\end{equation*}
Thus the excess momentum at first order is
\begin{equation*}
    P_r^{\mathrm{ex}}(t_0)
    =
    \int_{\mathbb R}dx\,
    \frac{
        \phi_s\partial_x\zeta_{r,\mathrm I}
        -(\partial_x\phi_s)\zeta_{r,\mathrm I}
    }{1-\phi_s^2}
    +O(\epsilon^2).
\end{equation*}

We now define the momentum fluctuation relative to the unperturbed
soliton:
\begin{equation*}
    \delta P_r
    \equiv
    P_r^{\mathrm{ex}}(t_0)-P_s,
    \qquad
    P_s=\int_{\mathbb R}dx\,p[\phi_s]=M_sv.
\end{equation*}
In the frame used here, $v=0$ and hence $P_s=0$. Therefore,
\begin{equation*}
    \boxed{
        P_r^{\mathrm{ex}}(t_0)=\delta P_r.
    }
\end{equation*}
This equality is exact by definition. Its perturbative expansion is
\begin{equation*}
    \delta P_r
    =
    \delta P_r^{(1)}+O(\epsilon^2),
\end{equation*}
where the first-order contribution is
\begin{equation*}
    \delta P_r^{(1)}
    =
    \int_{\mathbb R}dx\,
    \frac{
        \phi_s\partial_x\zeta_{r,\mathrm I}
        -(\partial_x\phi_s)\zeta_{r,\mathrm I}
    }{1-\phi_s^2}.
\end{equation*}

To express this fluctuation directly in terms of the noise quadrature,
we integrate the first term by parts:
\begin{equation*}
    \begin{aligned}
        \delta P_r^{(1)}
        ={}&
        \left[
            \frac{\phi_s\zeta_{r,\mathrm I}}{1-\phi_s^2}
        \right]_{-\infty}^{+\infty}
        \\
        &-
        \int_{\mathbb R}dx\,
        \left[
            \partial_x\left(\frac{\phi_s}{1-\phi_s^2}\right)
            +\frac{\partial_x\phi_s}{1-\phi_s^2}
        \right]\zeta_{r,\mathrm I}.
    \end{aligned}
\end{equation*}
The boundary term vanishes under the localization assumptions.
The derivative inside the integral is
\begin{equation*}
    \partial_x\left(\frac{\phi_s}{1-\phi_s^2}\right)
    =
    \frac{(1+\phi_s^2)\partial_x\phi_s}
    {(1-\phi_s^2)^2},
\end{equation*}
so that
\begin{equation*}
    \begin{aligned}
        \partial_x\left(\frac{\phi_s}{1-\phi_s^2}\right)
        +\frac{\partial_x\phi_s}{1-\phi_s^2}
        &=
        \frac{
            (1+\phi_s^2+1-\phi_s^2)\partial_x\phi_s
        }{(1-\phi_s^2)^2}
        \\
        &=
        \frac{2\partial_x\phi_s}{(1-\phi_s^2)^2}.
    \end{aligned}
\end{equation*}
Restoring the spatial arguments gives
\begin{equation}
    \boxed{
        \delta P_r^{(1)}
        =
        -2\int_{\mathbb R}dx\,
        \frac{\partial_x\phi_s(x-X_s)}
        {[1-\phi_s^2(x-X_s)]^2}
        \zeta_{r,\mathrm I}(x).
    }
    \label{app:momentum-linear}
\end{equation}

Using the derivative of the soliton profile,
\begin{equation*}
    \partial_x\phi_s(x-X_s)
    =
    -w^{\prime2}
    \operatorname{sech}\bigl[w'(x-X_s)\bigr]
    \operatorname{tanh}\bigl[w'(x-X_s)\bigr],
\end{equation*}
the first-order momentum fluctuation becomes
\begin{equation}
    \boxed{
        \delta P_r^{(1)}
        =
        2w^{\prime2}\int_{\mathbb R}dx\,
        \frac{
            \operatorname{sech}\bigl[w'(x-X_s)\bigr]
            \operatorname{tanh}\bigl[w'(x-X_s)\bigr]
        }{
            \bigl[
                1-w^{\prime2}
                \operatorname{sech}^2\bigl[w'(x-X_s)\bigr]
            \bigr]^2
        }
        \zeta_{r,\mathrm I}(x).
    }
    \label{app:momentum-linear_explicit}
\end{equation}

Since the noise is centered,
$\mathbb E[\delta P_r^{(1)}]=0$.
Using its covariance
\begin{equation*}
    \mathbb E[
        \zeta_{r,\mathrm I}(x)\zeta_{r,\mathrm I}(x')
    ]
    =
    \frac{\epsilon^2}{2}\delta(x-x'),
\end{equation*}
the variance follows by squaring the linear fluctuation and averaging:
\begin{equation*}
    \begin{aligned}
        \operatorname{Var}(\delta P_r^{(1)})
        =
        4\int_{\mathbb R}dx\int_{\mathbb R}dx'\,
        &\frac{\partial_x\phi_s(x-X_s)}
        {[1-\phi_s^2(x-X_s)]^2}
        \\
        {}\times&
        \frac{\partial_{x'}\phi_s(x'-X_s)}
        {[1-\phi_s^2(x'-X_s)]^2}
        \frac{\epsilon^2}{2}\delta(x-x').
    \end{aligned}
\end{equation*}
The delta function performs the $x'$ integration, yielding
\begin{equation}
    \boxed{
        \operatorname{Var}(\delta P_r^{(1)})
        =
        2\epsilon^2\int_{\mathbb R}dx\,
        \frac{[\partial_x\phi_s(x-X_s)]^2}
        {[1-\phi_s^2(x-X_s)]^4}.
    }
    \label{app:momentum-variance-finite-width}
\end{equation}
The momentum fluctuation retained here is of order $\epsilon$,
and its variance is of order $\epsilon^2$.
No small-$w'$ approximation has yet been introduced.

\subsection{Small-soliton coefficients and the joint Gaussian}

Now use $w'\ll1$. Equation~\eqref{app:soliton-charge} gives
$Q_s=2w'[1+O(w^{\prime2})]$, and the denominators in
Eqs.~\eqref{app:position-variance-finite-width} and
\eqref{app:momentum-variance-finite-width} differ from one by
$O(w^{\prime2})$. With $z=w'(x-X_s)$, the needed integrals are
\begin{align*}
 \int dx\,(x-X_s)^2\phi_s^2(x-X_s)
 &=\frac{1}{w'}\int dz\,z^2\operatorname{sech}^2z
 =\frac{\pi^2}{6w'},\\
 \int dx\,[\partial_x\phi_s(x-X_s)]^2
 &=w^{\prime3}\int dz\,\operatorname{sech}^2z\operatorname{tanh}^2z
 =\frac23w^{\prime3}.
\end{align*}
Thus, at the retained statistical order,
\begin{equation}
 \boxed{\sigma_X^2=\frac{\pi^2\epsilon^2}{12w^{\prime3}},\qquad
 \sigma_P^2=\frac43w^{\prime3}\epsilon^2.}
 \label{app:initial-variances}
\end{equation}
Each coefficient has relative corrections $O(w^{\prime2})$ already
within the linear-noise calculation. At fixed regularization and
fixed $w'$, the next contributions to the centered variances are of
order $\epsilon^4$: the product of a linear and a quadratic
fluctuation has zero expectation for centered Gaussian noise.
These statements concern two separate expansions and should not be
combined into an exact finite-$w'$ identity.

Position and momentum depend on different independent quadratures.
In particular their linear covariance is
\begin{equation*}
 \begin{split}
 \mathbb E[\delta X_r^{(1)}\delta P_r^{(1)}]
 =-\frac{4}{Q_s}\int dx\,dx'\,
 &\frac{(x-X_s)\phi_s(x-X_s)}{[1-\phi_s^2(x-X_s)]^2}\\
 {}\times&\frac{\partial_{x'}\phi_s(x'-X_s)}{[1-\phi_s^2(x'-X_s)]^2}
 \mathbb E[\zeta_{r,\mathrm R}(x)\zeta_{r,\mathrm I}(x')]=0.
 \end{split}
\end{equation*}
The vanishing covariance implies independence here because the
fluctuations are linear functionals of jointly Gaussian fields.

The uncertainty product at the retained order is
\begin{equation}
 \sigma_X\sigma_P=\frac{\pi\epsilon^2}{3}
 =\frac{\hbar_{\mathrm{eff}}}{2},\qquad
 \hbar_{\mathrm{eff}}=\frac{2\pi\epsilon^2}{3}.
 \label{app:noise-hbar}
\end{equation}

\subsection{Initial Statistics of the Translational Momentum}

The collective dynamics is formulated in terms of the translational
momentum $\Pi_r$, whereas the initial perturbative calculation above
determines the excess field momentum $P_r^{\mathrm{ex}}$.
We now establish their relation at the order considered.
All quantities in the following calculation are evaluated at $t_0$.

The exact momentum--velocity relation derived in the main text is
\begin{equation}
    P_r^{\mathrm{ex}}=M_r^{\mathrm{ex}}v_r,
    \qquad
    M_r^{\mathrm{ex}}=m'Q_r^{\mathrm{ex}},
    \qquad
    v_r\equiv\dot X_r(t_0).
    \label{app:excess-velocity}
\end{equation}
Here $v_r$ is the velocity of the collective center.
The momentum used in the effective description instead employs
the common reference mass $M_s=m'Q_s$:
\begin{equation}
    \Pi_r
    =
    M_sv_r
    =
    \frac{M_s}{M_r^{\mathrm{ex}}}P_r^{\mathrm{ex}}.
    \label{app:translational-momentum}
\end{equation}

The initial charge expansion gives
\begin{equation*}
    Q_r^{\mathrm{ex}}
    =
    Q_s+\delta Q_r^{(1)}+O(\epsilon^2).
\end{equation*}
Multiplication by $m'$ therefore yields
\begin{equation*}
    M_r^{\mathrm{ex}}
    =
    M_s+\delta M_r^{(1)}+O(\epsilon^2),
    \qquad
    \delta M_r^{(1)}=m'\delta Q_r^{(1)}.
\end{equation*}
For the real rest-frame profile used in the initial calculation,
the first-order mass fluctuation is explicitly
\begin{equation*}
    \delta M_r^{(1)}
    =
    2m'\int_{\mathbb R}dx\,
    \frac{\phi_s(x-X_s)}
    {[1-\phi_s^2(x-X_s)]^2}
    \zeta_{r,\mathrm R}(x).
\end{equation*}
Thus the mass fluctuation follows directly from the charge
fluctuation and is linear in the preparation noise.

To determine the corresponding velocity fluctuation, first retain
an arbitrary reference velocity $v_s$ in the kinematic relations.
Define
\begin{equation*}
    \delta M_r
    \equiv M_r^{\mathrm{ex}}-M_s,
    \qquad
    \delta P_r
    \equiv P_r^{\mathrm{ex}}-P_s,
    \qquad
    P_s=M_sv_s.
\end{equation*}
Equation~\eqref{app:excess-velocity} then gives
\begin{equation*}
    v_r
    =
    \frac{M_sv_s+\delta P_r}{M_s+\delta M_r}.
\end{equation*}
Subtracting the unperturbed velocity $v_s$ gives the exact identity
\begin{equation*}
    \begin{aligned}
        \delta v_r
        \equiv v_r-v_s
        &=
        \frac{
            M_sv_s+\delta P_r-v_s(M_s+\delta M_r)
        }{M_s+\delta M_r}
        \\
        &=
        \frac{\delta P_r-v_s\delta M_r}{M_s+\delta M_r}.
    \end{aligned}
\end{equation*}
The decomposition $v_r=v_s+\delta v_r$ therefore introduces no
independent assumption about a velocity noise: its fluctuation
is determined by those of the momentum and mass.

Provided that $|\delta M_r|/M_s\ll1$, the inverse mass expands as
\begin{equation*}
    \frac{1}{M_s+\delta M_r}
    =
    \frac{1}{M_s}
    -
    \frac{\delta M_r^{(1)}}{M_s^2}
    +O(\epsilon^2).
\end{equation*}
Since the numerator of $\delta v_r$ is already first order,
\begin{equation*}
    \delta v_r
    =
    \delta v_r^{(1)}+O(\epsilon^2),
    \qquad
    \delta v_r^{(1)}
    =
    \frac{\delta P_r^{(1)}-v_s\delta M_r^{(1)}}{M_s}.
\end{equation*}
In particular, in the initial rest frame $v_s=0$, substitution of
Eq.~\eqref{app:momentum-linear} gives
\begin{equation*}
    \delta v_r^{(1)}
    =
    \frac{\delta P_r^{(1)}}{M_s}
    =
    -\frac{2}{M_s}\int_{\mathbb R}dx\,
    \frac{\partial_x\phi_s(x-X_s)}
    {[1-\phi_s^2(x-X_s)]^2}
    \zeta_{r,\mathrm I}(x).
\end{equation*}
This expression explicitly establishes that the initial velocity
fluctuation is also linear in the background perturbation.

We can now compare the two momenta. Expanding their exact
definitions gives
\begin{equation*}
    \begin{aligned}
        P_r^{\mathrm{ex}}
        &=
        (M_s+\delta M_r)(v_s+\delta v_r)
        \\
        &=
        M_sv_s+M_s\delta v_r
        +v_s\delta M_r+\delta M_r\delta v_r,
        \\
        \Pi_r
        &=
        M_s(v_s+\delta v_r)
        =
        M_sv_s+M_s\delta v_r.
    \end{aligned}
\end{equation*}
Their difference is therefore
\begin{equation*}
    P_r^{\mathrm{ex}}-\Pi_r
    =
    v_s\delta M_r+\delta M_r\delta v_r.
\end{equation*}
In the rest frame, the term proportional to $v_s$ vanishes.
The remaining product is of order $\epsilon^2$, because both
$\delta M_r$ and $\delta v_r$ start at first order. Consequently,
\begin{equation}
    \boxed{
        P_r^{\mathrm{ex}}(t_0)
        =
        \Pi_r(t_0)+O(\epsilon^2).
    }
    \label{app:initial-momentum-matching}
\end{equation}

Defining the translational momentum fluctuation relative to the
same unperturbed momentum,
\begin{equation*}
    \delta\Pi_r\equiv\Pi_r-P_s,
\end{equation*}
we have $P_s=0$, $P_r^{\mathrm{ex}}=\delta P_r$ and
$\Pi_r=\delta\Pi_r$ in the rest frame. Their first-order terms
therefore coincide:
\begin{equation}
    \boxed{
        \delta\Pi_r^{(1)}=\delta P_r^{(1)}.
    }
    \label{app:initial-momentum-fluctuations}
\end{equation}
The initial variance and the covariance with the position
fluctuation calculated for $\delta P_r^{(1)}$ thus apply directly
to $\delta\Pi_r^{(1)}$:
\begin{equation*}
    \operatorname{Var}(\delta\Pi_r^{(1)})
    =
    \operatorname{Var}(\delta P_r^{(1)}),
    \qquad
    \mathbb E[
        \delta X_r^{(1)}\delta\Pi_r^{(1)}
    ]
    =
    0.
\end{equation*}
Hence the initial perturbative statistics obtained above
provide the statistics of the translational momentum required
for the collective ensemble dynamics.

\chapter{Gaussian Ensemble Dynamics and the Schr\"odinger Equation}
\label{app:gaussian}

Throughout this appendix, $f(X,\Pi,t)$ denotes the phase-space
density of the collective coordinates $(X,\Pi)$, $\rho(X,t)$
its position marginal, and $M_s$ the common translational mass.
For the ensemble considered here, the initial density is
\begin{equation}
    f(X,\Pi,t_0)
    =
    f_0(X,\Pi)
    =
    \frac{1}{2\pi\sigma_X\sigma_\Pi}
    \operatorname{exp}\left[
        -\frac{(X-X_0)^2}{2\sigma_X^2}
        -\frac{(\Pi-\Pi_0)^2}{2\sigma_\Pi^2}
    \right],
    \label{app:initial-joint}
\end{equation}
where $X_0$ and $\Pi_0$ are the initial ensemble means, and
$\sigma_\Pi^2=\sigma_P^2$ at the perturbative order retained
in the initial statistical calculation.

We first derive the Liouville equation and its local moment equations.
We then obtain the evolution of the ensemble means and covariances,
reconstruct the Gaussian density along the characteristics, and
calculate its conditional moments. Finally, these results allow us
to evaluate the full HJ expression and identify the temporal phase
used in the reconstruction of the SE in the main text.

\section{Liouville Dynamics and Local Moment Equations}
\label{app:liouville}

We derive the transport equation and its local moments without
assuming Gaussianity. All integrations by parts below assume
sufficient decay for the boundary terms to vanish.

The collective trajectories satisfy
\begin{equation*}
    \dot X_r=\frac{\Pi_r}{M_s},
    \qquad
    \dot\Pi_r=-U'(X_r).
\end{equation*}
For a smooth function $g(X,\Pi,t)$, its ensemble mean can first
be defined directly as an average over realizations:
\begin{equation*}
    \langle g\rangle(t)
    =
    \mathbb E_r\bigl[
        g(X_r(t),\Pi_r(t),t)
    \bigr].
\end{equation*}
In this representation, the same realizations are followed in time,
with the probabilities assigned at preparation. No phase-space
density appears explicitly in the average. Under the regularity
conditions allowing differentiation under the expectation,
\begin{equation*}
    \frac{d}{dt}\langle g\rangle
    =
    \mathbb E_r\biggl[
        \frac{d}{dt}g(X_r(t),\Pi_r(t),t)
    \biggr].
\end{equation*}
The chain rule along each trajectory gives
\begin{equation*}
    \frac{d}{dt}g(X_r(t),\Pi_r(t),t)
    =
    \partial_tg+\dot X_r\partial_Xg
    +\dot\Pi_r\partial_\Pi g.
\end{equation*}
Substituting the equations of motion and averaging therefore yields
\begin{equation}
    \frac{d}{dt}\langle g\rangle
    =
    \left\langle
        \partial_tg+\frac{\Pi}{M_s}\partial_Xg
        -U'(X)\partial_\Pi g
    \right\rangle.
    \label{app:weak-liouville}
\end{equation}
This step uses the fixed probabilities of the realizations;
it does not assume that the phase-space density is constant
along their trajectories.

The same mean can be represented using the time-dependent density:
\begin{equation*}
    \langle g\rangle(t)
    =
    \int_{\mathbb R^2}dX\,d\Pi\,
    f(X,\Pi,t)g(X,\Pi,t).
\end{equation*}
Differentiating this expression requires differentiating both factors:
\begin{equation*}
    \frac{d}{dt}\langle g\rangle
    =
    \int_{\mathbb R^2}dX\,d\Pi\,
    \left(
        g\,\partial_tf+f\,\partial_tg
    \right).
\end{equation*}
Equating the two expressions for the derivative of the mean,
the terms containing $\partial_tg$ cancel:
\begin{equation*}
    \int_{\mathbb R^2}dX\,d\Pi\,g\,\partial_tf
    =
    \int_{\mathbb R^2}dX\,d\Pi\,f
    \left(
        \frac{\Pi}{M_s}\partial_Xg
        -U'\partial_\Pi g
    \right).
\end{equation*}
Integration by parts, with vanishing boundary terms, gives
\begin{equation*}
    \int_{\mathbb R^2}dX\,d\Pi\,g
    \left[
        \partial_tf
        +\partial_X\left(\frac{\Pi}{M_s}f\right)
        -\partial_\Pi(U'f)
    \right]
    =0.
\end{equation*}
Since this holds for arbitrary smooth test functions $g$, the
quantity in square brackets vanishes. Expanding its spatial and
momentum derivatives gives
\begin{equation*}
    \partial_t f+\frac{\Pi}{M_s}\partial_X f-U'\partial_\Pi f
    +f\left[
        \partial_X\left(\frac{\Pi}{M_s}\right)
        +\partial_\Pi[-U'(X)]
    \right]=0.
\end{equation*}
Here $X$ and $\Pi$ are independent phase-space coordinates.
Thus $\partial_X(\Pi/M_s)=0$ and $\partial_\Pi[-U'(X)]=0$,
so the term in square brackets vanishes. We obtain
\begin{equation}
    \boxed{
        \partial_tf+\frac{\Pi}{M_s}\partial_Xf
        -U'(X)\partial_\Pi f=0.
    }
    \label{app:liouville-equation}
\end{equation}
The chain rule along a collective trajectory now gives
\begin{equation*}
    \frac{d}{dt}f(X_r(t),\Pi_r(t),t)
    =\partial_tf+\dot X_r\partial_Xf+\dot\Pi_r\partial_\Pi f=0.
\end{equation*}
Therefore $f$ is constant along each characteristic, although it
can change at a fixed phase-space point. This is a consequence of
the transport equation, rather than an assumption used to derive
Eq.~\eqref{app:weak-liouville}.

We next derive the local equations in position space.
Define
\begin{equation*}
    \rho(X,t)
    =
    \int_{\mathbb R}d\Pi\,f(X,\Pi,t),
    \qquad
    M_s\rho(X,t)u(X,t)
    =
    \int_{\mathbb R}d\Pi\,\Pi f(X,\Pi,t).
\end{equation*}
Integrating Eq.~\eqref{app:liouville-equation} over momentum gives
\begin{equation*}
    \partial_t\rho
    +\frac{1}{M_s}\partial_X
    \int_{\mathbb R}d\Pi\,\Pi f
    -U'\bigl[f\bigr]_{\Pi=-\infty}^{\Pi=+\infty}
    =0.
\end{equation*}
With sufficient decay in momentum, the boundary term vanishes.
The definition of $u$ then gives the continuity equation
\begin{equation}
    \partial_t\rho+\partial_X(\rho u)=0.
    \label{app:local-continuity}
\end{equation}

Multiplying the Liouville equation by $\Pi$ before integrating
instead yields
\begin{equation*}
    \partial_t(M_s\rho u)
    +\frac{1}{M_s}\partial_X
    \int_{\mathbb R}d\Pi\,\Pi^2f
    -U'\int_{\mathbb R}d\Pi\,\Pi\partial_\Pi f
    =0.
\end{equation*}
The last integral is evaluated by parts:
\begin{equation*}
    \int_{\mathbb R}d\Pi\,\Pi\partial_\Pi f
    =
    \bigl[\Pi f\bigr]_{-\infty}^{+\infty}
    -\int_{\mathbb R}d\Pi\,f
    =
    -\rho.
\end{equation*}
Therefore,
\begin{equation*}
    \partial_t(M_s\rho u)
    +\frac{1}{M_s}\partial_X
    \int_{\mathbb R}d\Pi\,\Pi^2f
    +\rho U'
    =0.
\end{equation*}

At positions where $\rho>0$, the conditional momentum density is
$f(\Pi\mid X,t)=f(X,\Pi,t)/\rho(X,t)$. Its mean is
\begin{equation*}
    \mathbb E[\Pi_r\mid X_r=X]
    =
    \frac{1}{\rho}\int_{\mathbb R}d\Pi\,\Pi f
    =
    M_su.
\end{equation*}
By the definition of conditional variance,
\begin{equation*}
    \mathbb E[\Pi_r^2\mid X_r=X]
    =
    M_s^2u^2
    +\operatorname{Var}(\Pi_r\mid X_r=X).
\end{equation*}
Hence
\begin{equation*}
    \int_{\mathbb R}d\Pi\,\Pi^2f
    =
    \rho M_s^2u^2
    +\rho\operatorname{Var}(\Pi_r\mid X_r=X).
\end{equation*}
Substitution into the momentum equation gives
\begin{equation*}
    \begin{split}
        M_s\left[
            \partial_t(\rho u)+\partial_X(\rho u^2)
        \right]
        +\frac{1}{M_s}\partial_X
        \left[
            \rho\operatorname{Var}(\Pi_r\mid X_r=X)
        \right]
        +\rho U'
        =0.
    \end{split}
\end{equation*}
The product rule gives
\begin{equation*}
    \partial_t(\rho u)+\partial_X(\rho u^2)
    =
    \rho(\partial_tu+u\partial_Xu)
    +u\bigl[\partial_t\rho+\partial_X(\rho u)\bigr].
\end{equation*}
The final bracket vanishes by continuity. Dividing by $M_s\rho$
therefore yields
\begin{equation}
    \boxed{
        \partial_tu+u\partial_Xu
        =
        -\frac{U'}{M_s}
        -\frac{1}{M_s^2\rho}\partial_X
        \left[
            \rho\operatorname{Var}(\Pi_r\mid X_r=X)
        \right].
    }
    \label{app:local-euler}
\end{equation}
These local equations do not require Gaussianity. The conditional
variance describes the spread of momenta among realizations
located at the same position and supplies the pressure contribution
to the equation for the local mean velocity.

\section[Moment Evolution]{Moment Evolution}
\label{app:moments}

From this point onwards, $U(X)$ is at most quadratic, so $U''$
is independent of position. Define
\begin{align*}
    \Sigma_X^2&=\langle X^2\rangle-\langle X\rangle^2,
    &\Sigma_\Pi^2&=\langle\Pi^2\rangle-\langle\Pi\rangle^2,\\
    C_{X\Pi}&=\langle X\Pi\rangle
        -\langle X\rangle\langle\Pi\rangle.
\end{align*}
These definitions and the following moment equations do not
require a Gaussian density.

Taking $g=X$ and $g=\Pi$ in Eq.~\eqref{app:weak-liouville} gives
\begin{equation*}
 \frac{d\langle X\rangle}{dt}=\frac{\langle\Pi\rangle}{M_s},\qquad
 \frac{d\langle\Pi\rangle}{dt}=-\langle U'(X)\rangle.
\end{equation*}
For an at-most-quadratic potential, its Taylor expansion terminates:
\begin{equation*}
    U'(X)
    =
    \left.\frac{dU}{dX}\right|_{X=\langle X\rangle}
    +
    \left.\frac{d^2U}{dX^2}\right|_{X=\langle X\rangle}
    \bigl(X-\langle X\rangle\bigr).
    \label{app:Uexp}
\end{equation*}
The first term is the first derivative evaluated at the mean
position. The second is the product of the second derivative,
evaluated at the same point, and the displacement
$X-\langle X\rangle$. Since the second derivative is constant,
we denote it simply by $U''$. Then 
\begin{equation*}
 U'(X)=U'(\langle X\rangle)+U''(X-\langle X\rangle).
\end{equation*}
The centered term has a zero mean, so
$\langle U'(X)\rangle=U'(\langle X\rangle)$ exactly. This identity
does not require a Gaussian density; its essential hypothesis is the
affine force.

The three raw second-moment equations follow from $g=X^2$, $\Pi^2$
and $X\Pi$:
\begin{align*}
 \frac{d\langle X^2\rangle}{dt}&=\frac{2\langle X\Pi\rangle}{M_s},\\
 \frac{d\langle\Pi^2\rangle}{dt}&=-2\langle\Pi U'(X)\rangle,\\
 \frac{d\langle X\Pi\rangle}{dt}
 &=\frac{\langle\Pi^2\rangle}{M_s}-\langle XU'(X)\rangle.
\end{align*}
Subtracting the derivatives of the products of means gives
\begin{align*}
 \frac{d\Sigma_X^2}{dt}
 &=\frac{2}{M_s}(\langle X\Pi\rangle
 -\langle X\rangle\langle\Pi\rangle),\\
 \frac{d\Sigma_\Pi^2}{dt}
 &=-2\left(\langle\Pi U'(X)\rangle
 -\langle\Pi\rangle\langle U'(X)\rangle\right),\\
 \frac{dC_{X\Pi}}{dt}
 &=\frac{\Sigma_\Pi^2}{M_s}
 -\left(\langle XU'(X)\rangle
 -\langle X\rangle\langle U'(X)\rangle\right).
\end{align*}
Inserting the affine expression for $U'$ now closes the system:
\begin{equation}
 \begin{aligned}
 \frac{d\langle X\rangle}{dt}&=\frac{\langle\Pi\rangle}{M_s},&
 \frac{d\langle\Pi\rangle}{dt}&=-U'(\langle X\rangle),\\
 \frac{d\Sigma_X^2}{dt}&=\frac{2C_{X\Pi}}{M_s},&
 \frac{d\Sigma_\Pi^2}{dt}&=-2U''C_{X\Pi},\\
 \frac{dC_{X\Pi}}{dt}&=\frac{\Sigma_\Pi^2}{M_s}-U''\Sigma_X^2.
 \end{aligned}
 \label{app:closed-moments}
\end{equation}

\section[Gaussian Density and Conditional Moments]{Gaussian Density from the Characteristics and Its Conditional Moments}
\label{app:gaussian-density}

For the trajectory passing through $(X,\Pi)$ at time $t$,
conservation along the characteristics gives
\begin{equation*}
    f(X,\Pi,t)
    =
    f_0\bigl(X_r(t_0),\Pi_r(t_0)\bigr).
\end{equation*}
Here $X_r(t_0)$ and $\Pi_r(t_0)$ are the initial coordinates
obtained by following that trajectory backwards. They are distinct
from the ensemble means $X_0$ and $\Pi_0$. We now express these
initial coordinates in terms of the final ones and use this
relation to determine the evolved density.

Introducing
\begin{equation*}
    \mathbf y=
    \begin{pmatrix}
        X\\
        \Pi
    \end{pmatrix},
    \qquad
    \langle\mathbf y\rangle=
    \begin{pmatrix}
        \langle X\rangle\\
        \langle\Pi\rangle
    \end{pmatrix},
\end{equation*}
Substituting the expansion in Eq.~\eqref{app:Uexp} into
Hamilton's equations and collecting the terms proportional
to $X$ and $\Pi$ gives
\begin{equation*}
    \dot X=\frac{\Pi}{M_s},
    \qquad
    \dot\Pi
    =
    -U''X-U'(\langle X\rangle)+U''\langle X\rangle.
\end{equation*}
we therefore obtain
\begin{equation*}
    \dot{\mathbf y}=B\mathbf y+\mathbf c,
    \qquad
    B=
    \begin{pmatrix}
        0&1/M_s\\
        -U''&0
    \end{pmatrix},
    \qquad
    \mathbf c=
    \begin{pmatrix}
        0\\
        -U'(\langle X\rangle)+U''\langle X\rangle
    \end{pmatrix}.
\end{equation*}
Although written in terms of the mean position, $\mathbf c$
is constant, since
\begin{equation*}
    \frac{d}{dt}
    \bigl[-U'(\langle X\rangle)+U''\langle X\rangle\bigr]
    =
    -U''\frac{d\langle X\rangle}{dt}
    +
    U''\frac{d\langle X\rangle}{dt}
    =
    0.
\end{equation*}

To construct the solution for an arbitrary initial condition,
first let $\mathbf b(t)$ solve
\begin{equation*}
    \dot{\mathbf b}
    =
    B\mathbf b+\mathbf c,
    \qquad
    \mathbf b(t_0)=\mathbf 0.
\end{equation*}
Subtracting this particular solution from $\mathbf y(t)$ gives
\begin{equation*}
    \mathbf h(t)
    :=
    \mathbf y(t)-\mathbf b(t),
    \qquad
    \dot{\mathbf h}=B\mathbf h,
    \qquad
    \mathbf h(t_0)=\mathbf y_0,
\end{equation*}
where
\begin{equation*}
    \mathbf y_0
    =
    \begin{pmatrix}
        X_r(t_0)\\
        \Pi_r(t_0)
    \end{pmatrix}.
\end{equation*}

Let $\mathbf a_1(t)$ and $\mathbf a_2(t)$ be the solutions
of the homogeneous system with initial conditions
\begin{equation*}
    \dot{\mathbf a}_j
    =
    B\mathbf a_j,
    \qquad j=1,2,
\end{equation*}
\begin{equation*}
    \mathbf a_1(t_0)
    =
    \begin{pmatrix}
        1\\
        0
    \end{pmatrix},
    \qquad
    \mathbf a_2(t_0)
    =
    \begin{pmatrix}
        0\\
        1
    \end{pmatrix}.
\end{equation*}
By linearity, the combination
\begin{equation*}
    \mathbf h(t)
    =
    X_r(t_0)\mathbf a_1(t)
    +
    \Pi_r(t_0)\mathbf a_2(t)
\end{equation*}
satisfies the homogeneous equation and has initial value
$\mathbf y_0$. Uniqueness therefore identifies it as the
required solution.

We define $A(t)$ by placing $\mathbf a_1(t)$ and
$\mathbf a_2(t)$ in its first and second columns, respectively:
\begin{equation*}
    A(t)
    =
    \begin{pmatrix}
        \mathbf a_1(t)&\mathbf a_2(t)\\
    \end{pmatrix}.
\end{equation*}
The preceding linear combination is then
$\mathbf h(t)=A(t)\mathbf y_0$, and hence
\begin{equation*}
    \boxed{
        \mathbf y(t)
        =
        A(t)\mathbf y_0+\mathbf b(t).
    }
\end{equation*}
By construction,
\begin{equation*}
    \dot A=BA,
    \qquad
    A(t_0)=I.
\end{equation*}
Both $A(t)$ and $\mathbf b(t)$ depend on the potential and
the elapsed time, but are common to all trajectories.
Only $\mathbf y_0$ varies between realizations.

The matrix $A(t)$ is invertible. Indeed, its determinant obeys
\begin{equation*}
    \frac{d}{dt}\operatorname{det}A
    =
    \operatorname{tr}(B)\operatorname{det}A
    =
    0.
\end{equation*}
Since $A(t_0)=I$, we obtain
\begin{equation*}
    \operatorname{det}A(t)=1.
\end{equation*}
This also expresses preservation of phase-space volume.

Taking the ensemble mean of the affine flow gives
\begin{equation*}
    \langle\mathbf y\rangle
    =
    A\langle\mathbf y_0\rangle+\mathbf b.
\end{equation*}
Subtracting this relation from the trajectory equation
eliminates $\mathbf b$. Define
\begin{equation*}
    \xi=X-\langle X\rangle(t),
    \qquad
    \eta=\Pi-\langle\Pi\rangle(t),
\end{equation*}
and introduce the centered vectors
\begin{equation*}
    \mathbf z
    =
    \begin{pmatrix}
        \xi\\
        \eta
    \end{pmatrix},
    \qquad
    \mathbf z_0
    =
    \begin{pmatrix}
        X_r(t_0)-X_0\\
        \Pi_r(t_0)-\Pi_0
    \end{pmatrix}.
\end{equation*}
Their evolution and its inverse are therefore
\begin{equation*}
    \boxed{
        \mathbf z=A\mathbf z_0,
        \qquad
        \mathbf z_0=A^{-1}\mathbf z.
    }
\end{equation*}

The same matrix determines the evolution of the covariance.
Since
\begin{equation*}
    \mathbf z\mathbf z^{T}
    =
    (A\mathbf z_0)(A\mathbf z_0)^{T}
    =
    A\mathbf z_0\mathbf z_0^{T}A^{T},
\end{equation*}
and $A$ does not depend on the realization, taking the
ensemble mean gives
\begin{equation*}
    \Sigma(t)
    :=
    \langle\mathbf z\mathbf z^{T}\rangle
    =
    A\Sigma_0A^{T}.
\end{equation*}
Here
\begin{equation*}
    \Sigma(t)
    =
    \begin{pmatrix}
        \Sigma_X^2&C_{X\Pi}\\
        C_{X\Pi}&\Sigma_\Pi^2
    \end{pmatrix},
    \qquad
    \Sigma_0
    =
    \begin{pmatrix}
        \sigma_X^2&0\\
        0&\sigma_\Pi^2
    \end{pmatrix}.
\end{equation*}
These are the covariances whose evolution is given in
Eq.~\eqref{app:closed-moments}. Inverting the matrix relation yields
\begin{equation*}
    \Sigma^{-1}
    =
    (A^{-1})^{T}\Sigma_0^{-1}A^{-1}.
\end{equation*}

The initial Gaussian in Eq.~\eqref{app:initial-joint} can be
written as
\begin{equation*}
    f_0\bigl(X_r(t_0),\Pi_r(t_0)\bigr)
    =
    \frac{1}{2\pi\sigma_X\sigma_\Pi}
    \operatorname{exp}\left[
        -\frac12\mathbf z_0^{T}\Sigma_0^{-1}\mathbf z_0
    \right],
\end{equation*}
because
\begin{equation*}
    \mathbf z_0^{T}\Sigma_0^{-1}\mathbf z_0
    =
    \frac{[X_r(t_0)-X_0]^2}{\sigma_X^2}
    +
    \frac{[\Pi_r(t_0)-\Pi_0]^2}{\sigma_\Pi^2}.
\end{equation*}
Substitution of the inverse flow gives
\begin{equation*}
    \begin{aligned}
        \mathbf z_0^{T}\Sigma_0^{-1}\mathbf z_0
        &=
        (A^{-1}\mathbf z)^{T}
        \Sigma_0^{-1}
        (A^{-1}\mathbf z)
        \\
        &=
        \mathbf z^{T}
        (A^{-1})^{T}\Sigma_0^{-1}A^{-1}
        \mathbf z
        \\
        &=
        \mathbf z^{T}\Sigma^{-1}\mathbf z.
    \end{aligned}
\end{equation*}
Thus the initial exponent remains quadratic in the final
coordinates, and the initial Gaussian remains Gaussian.

Define
\begin{equation*}
    \Delta
    :=
    \operatorname{det}\Sigma
    =
    \Sigma_X^2\Sigma_\Pi^2-C_{X\Pi}^2.
\end{equation*}
The inverse covariance matrix is
\begin{equation*}
    \Sigma^{-1}
    =
    \frac{1}{\Delta}
    \begin{pmatrix}
        \Sigma_\Pi^2&-C_{X\Pi}\\
        -C_{X\Pi}&\Sigma_X^2
    \end{pmatrix}.
\end{equation*}
Consequently,
\begin{equation*}
    \begin{aligned}
        \mathbf z^{T}\Sigma^{-1}\mathbf z
        &=
        \frac{1}{\Delta}
        \begin{pmatrix}
            \xi&\eta
        \end{pmatrix}
        \begin{pmatrix}
            \Sigma_\Pi^2\xi-C_{X\Pi}\eta\\
            -C_{X\Pi}\xi+\Sigma_X^2\eta
        \end{pmatrix}
        \\
        &=
        \frac{
            \Sigma_\Pi^2\xi^2
            -2C_{X\Pi}\xi\eta
            +\Sigma_X^2\eta^2
        }{\Delta}.
    \end{aligned}
\end{equation*}
Moreover, $\operatorname{det}A=1$ implies
\begin{equation*}
    \Delta
    =
    \operatorname{det}(A\Sigma_0A^{T})
    =
    (\operatorname{det}A)^2\operatorname{det}\Sigma_0
    =
    \sigma_X^2\sigma_\Pi^2.
    \label{app:conserved-determinant}
\end{equation*}
The characteristic solution therefore becomes
\begin{equation*}
    \boxed{
        f(X,\Pi,t)
        =
        \frac{1}{2\pi\sqrt{\Delta}}
        \operatorname{exp}\left[
            -\frac{
                \Sigma_\Pi^2\xi^2
                -2C_{X\Pi}\xi\eta
                +\Sigma_X^2\eta^2
            }{2\Delta}
        \right].
    }
\end{equation*}
This is the initial density expressed through the inverse flow,
with its coefficients written in terms of the evolved means
and covariances.

To obtain the marginal and conditional moments, complete
the momentum square:
\begin{equation*}
    \begin{aligned}
        &\Sigma_\Pi^2\xi^2
        -2C_{X\Pi}\xi\eta
        +\Sigma_X^2\eta^2
        \\
        &\qquad=
        \Sigma_X^2
        \left(
            \eta-\frac{C_{X\Pi}}{\Sigma_X^2}\xi
        \right)^2
        +
        \frac{\Delta}{\Sigma_X^2}\xi^2.
    \end{aligned}
\end{equation*}
The joint density consequently factorizes as
\begin{equation*}
    \begin{aligned}
        f(X,\Pi,t)
        ={}&
        \frac{1}{\sqrt{2\pi\Sigma_X^2}}
        \operatorname{exp}\left[
            -\frac{\xi^2}{2\Sigma_X^2}
        \right]
        \\
        &\times
        \frac{1}{\sqrt{2\pi\Delta/\Sigma_X^2}}
        \operatorname{exp}\left[
            -\frac{\Sigma_X^2}{2\Delta}
            \left(
                \eta-\frac{C_{X\Pi}}{\Sigma_X^2}\xi
            \right)^2
        \right].
    \end{aligned}
\end{equation*}
At fixed $X$, the second factor is a normalized Gaussian
in $\Pi$, so its integral over momentum equals one.
The first factor is therefore the position marginal:
\begin{equation}
    \rho(X,t)
    =
    \int_{\mathbb R}d\Pi\,f(X,\Pi,t)
    =
    \frac{1}{\sqrt{2\pi\Sigma_X^2(t)}}
    \operatorname{exp}\left[
        -\frac{\xi^2}{2\Sigma_X^2(t)}
    \right].
    \label{app:gaussian-marginal}
\end{equation}

Dividing $f$ by $\rho$ leaves the conditional momentum
density. Its mean and variance follow directly from
the second factor:
\begin{equation}
    M_su(X,t)
    =
    \mathbb E[\Pi_r\mid X_r=X]
    =
    \langle\Pi\rangle
    +
    \frac{C_{X\Pi}}{\Sigma_X^2}\xi,
    \label{app:conditional-mean}
\end{equation}
\begin{equation}
    \operatorname{Var}(\Pi_r\mid X_r=X)
    =
    \frac{\Delta}{\Sigma_X^2}
    =
    \frac{\hbar_{\mathrm{eff}}^2}{4\Sigma_X^2(t)}.
    \label{app:conditional-variance}
\end{equation}
The last equality uses
$\Delta=\sigma_X^2\sigma_\Pi^2=\hbar_{\mathrm{eff}}^2/4$.
The conditional variance is independent of position, although
it can vary with time. These expressions supply the local
moments appearing in Eqs.~\eqref{app:local-continuity}
and \eqref{app:local-euler}.

\section[Hamilton--Jacobi Residual]{The Full Hamilton--Jacobi Residual and the Temporal Phase}
\label{app:hj-residual}

Integrating Eq.~\eqref{app:conditional-mean} with respect to
$X$ at fixed time gives
\begin{equation*}
    S(X,t)=S_{\mathrm{sp}}(X,t)+C(t),
    \qquad
    S_{\mathrm{sp}}(X,t)
    =\langle\Pi\rangle\xi
      +\frac{C_{X\Pi}}{2\Sigma_X^2}\xi^2.
\end{equation*}
The function $C(t)$ is the spatial integration function; it is
distinct from the covariance $C_{X\Pi}(t)$.
For the Gaussian marginal in Eq.~\eqref{app:gaussian-marginal},
\begin{equation*}
    \frac{\partial_X^2\sqrt{\rho}}{\sqrt{\rho}}
    =-\frac{1}{2\Sigma_X^2}
      +\frac{\xi^2}{4(\Sigma_X^2)^2}.
\end{equation*}
Hence the quantum potential used in the main text is
\begin{equation*}
    Q[\rho]=-
    \frac{\hbar_{\mathrm{eff}}^2}{2M_s}
    \frac{\partial_X^2\sqrt{\rho}}{\sqrt{\rho}}
    =\frac{\hbar_{\mathrm{eff}}^2}{4M_s\Sigma_X^2}
     -\frac{\hbar_{\mathrm{eff}}^2\xi^2}
      {8M_s(\Sigma_X^2)^2}.
\end{equation*}
We now evaluate the HJ expression at arbitrary $X$.

Since $\xi=X-\langle X\rangle(t)$, its partial time derivative at
fixed $X$ is $-d\langle X\rangle/dt$. The derivatives of the spatial
primitive are therefore
\begin{align*}
 \partial_XS_{\mathrm{sp}}
 &=\langle\Pi\rangle+\frac{C_{X\Pi}}{\Sigma_X^2}\xi,\\
 \partial_tS_{\mathrm{sp}}
 &=\frac{d\langle\Pi\rangle}{dt}\xi
 -\langle\Pi\rangle\frac{d\langle X\rangle}{dt}
 -\frac{C_{X\Pi}}{\Sigma_X^2}\frac{d\langle X\rangle}{dt}\xi\\
 &\quad+\left[\frac{1}{2\Sigma_X^2}\frac{dC_{X\Pi}}{dt}
 -\frac{C_{X\Pi}}{2(\Sigma_X^2)^2}\frac{d\Sigma_X^2}{dt}\right]\xi^2.
\end{align*}
The kinetic term and the exact quadratic expansion of the potential are
\begin{align*}
 \frac{(\partial_XS_{\mathrm{sp}})^2}{2M_s}
 &=\frac{\langle\Pi\rangle^2}{2M_s}
 +\frac{\langle\Pi\rangle C_{X\Pi}}{M_s\Sigma_X^2}\xi
 +\frac{C_{X\Pi}^2}{2M_s(\Sigma_X^2)^2}\xi^2,\\
 U(X)&=U(\langle X\rangle)+U'(\langle X\rangle)\xi
 +\frac12U''\xi^2.
\end{align*}
We now collect the three powers of $\xi$ in
$\partial_tS_{\mathrm{sp}}+(\partial_XS_{\mathrm{sp}})^2/(2M_s)+U+Q$.
The constant coefficient, using the first mean equation, is
\begin{equation*}
 -\frac{\langle\Pi\rangle^2}{M_s}
 +\frac{\langle\Pi\rangle^2}{2M_s}
 +U(\langle X\rangle)+\frac{\hbar_{\mathrm{eff}}^2}{4M_s\Sigma_X^2}
 =-K(t),
\end{equation*}
where
\begin{equation}
 K(t)=\frac{\langle\Pi\rangle^2}{2M_s}
 -U(\langle X\rangle)-\frac{\hbar_{\mathrm{eff}}^2}{4M_s\Sigma_X^2}.
 \label{app:K}
\end{equation}
The coefficient of $\xi$ is
\begin{equation*}
 \frac{d\langle\Pi\rangle}{dt}+U'(\langle X\rangle)
 +\frac{C_{X\Pi}}{\Sigma_X^2}
 \left(\frac{\langle\Pi\rangle}{M_s}
 -\frac{d\langle X\rangle}{dt}\right)=0.
\end{equation*}
Before simplification, the coefficient of $\xi^2$ is
\begin{equation*}
 \frac{1}{2\Sigma_X^2}\frac{dC_{X\Pi}}{dt}
 -\frac{C_{X\Pi}}{2(\Sigma_X^2)^2}\frac{d\Sigma_X^2}{dt}
 +\frac{C_{X\Pi}^2}{2M_s(\Sigma_X^2)^2}
 +\frac{U''}{2}
 -\frac{\hbar_{\mathrm{eff}}^2}{8M_s(\Sigma_X^2)^2}.
\end{equation*}
Substitution of both covariance equations displays its cancellation:
\begin{equation*}
    \begin{aligned}
        &\frac{\Sigma_\Pi^2}{2M_s\Sigma_X^2}-\frac{U''}{2}
        -\frac{C_{X\Pi}^2}{M_s(\Sigma_X^2)^2}
        +\frac{C_{X\Pi}^2}{2M_s(\Sigma_X^2)^2}\\
        &\qquad+\frac{U''}{2}
        -\frac{\hbar_{\mathrm{eff}}^2}{8M_s(\Sigma_X^2)^2}\\
        &\quad=
        \frac{\Sigma_X^2\Sigma_\Pi^2-C_{X\Pi}^2
            -\hbar_{\mathrm{eff}}^2/4}
        {2M_s(\Sigma_X^2)^2}=0.
    \end{aligned}
\end{equation*}
Thus the disappearance of all spatial dependence follows from the
mean equations and the conserved determinant, not from evaluating
the expression at a specially selected position.

Restoring the temporal integration function gives the full result
\begin{equation}
 \boxed{\partial_tS+\frac{(\partial_XS)^2}{2M_s}+U+Q[\rho]
 =\dot C(t)-K(t)=c(t).}
 \label{app:full-hj-residual}
\end{equation}
It is now legitimate to evaluate this expression at
$X=\langle X\rangle(t)$ as a check. The derivatives must still be
taken first: even though $S_{\mathrm{sp}}$ is zero along that moving
point, its partial time derivative there is
$-\langle\Pi\rangle^2/M_s$, not zero.

Finally, the phase convention used for the SE is obtained by
removing this spatially uniform residual:
\begin{equation*}
    \widetilde S(X,t)
    =S(X,t)-\int_{t_0}^{t}c(t')\,dt'
    =S_{\mathrm{sp}}(X,t)+C(t_0)+\int_{t_0}^{t}K(t')\,dt'.
\end{equation*}
Thus $\partial_X\widetilde S=\partial_XS=M_su$, while
\begin{equation*}
    \partial_t\widetilde S
    +\frac{(\partial_X\widetilde S)^2}{2M_s}
    +U+Q[\rho]=0.
\end{equation*}
If the spatially uniform wave-function phase is written as
$\gamma(t)$, this construction gives
$\hbar_{\mathrm{eff}}\gamma(t)=C(t_0)+\int_{t_0}^{t}K(t')\,dt'$,
and therefore $\hbar_{\mathrm{eff}}\dot\gamma=K$.
The spatial integration alone leaves $C(t)$ undetermined;
the redefinition above selects the phase convention for the
specified Hamiltonian. The constant $C(t_0)$ fixes the initial
global phase.

\newpage
\null
\thispagestyle{empty}
\newpage
\addcontentsline{toc}{chapter}{\bibname}

\backmatter
\bibliographystyle{unsrtnat-initials}
\bibliography{bibliography}

\begin{thebibliography}{32}
\providecommand{\natexlab}[1]{#1}
\providecommand{\url}[1]{\texttt{#1}}
\expandafter\ifx\csname urlstyle\endcsname\relax
  \providecommand{\doi}[1]{doi: #1}\else
  \providecommand{\doi}{doi: \begingroup \urlstyle{rm}\Url}\fi

\bibitem[Manton and Sutcliffe(2004)]{MantonSutcliffe_2004}
N.~Manton and P.~Sutcliffe.
\newblock \emph{Topological Solitons}.
\newblock Cambridge Monographs on Mathematical Physics. Cambridge University
  Press, 2004.
\newblock ISBN 9780521838368.
\newblock \doi{10.1017/CBO9780511617034}.

\bibitem[Weinberg(2012)]{Weinberg_ClassicalSolutions_2012}
E.~J. Weinberg.
\newblock \emph{Classical Solutions in Quantum Field Theory: Solitons and
  Instantons in High Energy Physics}.
\newblock Cambridge Monographs on Mathematical Physics. Cambridge University
  Press, 2012.
\newblock ISBN 9780521114639.
\newblock \doi{10.1017/CBO9781139017787}.

\bibitem[Manton(2008)]{Manton_Particles_2008}
N.~S. Manton.
\newblock Solitons as elementary particles: a paradigm scrutinized.
\newblock \emph{Nonlinearity}, 21\penalty0 (11):\penalty0 T221--T232, 2008.
\newblock \doi{10.1088/0951-7715/21/11/T01}.

\bibitem[Yamamoto(1977)]{Yamamoto_Spinor_1977}
H.~Yamamoto.
\newblock Spinor soliton as an elementary particle.
\newblock \emph{Progress of Theoretical Physics}, 58\penalty0 (3):\penalty0
  1014--1023, 1977.
\newblock \doi{10.1143/PTP.58.1014}.

\bibitem[Dieli et~al.(2026)Dieli, Pierangeli, Baronio, Trillo, and
  Conti]{Dieli_Lump_2026}
L.~Dieli, D.~Pierangeli, F.~Baronio, S.~Trillo, and C.~Conti.
\newblock Observation of lump solitons.
\newblock \emph{Physical Review Letters}, 136\penalty0 (5):\penalty0 053804,
  2026.
\newblock \doi{10.1103/ggbs-y21w}.

\bibitem[Wheeler(1955)]{Wheeler_Geons_1955}
J.~A. Wheeler.
\newblock Geons.
\newblock \emph{Physical Review}, 97\penalty0 (2):\penalty0 511--536, 1955.
\newblock \doi{10.1103/PhysRev.97.511}.

\bibitem[Del~Grosso et~al.(2023)Del~Grosso, Franciolini, Pani, and
  Urbano]{DelGrosso_FermionStars_2023}
L.~Del~Grosso, G.~Franciolini, P.~Pani, and A.~Urbano.
\newblock Fermion soliton stars.
\newblock \emph{Physical Review D}, 108\penalty0 (4):\penalty0 044024, 2023.
\newblock \doi{10.1103/PhysRevD.108.044024}.

\bibitem[Del~Grosso and Pani(2023)]{DelGrossoPani_Asymmetric_2023}
L.~Del~Grosso and P.~Pani.
\newblock Fermion soliton stars with asymmetric vacua.
\newblock \emph{Physical Review D}, 108\penalty0 (6):\penalty0 064042, 2023.
\newblock \doi{10.1103/PhysRevD.108.064042}.

\bibitem[Dieli and Conti(2024)]{DieliConti_DarkMatter_2024}
L.~Dieli and C.~Conti.
\newblock Dark matter condensates as highly nonlocal solitons: instability in
  the {Schwarzschild} metric and laboratory analog.
\newblock \emph{New Journal of Physics}, 26\penalty0 (3):\penalty0 033001,
  2024.
\newblock \doi{10.1088/1367-2630/ad2bb0}.

\bibitem[Valentini(2025{\natexlab{a}})]{Valentini_BeyondQuantum_2025}
A.~Valentini.
\newblock \emph{Beyond the Quantum: A Quest for the Origin and Hidden Meaning
  of Quantum Mechanics}.
\newblock Oxford University Press, 2025{\natexlab{a}}.
\newblock ISBN 9780198853749.
\newblock \doi{10.1093/oso/9780198853749.001.0001}.

\bibitem[Valentini(2025{\natexlab{b}})]{Valentini_Trouble_2025}
A.~Valentini.
\newblock The trouble with pilot-wave theory: a critical evaluation.
\newblock In A.~Oldofredi, editor, \emph{Guiding Waves in Quantum Mechanics:
  One Hundred Years of {de Broglie--Bohm} Pilot-Wave Theory}, pages 67--101.
  Oxford University Press, 2025{\natexlab{b}}.
\newblock \doi{10.1093/oso/9780198901853.003.0003}.
\newblock URL \url{https://arxiv.org/abs/2408.05403}.

\bibitem[Hall and Reginatto(2002)]{Hall_Reginatto}
M.~J.~W. Hall and M.~Reginatto.
\newblock Schr{\"o}dinger equation from an exact uncertainty principle.
\newblock \emph{Journal of Physics A: Mathematical and General}, 35\penalty0
  (14):\penalty0 3289--3307, 2002.
\newblock \doi{10.1088/0305-4470/35/14/310}.

\bibitem[Gomila(2025)]{Gomila_solitons}
D.~Gomila.
\newblock {Solitons, chaos, and quantum phenomena: A deterministic approach to
  the Sch\"odinger equation}.
\newblock \emph{Phys. Rev. Res.}, 7\penalty0 (3):\penalty0 033276, Sep 2025.
\newblock \doi{10.1103/PhysRevResearch.7.033276}.

\bibitem[Weinberg(1989)]{Weinberg1989}
S.~Weinberg.
\newblock Testing quantum mechanics.
\newblock \emph{Annals of Physics}, 194:\penalty0 336--386, 1989.
\newblock \doi{10.1016/0003-4916(89)90276-5}.

\bibitem[Teschl(2014)]{Teschl2014}
G.~Teschl.
\newblock \emph{Mathematical Methods in Quantum Mechanics: With Applications to
  Schr{\"o}dinger Operators}, volume 157 of \emph{Graduate Studies in
  Mathematics}.
\newblock American Mathematical Society, Providence, RI, 2 edition, 2014.
\newblock ISBN 978-1-4704-1704-8.
\newblock URL
  \url{https://www.mat.univie.ac.at/~gerald/ftp/book-schroe/schroe2.pdf}.

\bibitem[Cannas~da Silva(2007)]{CannasDaSilva2007}
A.~Cannas~da Silva.
\newblock Introduction to symplectic and hamiltonian geometry.
\newblock Notes for a short course at IMPA, Rio de Janeiro, 2007.
\newblock URL \url{https://www.math.ist.utl.pt/~acannas/Books/impa.pdf}.
\newblock Course held in February 2002; revised November 2007.

\bibitem[Woit(2017)]{Woit2017}
P.~Woit.
\newblock \emph{Quantum Theory, Groups and Representations: An Introduction}.
\newblock Springer, Cham, 2017.
\newblock \doi{10.1007/978-3-319-64612-1}.
\newblock URL \url{https://doi.org/10.1007/978-3-319-64612-1}.

\bibitem[Peres(1989)]{Peres1989}
A.~Peres.
\newblock {Nonlinear variants of Schrodinger's equation violate the second law
  of thermodynamics}.
\newblock \emph{Phys. Rev. Lett.}, 63:\penalty0 1114--1114, 1989.
\newblock \doi{10.1103/PhysRevLett.63.1114}.

\bibitem[Preskill(2015)]{PreskillNotes}
J.~Preskill.
\newblock Lecture notes for ph219/cs219: Quantum information. chapter 2:
  Foundations i: States and ensembles.
\newblock California Institute of Technology, lecture notes, 2015.
\newblock URL \url{https://www.preskill.caltech.edu/ph219/chap2_15.pdf}.
\newblock Updated July 2015.

\bibitem[Bohm(1952{\natexlab{a}})]{Bohm_I}
D.~Bohm.
\newblock A suggested interpretation of the quantum theory in terms of "hidden"
  variables. i.
\newblock \emph{Phys. Rev.}, 85:\penalty0 166--179, Jan 1952{\natexlab{a}}.
\newblock \doi{10.1103/PhysRev.85.166}.
\newblock URL \url{https://link.aps.org/doi/10.1103/PhysRev.85.166}.

\bibitem[Bohm(1952{\natexlab{b}})]{Bohm_II}
D.~Bohm.
\newblock A suggested interpretation of the quantum theory in terms of "hidden"
  variables. ii.
\newblock \emph{Phys. Rev.}, 85:\penalty0 180--193, Jan 1952{\natexlab{b}}.
\newblock \doi{10.1103/PhysRev.85.180}.
\newblock URL \url{https://link.aps.org/doi/10.1103/PhysRev.85.180}.

\bibitem[Goldstein et~al.(2002)Goldstein, Poole, and
  Safko]{Goldstein_classical}
H.~Goldstein, C.~P. Poole, and J.~L. Safko.
\newblock \emph{Classical Mechanics}.
\newblock Addison-Wesley, 3rd edition, 2002.

\bibitem[Holland(1995)]{Holland_QTM}
P.~R. Holland.
\newblock \emph{The Quantum Theory of Motion: An Account of the de Broglie-Bohm
  Causal Interpretation of Quantum Mechanics}.
\newblock Cambridge University Press, 1995.

\bibitem[Heifetz and Plochotnikov(2020)]{HeifetzPlochotnikov2020}
E.~Heifetz and I.~Plochotnikov.
\newblock Effective classical stochastic theory for quantum tunneling.
\newblock \emph{Physics Letters A}, 384:\penalty0 126511, 2020.
\newblock \doi{10.1016/j.physleta.2020.126511}.

\bibitem[Cover and Thomas(2006)]{CoverThomas2006}
T.~M. Cover and J.~A. Thomas.
\newblock \emph{Elements of Information Theory}.
\newblock John Wiley \& Sons, 2 edition, 2006.

\bibitem[Bowcock et~al.(2009)Bowcock, Foster, and Sutcliffe]{Bowcock_Qballs}
P.~Bowcock, D.~Foster, and P.~Sutcliffe.
\newblock {Q-balls, Integrability and Duality}.
\newblock \emph{J. Phys. A}, 42:\penalty0 085403, 2009.
\newblock \doi{10.1088/1751-8113/42/8/085403}.

\bibitem[de~Melo et~al.(2016)de~Melo, de~Montigny, Pinfold, and
  Tuszynski]{deMelo_GCSGE}
G.~R. de~Melo, M.~de~Montigny, J.~Pinfold, and J.~A. Tuszynski.
\newblock Symmetries and soliton solutions of the galilean complex sine-gordon
  equation.
\newblock \emph{Physics Letters A}, 380\penalty0 (16):\penalty0 1435--1440,
  2016.
\newblock \doi{10.1016/j.physleta.2016.02.023}.

\bibitem[Srednicki(2007)]{Srednicki}
M.~Srednicki.
\newblock \emph{Quantum Field Theory}.
\newblock Cambridge University Press, 2007.
\newblock ISBN 978-0-521-86449-7.
\newblock \doi{10.1017/CBO9780511813917}.

\bibitem[Nelson(1966)]{Nelson_1966}
E.~Nelson.
\newblock Derivation of the {Schr\"odinger} equation from newtonian mechanics.
\newblock \emph{Physical Review}, 150\penalty0 (4):\penalty0 1079--1085, 1966.
\newblock \doi{10.1103/PhysRev.150.1079}.

\bibitem[Tsekov et~al.(2017)Tsekov, Heifetz, and Cohen]{Tsekov_2017}
R.~Tsekov, E.~Heifetz, and E.~Cohen.
\newblock Derivation of the local-mean stochastic quantum force.
\newblock \emph{Fluctuation and Noise Letters}, 16\penalty0 (3):\penalty0
  1750028, 2017.
\newblock \doi{10.1142/S0219477517500286}.

\bibitem[Folli and Conti(2013)]{FolliConti_RandomWalk_2013}
V.~Folli and C.~Conti.
\newblock Random walk of solitary and shock waves in nonlocal disordered media.
\newblock \emph{New Journal of Physics}, 15\penalty0 (8):\penalty0 085026,
  2013.
\newblock \doi{10.1088/1367-2630/15/8/085026}.

\bibitem[Conti(2023)]{Conti_RandomWalk_2023}
C.~Conti.
\newblock Random walk and non-{Gaussianity} of the {3D} second-quantized
  {Schr\"odinger--Newton} nonlocal soliton.
\newblock \emph{New Journal of Physics}, 25\penalty0 (2):\penalty0 023026,
  2023.
\newblock \doi{10.1088/1367-2630/acbb1d}.

\end{thebibliography}

\end{document}